\documentclass[english,11pt,aps,letter,superscriptaddress,nofootinbib,pre]{revtex4}
\usepackage[T1]{fontenc}
\usepackage[latin9]{inputenc}
\usepackage{xcolor}
\usepackage{pdfcolmk}
\usepackage{verbatim}
\usepackage{amsmath}
\usepackage{amssymb}
\usepackage{graphicx}
\PassOptionsToPackage{normalem}{ulem}
\usepackage{ulem}

\makeatletter

\providecolor{lyxadded}{rgb}{1,0,0}
\providecolor{lyxdeleted}{rgb}{0,0,1}

\@ifundefined{textcolor}{}
{%
 \definecolor{BLACK}{gray}{0}
 \definecolor{WHITE}{gray}{1}
 \definecolor{RED}{rgb}{1,0,0}
 \definecolor{GREEN}{rgb}{0,1,0}
 \definecolor{BLUE}{rgb}{0,0,1}
 \definecolor{CYAN}{cmyk}{1,0,0,0}
 \definecolor{MAGENTA}{cmyk}{0,1,0,0}
 \definecolor{YELLOW}{cmyk}{0,0,1,0}
}

\usepackage{ae,aecompl}

\newcommand{\sfrac}[2]{\mathchoice
  {\kern0em\raise.5ex\hbox{\the\scriptfont0 #1}\kern-.15em/
   \kern-.15em\lower.25ex\hbox{\the\scriptfont0 #2}}
  {\kern0em\raise.5ex\hbox{\the\scriptfont0 #1}\kern-.15em/
   \kern-.15em\lower.25ex\hbox{\the\scriptfont0 #2}}
  {\kern0em\raise.5ex\hbox{\the\scriptscriptfont0 #1}\kern-.2em/
   \kern-.15em\lower.25ex\hbox{\the\scriptscriptfont0 #2}}
  {#1\!/#2}}

\makeatletter
\DeclareMathSizes{\@xipt}{10}{6}{5}
\makeatother

\makeatother

\usepackage{babel}
\begin{document}
\global\long\def\V#1{\boldsymbol{#1}}
\global\long\def\M#1{\boldsymbol{#1}}
\global\long\def\Set#1{\mathbb{#1}}

\global\long\def\D#1{\Delta#1}
\global\long\def\d#1{\delta#1}

\global\long\def\norm#1{\left\Vert #1\right\Vert }
\global\long\def\abs#1{\left|#1\right|}

\global\long\def\grad{\M{\nabla}}
\global\long\def\avv#1{\langle#1\rangle}
\global\long\def\av#1{\left\langle #1\right\rangle }

\global\long\def\myhalf{\sfrac{1}{2}}
\global\long\def\mythreehalves{\sfrac{3}{2}}

\title{Low Mach Number Fluctuating Hydrodynamics of Multispecies Liquid
Mixtures}

\author{Aleksandar Donev}

\email{donev@courant.nyu.edu}

\selectlanguage{english}%

\affiliation{Courant Institute of Mathematical Sciences, New York University,
New York, NY 10012}

\author{Andy Nonaka}

\affiliation{Center for Computational Science and Engineering, Lawrence Berkeley
National Laboratory, Berkeley, CA, 94720}

\author{Amit Kumar Bhattacharjee}

\affiliation{Courant Institute of Mathematical Sciences, New York University,
New York, NY 10012}

\author{Alejandro L. Garcia }

\affiliation{Department of Physics and Astronomy, San Jose State University, San
Jose, California, 95192}

\author{John B. Bell}

\affiliation{Center for Computational Science and Engineering, Lawrence Berkeley
National Laboratory, Berkeley, CA, 94720}
\begin{abstract}
We develop a low Mach number formulation of the hydrodynamic equations
describing transport of mass and momentum in a multispecies mixture
of incompressible miscible liquids at specified temperature and pressure
that generalizes our prior work on ideal mixtures of ideal gases {[}\emph{K.
Balakrishnan, A. L. Garcia, A. Donev and J. B. Bell, Phys. Rev. E
89:013017, 2014}{]} and binary liquid mixtures {[}\emph{A. Donev,
A. J. Nonaka, Y. Sun, T. G. Fai, A. L. Garcia and J. B. Bell, CAMCOS,
9-1:47-105, 2014}{]}. In this formulation we combine and extend a
number of existing descriptions of multispecies transport available
in the literature. The formulation applies to non-ideal mixtures of
arbitrary number of species, without the need to single out a ``solvent''
species, and includes contributions to the diffusive mass flux due
to gradients of composition, temperature and pressure. Momentum transport
and advective mass transport are handled using a low Mach number approach
that eliminates fast sound waves (pressure fluctuations) from the
full compressible system of equations and leads to a quasi-incompressible
formulation. Thermal fluctuations are included in our fluctuating
hydrodynamics description following the principles of nonequilibrium
thermodynamics. We extend the semi-implicit staggered-grid finite-volume
numerical method developed in our prior work on binary liquid mixtures
{[}\emph{A. J. Nonaka, Y. Sun, J. B. Bell and A. Donev, 2014, ArXiv:1410.2300}{]},
and use it to study the development of giant nonequilibrium concentration
fluctuations in a ternary mixture subjected to a steady concentration
gradient. We also numerically study the development of diffusion-driven
gravitational instabilities in a ternary mixture, and compare our
numerical results to recent experimental measurements {[}\emph{J.
Carballido-Landeira, P. M.J. Trevelyan, C. Almarcha and A. De Wit,
Physics of Fluids, 25:024107, 2013}{]} in a Hele-Shaw cell. We find
that giant nonequilibrium fluctuations can trigger the instability
but are eventually dominated by the deterministic growth of the unstable
mode, in both quasi two-dimensional (Hele-Shaw), and fully three-dimensional
geometries used in typical shadowgraph experiments.
\end{abstract}
\maketitle

\section{Introduction}

The fluctuating hydrodynamic description of binary mixtures of miscible
fluids is well-known \cite{FluctHydroNonEq_Book,Bell:09}, and has
been used successfully to study long-ranged non-equilibrium correlations
in the fluctuations of concentration and temperature \cite{FluctHydroNonEq_Book}.
Much less is known about mixtures of more than two species (multicomponent
mixtures), both theoretically and experimentally, despite their ubiquity
in nature and technological processes. Part of the difficulty is in
the increased complexity of the formulation of multispecies diffusion
and the increased difficulty of obtaining analytical results, as well
as the far greater complexity of experimentally measuring transport
coefficients in multispecies mixtures. In fact, experimental efforts
to characterize the thermo-physical properties of ternary mixtures
are quite recent and rather incomplete \cite{DCMIX2}.

At the same time, many interesting physical phenomena occur only in
mixtures of more than two species. Examples include diffusion-driven
gravitational instabilities that only occur when there are at least
two distinct diffusion coefficients \cite{DiffusiveInstability_Porous,MixedDiffusiveInstability},
as well as reverse diffusion, in which one of the species in a mixture
of more than two species diffuses in the direction opposite to its
concentration gradient. Another motivation for this work is to extend
our models and numerical studies to chemically-reactive liquid mixtures
\cite{DiffusiveInstability_Chemistry_PRL}; interesting chemical reaction
networks typically involve many more than two species. Giant nonequilibrium
thermal fluctuations \cite{GiantFluctuations_Nature,FluctHydroNonEq_Book,FractalDiffusion_Microgravity}
are expected to exhibit qualitatively new phenomena in multispecies
mixtures due to their coupling with phenomena such as diffusion- and
buoyancy-driven instabilities. Due to the difficulty in obtaining
analytical results in multispecies mixtures, it is important to develop
computational tools for modeling complex flows of multispecies mixtures.
In previous work, we developed a fluctuating hydrodynamics finite-volume
solver for ideal mixtures of ideal gases \cite{MultispeciesCompressible},
and studied giant fluctuations, diffusion-driven instabilities, and
reverse diffusion in gas mixtures. In practice, however, such phenomena
are much more commonly observed and measured experimentally in non-ideal
mixtures of liquids. It is therefore important to develop fluctuating
hydrodynamics codes that take into account the large speed of sound
(small compressibility) of liquids, as well as the non-ideal nature
of most liquid solutions and mixtures. While thermal transport does
of course play a role in liquid mixtures as well, it is often the
case that experimental measurements are made isothermally or in the
presence of steady temperature gradients, or that temperature and
concentration fluctuations essentially decouple from each other \cite{GiantFluctFiniteEffects}.
In this work we extend our work on isothermal low Mach number fluctuating
hydrodynamics for non-ideal binary mixtures of liquids \cite{LowMachExplicit,LowMachImplicit}
to multispecies mixtures.

Transport in multispecies fluid mixtures is a topic of great fundamental
and engineering importance, and has been studied in the field of nonequilibrium
thermodynamics \cite{IrrevThermoBook_Mazur} and chemical engineering
\cite{MulticomponentBook_KT} for a long time. While the case of a
binary fluid mixture is well-understood and the complete hydrodynamic
equations are well-known, including thermal fluctuations \cite{FluctHydroNonEq_Book,Bell:09},
there are some inconsistencies in the literature regarding the treatment
of multispecies diffusion. Here we combine several sources together
to obtain a formulation that is, to our knowledge, new, although all
of the required pieces are known. In particular, our focus here is
on formulating the equations in a way that: (1) fits the framework
of nonequilibrium thermodynamics, notably, the GENERIC framework \cite{OttingerBook};
(2) is amenable to computer simulations for large numbers of species;
(3) allows for straightforward inclusion of thermal fluctuations;
(4) includes all standard mass transport processes (Fickian diffusion,
thermodiffusion, barodiffusion) and applies to non-ideal mixtures
of non-ideal fluids; (5) expresses diffusive fluxes in a barycentric
(center of mass) frame to allow seamless integration with the Navier-Stokes
equations; and (6) is based on the Maxwell-Stefan (rather than Fickian)
formulation of diffusion \cite{IrrevThermoBook_Kuiken,MulticomponentBook_KT,NonEqThermo_Bedeaux}.
These requirements inform our choice of the different elements of
the formulation from different sources.

Our primary source is the recent monograph by Kuiken \cite{IrrevThermoBook_Kuiken},
which contains a nearly complete formulation except for some confusion
between ideal and non-ideal mixtures that we clarify later on. Another
primary source we rely on is the book of Krishna and Taylor \cite{MulticomponentBook_KT}.
This book, like many other sources in chemical engineering, rely on
separating one of the $N$ species as a special ``solvent'' species
in order to eliminate the redundancy in the description. While this
simplifies the analytical formulation to some extent, it breaks the
inherent symmetry of the problem by singling out a species. This complicates
the numerical implementation and does not work well when the reference
species vanishes. Following our prior work on mixtures of ideal gases
\cite{MultispeciesCompressible}, we rely on the monograph by Giovangigli
\cite{MulticomponentBook_Giovangigli} to develop a formulation that
treats all species equally and deals with the redundancy by using
linear algebra techniques, see also a recent mathematical analysis
and generalization to non-ideal gas mixtures \cite{MulticomponentNonideal_GG}.
Here we amend this formulation to account for non-ideality of the
mixture, based in large part on \cite{MulticomponentBook_KT} but
now rewritten in terms of $N$ rather than $N-1$ variables. For thermodiffusion
we rely on the formulation of Kjelstrup \emph{et al.} summarized in
the book \cite{NonEqThermo_Bedeaux}, instead of that of Kuiken, in
order to be in agreement to the standard definition of Soret and thermodiffusion
coefficients for binary mixtures. Thermal fluctuations are formulated
by fitting the formulation into the GENERIC framework, relying closely
on the work of Ottinger \cite{GENERIC_Mixtures}.

In the remained of this section, we will define some notation and
introduce the low Mach number formulation. In section \eqref{sec:Diffusion}
we discuss the formulation of both the deterministic and stochastic
mass fluxes. By combining the proposed formulation of the mass fluxes
with the low Mach number Navier-Stokes equations we obtain a description
of the fluctuating hydrodynamics of quasi-incompressible mixtures
of incompressible miscible liquids. In section \eqref{sec:NumericalAlgorithm},
we introduce and validate a discretization of the resulting system
of equations using finite-volume methods that exactly enforces the
low Mach number quasi-incompressibility constraint \cite{Cahn-Hilliard_QuasiIncomp}.
The method treats viscosity implicitly, allowing us to study flows
over a broad range of Reynolds numbers, including steady Stokes flow.
In section \eqref{sec:Instabilities}, we use our algorithms to numerically
study the development of diffusion-driven gravitational instabilities
in a ternary solution of sugar and salt in water. We find a favorable
comparison between our numerical results and recent experimental measurements
in a Hele-Shaw cell \cite{MixedDiffusiveInstability}. We propose
that shadowgraph measurements in a different geometry may yield additional
information about the possible coupling between the nonlinear instability
and the giant concentration fluctuations that develop due to the presence
of sharp gradients at the fluid-fluid interface.

\subsection{Basic Notation}

Our notation is based closely, though not entirely, on the work of
Kuiken. We avoid the use of molar quantities and instead rely on ``per
molecule'' equivalents, and also prefer the term ``species'' over
``component''. Vectors (both in the geometrical and in the linear
algebra sense), matrices (and tensors) and operators are denoted with
bold letters. A diagonal matrix whose diagonal is given by a vector
is denoted by the corresponding capital letter, for example, $\V X=\text{Diag}\left\{ \V x\right\} $
implies $X_{ij}=x_{i}\delta_{ij}$. The vector of $N$ partial mass
densities is $\rho\V w=\left(\rho_{1},\dots,\rho_{N}\right)$, giving
the total mass density $\rho=\sum_{i=1}^{N}\rho_{i}$. The partial
number densities are denoted with $n_{k}=\rho_{k}/m_{k}$, where $m_{k}$
is the molecular mass of species $k$, with total number density $n=\sum_{i=1}^{N}n_{i}$.

The mass fractions are denoted with $\V w$, $w_{k}=\rho_{k}/\rho$,
while the number or mole fractions are denoted with $\V x$, $x_{k}=n_{k}/n$;
both the mass and number fractions must sum to unity, $\V 1^{T}\V x=\V 1^{T}\V w=1$,
where $\V 1$ denotes a vector of ones. One can transform between
mass and number fractions by
\[
x_{k}=\frac{\bar{m}}{m_{k}}w_{k}=\left(\sum_{i=1}^{N}\frac{w_{i}}{m_{i}}\right)^{-1}\frac{w_{k}}{m_{k}},
\]
where the mixture-averaged molecular mass is 
\[
\bar{m}=\frac{\rho}{n}=\left(\sum_{i=1}^{N}\frac{w_{i}}{m_{i}}\right)^{-1}.
\]
A useful formula that we will use later is the Jacobian of transforming
from mass to number fractions
\begin{equation}
\frac{\partial\V x}{\partial\V w}=\left(\M X-\V x\V x^{T}\right)\M W^{-1},\label{eq:dx_dw}
\end{equation}
where $\V W=\text{Diag}\left\{ \V w\right\} $.

\subsection{Low Mach Number Hydrodynamic Equations}

The hydrodynamics of miscible mixtures of incompressible liquids can
be described using low Mach number equations, as explained in more
detail in our prior works \cite{LowMachExplicit,LowMachImplicit}.
The low Mach number equations can be obtained by making the ansatz
that the thermodynamic behavior of the system is captured by a reference
pressure $P\left(\V r,t\right)=P_{0}\left(\V r\right)$, with the
additional pressure contribution $\pi\left(\V r,t\right)=O\left(\mbox{Ma}^{2}\right)$
capturing the mechanical behavior while not affecting the thermodynamics,
where $\text{Ma}$ is the Mach number. The reference pressure is determined
from the condition of hydrostatic equilibrium in the absence of flow.
In a gravitational field, the reference state is stratified and the
reference pressure is in hydrostatic balance, $\nabla P=\rho\V g$,
where $\V g$ is the gravitational acceleration (see \cite{AnelasticApproximation}
for details of the construction of these types of models). In this
work we will assume that the reference pressure gradients are weak
so that the thermodynamic properties of the system can be evaluated
at a reference pressure $P_{0}$ that does not depend on space and
time.

We will focus here on systems for which the temperature $T\left(\V r,t\right)=T_{0}\left(\V r\right)$
is \emph{specified} and not modeled explicitly. A constant-temperature
model is appropriate, for example, if the system is in contact with
an external heat reservoir and the thermal conductivity is sufficiently
large to ensure a constant temperature (e.g., a constant temperature
gradient) is maintained, and the Dufour effect is negligible. In the
context of fluctuating hydrodynamics, one can often argue that temperature
(more precisely, energy) fluctuations decouple from concentration
fluctuations \cite{FluctHydroNonEq_Book}, and one can model mass
and energy transport separately. It is possible to extend our low
Mach number formulation to include energy transport \cite{LowMachAdaptive},
as we will consider in future work. Here we simply account for mass
transport due to imposed fixed thermodynamic pressure and temperature
gradients in the form of barodiffusion and Soret mass fluxes, but
do not model the evolution of the thermodynamic pressure and temperature
explicitly.

In low Mach number models the total mass density $\rho\left(\V w;T_{0},P_{0}\right)$
is a specified function of the local composition at the given reference
pressure and temperature. Here we consider mixtures of incompressible
liquids that do not change density upon mixing. A straightforward
multispecies generalization of the binary formulation we proposed
in Ref. \cite{LowMachExplicit,LowMachImplicit} is given in Eq. (2.1)
in \cite{CHNS_Ternary}, and takes the form of an equation of state
(EOS) constraint
\begin{equation}
\sum_{i=1}^{N}\frac{\rho_{i}}{\bar{\rho}_{i}}=\rho\sum_{i=1}^{N}\frac{w_{i}}{\bar{\rho}_{i}}=1,\label{eq:EOS}
\end{equation}
where $\bar{\rho}_{i}$ are the pure-component densites, which we
will assume to be specified constants. We note that even if the specific
EOS (\ref{eq:EOS}) is not a very good approximation over the entire
range of concentrations, it may nevertheless be a very good approximation
over the range of concentrations of interest if the densities $\bar{\rho}_{i}$
are adjusted accordingly. In this case, $\bar{\rho}_{i}$ are to be
interpreted not necessarily as pure component densities, since some
of the components may not even exist as fluid phases at the reference
pressure and temperature, but rather, as numerical parameters describing
locally the dependence of the mass density on the composition at the
specified reference pressure and temperature.

The\emph{ low Mach number} equations for the center of mass velocity
$\V v\left(\V r,t\right)$, the mechanical component of the pressure
$\pi\left(\V r,t\right)$, and the partial densities $\left\{ \rho_{1}\left(\V r,t\right),\dots,\rho_{N}\left(\V r,t\right)\right\} $
of a multispecies mixture of $N$ fluids can be written in conservation
form as \cite{LowMachExplicit},
\begin{align}
\partial_{t}\left(\rho\V v\right)+\nabla\pi= & -\grad\cdot\left(\rho\V v\V v^{T}\right)+\grad\cdot\left(\eta\bar{\grad}\V v+\M{\Sigma}\right)+\rho\V g\label{eq:momentum_eq}\\
\partial_{t}\rho_{k}= & -\grad\cdot\left(\rho_{k}\V v\right)-\grad\cdot\V F_{k},\quad k=1,\dots,N\label{eq:rho_part_eq}\\
\grad\cdot\V v= & -\grad\cdot\left(\sum_{i=1}^{N}\frac{\V F_{i}}{\bar{\rho}_{i}}\right),\label{eq:div_v_constraint}
\end{align}
where $\eta\left(\V w;T_{0},P_{0}\right)$ is the viscosity, $\bar{\grad}=\grad+\grad^{T}$
is a symmetric gradient, and $\V g$ is the gravitational acceleration.
Note that the bulk viscosity terms have been absorbed into the pressure
$\pi$ in the low Mach formulation \cite{LowMachExplicit}. Thermal
fluctuations are accounted for through the stochastic momentum flux
$\M{\Sigma}$, formally modeled as \cite{Landau:Fluid,FluctHydroNonEq_Book}
\begin{equation}
\M{\Sigma}=\sqrt{\eta k_{B}T}\left(\M{\mathcal{W}}+\M{\mathcal{W}}^{T}\right)\label{eq:stoch_flux_covariance}
\end{equation}
where $k_{B}$ is Boltzmann's constant and $\M{\mathcal{W}}(\V r,t)$
is a standard white noise Gaussian tensor field with uncorrelated
components,
\[
\avv{\mathcal{W}_{ij}(\V r,t)\mathcal{W}_{kl}(\V r^{\prime},t')}=\delta_{ik}\delta_{jl}\;\delta(t-t^{\prime})\delta(\V r-\V r^{\prime}).
\]

Here $\V F=\left\{ \V F_{1},\dots,\V F_{N}\right\} $ is a composite
vector of diffusive deterministic, $\overline{\V F}$, and stochastic,
$\widetilde{\V F}$, fluxes in the barycentric (center of mass) frame,
where $\V F_{i}=\overline{\V F}_{i}+\widetilde{\V F}_{i}$ is the
flux for species $i$. We will use a compact matrix notation in which
we can write the mass conservation equations without subscripts,
\[
\partial_{t}\left(\rho\V w\right)=-\grad\cdot\left(\rho\V w\V v\right)-\grad\cdot\V F.
\]
The diffusive fluxes preserve mass conservation because they sum to
zero ($\V 1^{T}\V F=0$ in matrix notation), 
\begin{equation}
\sum_{i=1}^{N}\V F_{i}=0,\label{eq:sum_F_i}
\end{equation}
which ensures that the total mass density obeys the usual continuity
equation
\begin{equation}
\partial_{t}\rho=-\grad\cdot\left(\rho\V v\right).\label{eq:rho_eq}
\end{equation}
Differentiating the EOS constraint (\ref{eq:EOS}) in time we get
\[
\sum_{i=1}^{N}\frac{\partial_{t}\left(\rho_{i}\right)}{\bar{\rho}_{i}}=-\sum_{i=1}^{N}\frac{\grad\cdot\V F_{i}}{\bar{\rho}_{i}}-\left(\sum_{i=1}^{N}\frac{\rho_{i}}{\bar{\rho}_{i}}\right)\grad\cdot\V v-\V v\cdot\grad\left(\sum_{i=1}^{N}\frac{\rho_{i}}{\bar{\rho}_{i}}\right)=-\grad\cdot\left(\sum_{i=1}^{N}\frac{\V F_{i}}{\bar{\rho}_{i}}\right)-\grad\cdot\V v=0,
\]
giving the velocity constraint (\ref{eq:div_v_constraint}). Only
if the diffusive fluxes vanish or all of the species have the same
pure densities does one recover the more familiar incompressible flow
limit $\grad\cdot\V v=0$ \cite{Cahn-Hilliard_QuasiIncomp}.

\section{\label{sec:Diffusion}Diffusion}

In this section, we develop a formulation of the diffusive fluxes
in the barycentric frame for a non-ideal mixture, in a manner suitable
for numerical modeling. Nonequilibrium thermodynamics expresses the
deterministic diffusive fluxes in terms of the thermodynamic driving
force (gradients of the chemical potential); in short-hand matrix
notation \cite{MultispeciesCompressible},
\begin{equation}
\overline{\V F}=-\M L\left(\frac{\grad_{T}\V{\mu}}{T}+\V{\xi}\frac{\grad T}{T^{2}}\right),\label{eq:flux_Onsager}
\end{equation}
where $\mu_{k}\left(\V x,T,P\right)$ is the chemical potential of
species $k$, and $\grad_{T}$ denotes a gradient at constant temperature.
It is important to note that we use chemical potential per unit mass
\cite{IrrevThermoBook_Mazur}, which differs from the more commonly
used chemical potential per mole \cite{MulticomponentBook_KT} by
a factor of $m_{k}N_{A}$, where $N_{A}$ is Avogadro's number. The
matrix of Onsager coefficients $\M L$ is symmetric (by Onsager's
reciprocity) and positive-semidefinite (to ensure dissipation, i.e.,
positive entropy production), and has zero row and column sums (to
ensure mass conservation (\ref{eq:sum_F_i})), see (\ref{eq:Onsager_L})
for the explicit form. The vector of thermal diffusion ratios $\V{\xi}$
also sums to zero to ensure mass conservation (\ref{eq:sum_F_i}),
see (\ref{eq:xi_Onsager}) for the explicit form.

In order to make (\ref{eq:flux_Onsager}) suitable for computation,
we need to express the gradients of the chemical potential and the
Onsager matrix in terms of more readily-computable quantities. The
gradient of chemical potential at constant temperature can be expressed
in terms of the gradient of composition and pressure using the chain
rule.
\begin{equation}
\grad_{T}\V{\mu}=\grad_{T,P}\V{\mu}+\left(\frac{\partial\V{\mu}}{\partial P}\right)\grad P=\left(\frac{\partial\V{\mu}}{\partial\V x}\right)\grad\V x+\left(\frac{\partial\V{\mu}}{\partial P}\right)\grad P.\label{eq:gradT_mu}
\end{equation}
We now explain how to relate $\partial\V{\mu}/\partial\V x$, $\partial\V{\mu}/\partial P$
and $\M L$ to more familiar thermodynamic and transport quantities.
This will allow us to express the deterministic component of $\V F$
as a function of the local gradients of composition, temperature and
pressure (see (\ref{eq:diff_fluxes}) for the final result), and will
provide us with a model for the stochastic mass fluxes (see (\ref{eq:species_eq})
for the final result).

\subsection{Chemical potentials}

We use the specific (per mass) Gibbs density $g\left(\V w,T,P\right)=u-Ts+Pv$
as the thermodynamic potential, where $u$ is the specific internal
energy density, $s$ the specific entropy density, and $v=\rho^{-1}$
is the specific volume. The chemical potentials per unit mass are
$\V{\mu}=\partial g/\partial\V w$. For non-ideal mixtures, we can
express chemical potentials as a sum of ideal and excess contributions,
\[
\mu_{k}\left(\V x,T,P\right)=\mu_{k}^{(\text{id})}+\mu_{k}^{(\text{ex})}=\left(\mu_{k}^{0}\left(T,P\right)+\frac{k_{B}T}{m_{k}}\ln\left(x_{k}\right)\right)+\frac{k_{B}T}{m_{k}}\ln\left(\gamma_{k}\right),
\]
where $\mu_{k}^{0}\left(T,P\right)$ is a reference chemical potential
(e.g., pure liquid state at standard conditions), and $\gamma_{k}\left(\V x,T,P\right)$
is the activity coefficient of species $k$; for an ideal mixture
$\gamma_{k}=1$. In the low Mach number setting we consider here,
the chemical potentials depends on pressure \emph{only} through the
reference state. This is always true for an ideal mixture, but may
be assumed more generally so long as the activities only depend on
composition and \emph{not} on pressure.

\subsubsection{Thermodynamic Factors}

Note that all thermodynamic functions are in principle only defined
for valid compositions, $\V 1^{T}\V w=1$, however, any analytical
extension of these functions can be used to work with unconstrained
derivatives instead of the more traditional constrained derivatives.
The matrix of thermodynamic factors $\M{\Gamma}$ is defined via the
\emph{unconstrained} derivatives
\[
\M{\Gamma}=\frac{\bar{m}}{k_{B}T}\M W\left(\frac{\partial\V{\mu}}{\partial\V x}\right)=\M I+\frac{\bar{m}}{k_{B}T}\M W\left(\frac{\partial\V{\mu}^{(\text{ex})}}{\partial\V x}\right),
\]
which we can write in component form as
\begin{equation}
\Gamma_{ij}=\frac{\rho_{i}}{nk_{B}T}\left(\frac{\partial\mu_{i}}{\partial x_{j}}\right)=\delta_{ij}+\frac{x_{i}}{x_{j}}\left(\frac{\partial\ln\gamma_{i}}{\partial\ln x_{j}}\right).\label{eq:Gamma}
\end{equation}
We note that the thermodynamic factors are incorrectly defined in
the book by Kuiken \cite{IrrevThermoBook_Kuiken} to have pressure
$P$ in the denominator instead of $nk_{B}T$ (the two are of course
equal for ideal gases). Using the matrix of thermodynamic factors
we can express the contribution to the gradient of the chemical potentials
in terms of gradients of composition, 
\begin{equation}
\grad_{T,P}\V{\mu}=\left(\frac{\partial\V{\mu}}{\partial\V x}\right)\grad\V x=\frac{k_{B}T}{\bar{m}}\M W^{-1}\M{\Gamma}\grad\V x.\label{eq:grad_mu}
\end{equation}
Note that the constraint $\sum_{i=1}^{N}x_{i}=1$ is automatically
taken into account since $\sum_{i=1}^{N}\grad x_{i}=0$ and the component
of the unconstrained derivatives normal to the constraint does not
actually matter.

For nonideal (dense) gas mixtures it is possible to relate $\M{\Gamma}$
to the equation of state, see \cite{MulticomponentNonideal_GG} for
example calculations for a dense-gas EOS. In order to model the thermodynamic
factors as a function of composition in liquid mixtures, several different
models have been defined and experimentally parameterized, such as
the Wilson, NTLR, or UNIQUAC models, as described in more detail in
Appendix D of the book by Krishna and Taylor \cite{MulticomponentBook_KT}.
These models are all based on the normalized excess Gibbs energy density
per particle $\tilde{g}_{\text{ex}}\left(\V x,T,P\right)$. Converting
this to specific excess Gibbs energy density, we can write
\[
\V{\mu}=\V{\mu}^{(\text{id})}+\frac{k_{B}T}{\bar{m}}\left(\frac{\partial\tilde{g}_{\text{ex}}}{\partial\V w}\right),
\]
giving the thermodynamic factors in the form
\[
\M{\Gamma}=\frac{\bar{m}}{k_{B}T}\M W\left(\frac{\partial\V{\mu}}{\partial\V x}\right)=\frac{\bar{m}}{k_{B}T}\M W\left(\frac{\partial\V{\mu}}{\partial\V w}\right)\left(\frac{\partial\V w}{\partial\V x}\right)=\M I+\M W\left(\frac{\partial^{2}\tilde{g}_{\text{ex}}}{\partial\V w^{2}}\right)\left(\frac{\partial\V w}{\partial\V x}\right).
\]
By converting the second-order derivatives with respect to $\V w$
to the more traditional derivatives with respect to $\V x$ by using
the Jacobian \eqref{eq:dx_dw}, we obtain the final relation
\begin{equation}
\M{\Gamma}=\M I+\left(\M X-\V x\V x^{T}\right)\left(\frac{\partial^{2}g_{\text{ex}}}{\partial\V x^{2}}\right)=\M I+\left(\M X-\V x\V x^{T}\right)\M H,\label{eq:Gamma_F}
\end{equation}
where the symmetric matrix $\M H=\partial^{2}g_{\text{ex}}/\partial\V x^{2}$
is the Hessian of the excess free energy per particle.

In the neighborhood of a stable point (far from phase separation),
the total Gibbs energy density is locally a convex function of composition.
By also including the ideal contribution to the free energy, it is
not hard to show that this implies the stability condition
\begin{equation}
\left(\M X-\V x\V x^{T}\right)+\left(\M X-\V x\V x^{T}\right)\M H\left(\M X-\V x\V x^{T}\right)\succeq\M 0,\label{eq:stability_cond}
\end{equation}
where one of the eigenvalues is always zero and the rest must be non-negative.
The physically key quantity required to model the thermodynamics of
non-ideal mixtures is $\M H$, rather than the more traditional $\M{\Gamma}$.
To avoid a large departure from the literature we continue to use
$\M{\Gamma}$ but we note that in our numerical codes the input is
$\M H$ and $\M{\Gamma}$ is calculated from (\ref{eq:Gamma_F}).
If one tries to model $\M{\Gamma}$ directly, it is difficult to ensure
the correct symmetry structure, which is obscured in $\M{\Gamma}$
but directly evident in $\M H$. We therefore disagree with statements
in the literature that it is more accurate to use models for activities
(first derivatives) than to use models for the excess free energy
and then take second derivatives. While the former may indeed be more
accurate it may also lead to inconsistent thermodynamics; for thermodynamic
consistency one must model the excess free energy as a function of
composition.

\subsubsection{Partial Volumes}

In order to compute \eqref{eq:gradT_mu}, we express the partial volumes
$\V{\theta}=\partial\V{\mu}/\partial P$ by using a Maxwell relation,
\[
\frac{\partial\mu_{k}}{\partial P}=\theta_{k}\left(\V x,T,P\right)=\left(\frac{\partial v}{\partial w_{k}}\right)=-\rho^{-2}\left(\frac{\partial\rho}{\partial w_{k}}\right)_{T,P},
\]
where $v\left(\V w,T,P\right)=\rho^{-1}$ is the specific volume.
For a mixture of incompressible liquids given by the EOS \eqref{eq:EOS},
the above relates $\theta_{k}=\bar{\rho}_{k}^{-1}$ to the pure-component
densities. Instead of partial volumes we will use the volume fractions
$\varphi_{k}=\rho_{k}\theta_{k}$. In ideal gas mixtures $\varphi_{i}=x_{i}$,
and in the low Mach number setting $\varphi_{k}=\rho_{k}/\bar{\rho}_{k}$;
note that $\sum_{i=1}^{N}\varphi_{i}=1$.

\subsection{Diffusion Driving Force}

The thermodynamic driving forces for diffusion are the chemical potential
gradients, $\grad_{T}\V{\mu}$. Note however, that the definition
of the thermodynamic force is not unique, as becomes evident when
we consider the average local entropy production rate due to mass
diffusion,
\[
\frac{ds}{dt}=-\frac{1}{T}\sum_{i=1}^{N}\left(\grad_{T}\mu_{i}-\V a_{i}\right)\cdot\overline{\V F}_{i},
\]
where $\V a_{i}$ is the acceleration of the particles of species
$i$ due to external fields (e.g., gravity or electric fields). Because
the fluxes add to zero, we can add an arbitrary vector $\V{\alpha}$
to all of the chemical potential gradients without changing the entropy
production rate. Let us therefore write the entropy production rate
as
\[
\frac{ds}{dt}=-\frac{1}{T}\sum_{i=1}^{N}\left(\grad_{T}\mu_{i}-\V a_{i}+\V{\alpha}\right)\cdot\overline{\V F}_{i}=-\frac{k_{B}}{\bar{m}}\sum_{i=1}^{N}\frac{\V d_{i}\cdot\overline{\V F}_{i}}{w_{i}},
\]
where the above defines the thermodynamic driving force $\V d_{k}$
for the diffusion of the $k$-th species. Thermodynamic equilibrium
corresponding to a vanishing of the entropy production rate, more
specifically, to a vanishing of both the driving forces and the fluxes,
$\V d_{\text{eq}}=0$ and $\overline{\V F}_{\text{eq}}=0$.

The fluxes above are defined in the barycentric frame. In order to
determine the appropriate value of $\V{\alpha}$, let us consider
transforming to a frame of reference that is moving with velocity
$\V v_{\text{ref}}$ relative to the center of mass of the mixture.
This changes the fluxes to $\V F_{k}\mapsto\V F_{k}-\rho w_{k}\V v_{\text{ref}}$
and changes the entropy production rate by $\left(\rho/T\right)\V v_{\text{ref}}\cdot\left(\sum_{i=1}^{N}\V d_{i}\right)$.
This implies that if we want to have Galilean invariance of the entropy
production rate, we should ensure that the driving forces sum to zero,
$\V 1^{T}\V d=0$. 

If we take a gradient at constant temperature of both sides of the
Gibbs-Duhem relation,
\begin{equation}
-sdT+vdP=\sum_{i=1}^{N}w_{i}d\mu_{i},\label{eq:GibbsDuhem}
\end{equation}
we get the relation
\begin{equation}
\sum_{i=1}^{N}w_{i}\grad_{T}\mu_{i}=v\grad P=\rho^{-1}\grad P.\label{eq:GD_alt}
\end{equation}
which shows that $\V 1^{T}\V d=0$ implies
\[
\V{\alpha}=\sum_{i=1}^{N}w_{i}\V a_{i}-\rho^{-1}\grad P.
\]
In this work we will only consider gravity, for which all species
accelerations are equal to the gravitational acceleration, $\V a_{k}=\sum_{i=1}^{N}w_{i}\V a_{i}=\V g$. 

This leads us to define the diffusion driving force as \cite{IrrevThermoBook_Kuiken,MulticomponentBook_Giovangigli,MulticomponentBook_KT}
\begin{equation}
\V d=\M{\Gamma}\grad\V x+\left(\V{\phi}-\V w\right)\left(\frac{\grad P}{nk_{B}T}\right)\label{eq:d_def}
\end{equation}
Note that Kuiken \cite{IrrevThermoBook_Kuiken} puts pressure $P$
in the denominator of the barodiffusion term instead of $nk_{B}T$,
which leads to an inconsistency with the majority of the literature
and the standard definition of the Maxwell-Stefan (MS) diffusion coefficients
\cite{NonEqThermo_Bedeaux}. Note that each of the two terms in the
driving force separately sums to zero, since $\V 1^{T}\V{\phi}=\V 1^{T}\V w=1$
and 
\[
\V 1^{T}\M{\Gamma}\grad\V x=\V 1^{T}\left(\M X-\V x\V x^{T}\right)\M H\grad\V x=\left(\V x-\V x\right)^{T}\M H\grad\V x=0.
\]

\subsection{Maxwell-Stefan Description of Diffusion}

In order to compute the diffusive fluxes using \eqref{eq:flux_Onsager},
we need to relate the Onsager matrix $\M L$ to the more familiar
Maxwell-Stefan (MS) diffusion coefficients. The Maxwell-Stefan relations
are obtained by equating the driving force to the frictional force
on a species due to the difference in its velocity relative to other
species, 
\begin{equation}
\V d_{i}=\sum_{j\neq i=1}^{N}\frac{x_{i}x_{j}}{D_{ij}}\left(\V v_{i}-\V v_{j}\right),\label{eq:d_SM-K}
\end{equation}
where 
\[
\V v_{k}=\frac{\overline{\V F}_{k}}{\rho_{k}}+D_{k}^{(T)}\frac{\grad T}{T}
\]
is the mass-averaged velocity of species $k$ augmented by the thermodiffusion
``slip''. Here the symmetric matrix of MS \emph{binary} diffusion
coefficients $\M D$ has zero diagonal, $D_{kk}=0$, and the off-diagonal
elements are positive (there may be exceptions to this rule for ionic
solutions \cite{MS_diffusion_ionic}) diffusion coefficients that
have a physical interpretation of suitably dimensionalized inverse
friction coefficients between \emph{pairs} of species. This positivity
of $\M D$ ensures a positive entropy production, and thus consistency
with the second law. It is observed that the MS diffusion coefficients
show less variation with changes in composition than alternatives
such as Fickian diffusion coefficients \cite{IrrevThermoBook_Kuiken,NonEqThermo_Bedeaux}.
The off-diagonal elements of $\M D$ can therefore be interpolated
as a function of composition relatively easily \cite{Diffusion_InfiniteDilution,Darken_MS_diffusion,MS_diffusion_NMR}.
The thermodiffusive fluxes are expressed in terms of the thermodiffusion
coefficients $\V D^{(T)}$. Since only differences of $D_{k}^{(T)}$'s
appear, there are only $N-1$ thermodiffusion coefficients; in order
to ensure that the mass fluxes sum to zero, $\sum_{i=1}^{N}\rho_{i}\V v_{i}=0$,
we require that $\sum_{i=1}^{N}\rho_{i}D_{i}^{(T)}=0$. This gives
the constraint 
\begin{equation}
\sum_{i=1}^{N}w_{i}D_{i}^{(T)}=0,\label{eq:D_T_constraint}
\end{equation}
which removes the redundancy in the specification of the thermodiffusion
coefficients %
\footnote{We thank an anonymous reviewer for pointing out this relation.%
}.

We can write (\ref{eq:d_SM-K}) in matrix form as
\begin{equation}
\V d=-\rho^{-1}\M{\Lambda}\M W^{-1}\overline{\V F}-\frac{\grad T}{T}\V{\zeta},\label{eq:d_def_MS}
\end{equation}
where the symmetric matrix $\M{\Lambda}$ is defined via 
\begin{equation}
\Lambda_{ij}=-\frac{x_{i}x_{j}}{D_{ij}}\mbox{ if }i\neq j,\mbox{ and }\Lambda_{ii}=-\sum_{j\neq i=1}^{N}\Lambda_{ij}.\label{eq:Lambda_ij}
\end{equation}
It is relatively straightforward to show that $\M{\Lambda}$ is positive
semidefinite if $D_{ij}>0$ for $i\neq j$. Here we introduced the
vector of thermal diffusion ratios 
\begin{equation}
\zeta_{i}=-\sum_{j\neq i=1}^{N}\frac{x_{i}x_{j}}{D_{ij}}\left(D_{i}^{(T)}-D_{j}^{(T)}\right),\label{eq:zeta_def}
\end{equation}
where $\sum_{i=1}^{N}\zeta_{i}=0$ by construction. In our algorithm,
the primary input are the MS diffusion coefficients $\M D$ and the
thermodiffusion coefficients $\V D^{(T)}$; $\M{\Lambda}$ and $\V{\zeta}$
are calculated from them.

By combining (\ref{eq:d_def}) and (\ref{eq:d_def_MS}), we obtain
\begin{equation}
-\rho^{-1}\M{\Lambda}\M W^{-1}\overline{\V F}-\frac{\grad T}{T}\V{\zeta}=\M{\Gamma}\grad\V x+\left(\V{\phi}-\V w\right)\left(\frac{\grad P}{nk_{B}T}\right)\label{eq:d_eq}
\end{equation}
 which now relates the deterministic diffusive fluxes $\overline{\V F}$
with the gradients in composition, pressure and temperature. By solving
the above linear system for $\overline{\V F}$ subject to the condition
$\sum_{i=1}^{N}\overline{\V F}_{i}=0$, we can obtain a formula for
the fluxes in terms of gradients of $\V x$, $P$ and $T$. In order
to carry out this computation we follow Giovangigli \cite{MulticomponentBook_Giovangigli},
where linear algebra tools are developed to solve the linear system
\eqref{eq:d_eq}.

\subsection{\label{sub:Fickian}Fick's Law}

Let us introduce %
\footnote{Giovangigli \cite{MulticomponentBook_Giovangigli} attributes the
introduction of $\M{\chi}$ to \cite{DiffusionMatrix_Gases}.%
} the symmetric positive-semidefinite \emph{diffusion matrix} $\M{\chi}$
as a pseudo-inverse of $\M{\Lambda}$ \cite{MulticomponentBook_Giovangigli},
see Appendix \ref{sec:DiffusionMatrix} for more details %
\footnote{Note that the constant $\text{Trace}\left(\M{\Lambda}\right)$ is
arbitrarily chosen.%
},
\begin{equation}
\M{\chi}=\left(\M{\Lambda}+\alpha\,\V w\V w^{T}\right)^{-1}-\alpha^{-1}\,\V 1\V 1^{T},\label{eq:Lambda_to_chi}
\end{equation}
where $\alpha>0$ is an arbitrary constant, for example, the choice
$\alpha=\text{Trace}\left(\M{\Lambda}\right)$ guards against roundoff
errors. One can directly compute $\M{\chi}$ using the above formula,
but we will discuss numerical alternatives in Appendix \ref{sec:DiffusionMatrix}.
While the MS diffusion coefficients are binary friction coefficients,
the matrix $\M{\chi}$ is a multispecies construct that takes into
account the composition of the mixture. One can in fact start the
formulation from the matrix $\M{\chi}$, however, we prefer to use
the more-standard MS coefficients as input and compute $\M{\chi}$
from them. The reason behind this choice is the belief that the MS
diffusion coefficients change more slowly with composition and thus
are easier to tabulate and interpolate, than would be $\M{\chi}$.

It can be shown \cite{MulticomponentBook_Giovangigli} that the solution
of this linear system of equations (\ref{eq:d_eq}) can be written
in the Fickian form
\begin{equation}
\overline{\V F}=-\rho\M W\M{\chi}\left[\M{\Gamma}\grad\V x+\left(\V{\phi}-\V w\right)\frac{\grad P}{nk_{B}T}+\V{\zeta}\frac{\grad T}{T}\right].\label{eq:diff_fluxes}
\end{equation}
This expression will be used in our numerical codes to compute the
fluxes from the gradients in composition, pressure and temperature.
We use gradients of number fractions (composition) rather than gradients
of chemical potential since the later is numerically ill-behaved due
to the logarithmic divergence of the chemical potentials for nearly
vanishing species. It is, however, also possible to isolate the singularity
of the chemical potentials and to use the gradient of the non-singular
part of the chemical potentials directly, instead of using $\M{\Gamma}$
to convert to gradients of composition, as done in Ref. \cite{Flames_Giovangigli}.
Observe that the fluxes automatically add up to zero, $\V 1^{T}\overline{\V F}=0$,
since $\V 1^{T}\M W\M{\chi}=\V w^{T}\M{\chi}=\left(\M{\chi}\V w\right)^{T}=0$.
Note that the expression inside the brackets in (\ref{eq:diff_fluxes})
adds to zero over all species, since $\V 1^{T}\V{\zeta}=0$ and $\V 1^{T}\V d=0$.
It is important to preserve these ``sum to zero'' properties in
spatial discretizations of Fick's law, as we discuss in more detail
in Section \ref{sub:DiffusionNumerics}.

\subsection{Thermal Fluctuations}

To formulate the stochastic mass flux, let us first relate the diffusion
matrix $\M{\chi}$ to the more familiar Onsager matrix. By comparing
the Onsager and the Maxwell-Stefan expressions for the fluxes
\begin{eqnarray*}
\overline{\V F} & = & -\M L\left(\frac{\grad_{T}\V{\mu}}{T}+\V{\xi}\frac{\grad T}{T^{2}}\right)=-\rho\M W\M{\chi}\left(\V d+\V{\zeta}\frac{\grad T}{T}\right)\\
 & = & -\rho\M W\M{\chi}\left(\frac{\bar{m}}{k_{B}T}\M W\grad_{T}\V{\mu}+\V{\zeta}\frac{\grad T}{T}\right),
\end{eqnarray*}
we can directly identify (see (2.17) in \cite{MulticomponentNonideal_GG})
\begin{equation}
\M L=\frac{\bar{m}\rho}{k_{B}}\M W\M{\chi}\M W\quad\mbox{i.e.}\quad L_{ij}=\frac{\bar{m}\rho}{k_{B}}w_{i}w_{j}\chi_{ij}\label{eq:Onsager_L}
\end{equation}
which makes it clear that the Onsager matrix is symmetric positive
semidefinite (SPD) since $\M{\chi}$ is SPD. For the Soret effect,
we can identify $\V{\xi}$ and $\V{\zeta}$ as rescaled versions of
each other
\begin{equation}
\xi_{k}=\frac{k_{B}T}{\bar{m}w_{k}}\zeta_{k}.\label{eq:xi_Onsager}
\end{equation}

The fact $\M L$ is SPD by construction is crucial for adding thermal
fluctuations (stochastic mass fluxes), since that requires the ``square
root'' of the Onsager matrix, notably, a matrix $\M L_{\frac{1}{2}}$
that satisfies $\M L_{\frac{1}{2}}\M L_{\frac{1}{2}}^{\star}=\M L$
where star denotes an adjoint (transpose for real matrices or conjugate
transpose for complex matrices) \cite{OttingerBook,FluctHydroNonEq_Book,MultispeciesCompressible}.
It is easy to see that 
\begin{equation}
\M L_{\frac{1}{2}}=\left(\frac{\bar{m}\rho}{k_{B}}\right)^{\frac{1}{2}}\M W\M{\chi}_{\frac{1}{2}}\label{eq:sqrt_L}
\end{equation}
meets this criterion, where $\M{\chi}_{\frac{1}{2}}\M{\chi}_{\frac{1}{2}}^{\star}=\M{\chi}$;
for example, $\M{\chi}_{\frac{1}{2}}$ can be taken to the lower-triangular
Cholesky factor of $\M{\chi}$. In fluctuating hydrodynamics we simply
add a stochastic contribution to the mass flux of the form
\begin{equation}
\widetilde{\V F}=\sqrt{2k_{B}}\,\M L_{\frac{1}{2}}\V{\mathcal{Z}}=\sqrt{2\bar{m}\rho}\,\M W\M{\chi}_{\frac{1}{2}}\V{\mathcal{Z}},\label{eq:stoch_flux}
\end{equation}
where $\V{\mathcal{Z}}$ denotes a collection of $N$ spatio-temporal
white noise random fields (note that only $N-1$ random fields are
actually required since one of the eigenvalues of $\M L$ is zero),
i.e., a random Gaussian field with correlations,
\[
\av{\mathcal{Z}_{i}\left(\V r,t\right)\mathcal{Z}_{j}\left(\V r^{\prime},t^{\prime}\right)}=\delta_{ij}\delta\left(\V r-\V r^{\prime}\right)\delta\left(t-t^{\prime}\right).
\]
Observe that the stochastic fluxes sum to zero, $\V 1^{T}\widetilde{\V F}=0$
because $\M{\chi}_{\frac{1}{2}}\V w=0$ follows from $\M{\chi}\V w=0$.

We are finally in a position to write the complete equation for the
mass fractions (\ref{eq:rho_part_eq}),
\begin{equation}
\partial_{t}\left(\rho\V w\right)+\grad\cdot\left(\rho\V w\V{\V v}\right)=\grad\cdot\left\{ \rho\M W\left[\M{\chi}\left(\M{\Gamma}\grad\V x+\left(\V{\phi}-\V w\right)\frac{\grad P}{nk_{B}T}+\V{\zeta}\frac{\grad T}{T}\right)+\sqrt{\frac{2}{n}}\,\M{\chi}_{\frac{1}{2}}\V{\mathcal{Z}}\right]\right\} .\label{eq:species_eq}
\end{equation}
In Appendix \ref{sec:MS_Fluct} we demonstrate that the stochastic
mass fluxes \eqref{eq:stoch_flux} can also be derived by following
the Maxwell-Stefan construction and augmenting the dissipative frictional
forces between pairs of species by corresponding (Langevin) fluctuating
forces. That formulation gives another physical interpretation to
the stochastic mass fluxes, but is not as useful for computational
purposes because the number of pairs of species (and thus stochastic
forces) is much larger than the number of species, so in computations
we use \eqref{eq:species_eq}.

Important quantities that can be derived from the fluctuating equation
(\ref{eq:species_eq}) are the spectrum of the time correlation functions
and the amplitude of the fluctuations at thermodynamic equilibrium,
referred to as the dynamic and static structure factors, respectively.
The matrix of equilibrium structure factors can be expressed either
in terms of mass or mole fractions. Here we define the matrix of static
covariances in terms of the fluctuations in the mass fractions $\d{\V w}$
around the equilibrium concentrations $\V w$. The dynamic structure
factor matrix $\M S_{w}\left(\V k,t\right)$ is defined as
\begin{equation}
S_{w}^{\left(i,j\right)}\left(\V k,t\right)=\av{\left(\widehat{\d w}_{i}\left(\V k,t\right)\right)\left(\widehat{\d w}_{j}\left(\V k,0\right)\right)^{\star}},\label{eq:S_k_w}
\end{equation}
where $i$ and $j$ are two species (including $i=j$), $\V k$ is
the wavevector, hat denotes a Fourier transform, and star denotes
a complex conjugate. The equal time covariance in Fourier space is
the static structure factor $\M S_{w}\left(\V k\right)=\av{\left(\widehat{\d{\V w}}\right)\left(\widehat{\d{\V w}}\right)^{\star}}$,
\begin{equation}
S_{w}^{\left(i,j\right)}\left(\V k\right)=S_{w}^{\left(i,j\right)}\left(\V k,t=0\right)=\av{\left(\widehat{\d w}_{i}\right)\left(\widehat{\d w}_{j}\right)^{\star}}.\label{eq:S_k}
\end{equation}

The equilibrium static factors were computed for a ternary mixture
in \cite{TernaryEquilibriumFluct}. In Appendix \ref{sub:S_k} we
use \eqref{eq:species_eq} to obtain the equilibrium static structure
factor for a mixture with an arbitrary number of species,
\begin{equation}
\M S_{w}=\frac{\bar{m}}{\rho}\left(\M W-\V w\V w^{T}\right)\left[\left(\M X-\V x\V x^{T}\right)+\left(\M X-\V x\V x^{T}\right)\M H\left(\M X-\V x\V x^{T}\right)+\V 1\V 1^{T}\right]^{-1}\left(\M W-\V w\V w^{T}\right).\label{eq:S_w}
\end{equation}
If the stability condition (\ref{eq:stability_cond}) is satisfied
then $\M S_{w}\succeq\V 0$ will be symmetric positive semidefinite,
as it must be since it is a covariance matrix. If the mixture is unstable
then the above calculation is invalid because the fluctuations around
the mean will not be small and linearized fluctuating hydrodynamics
will not apply. In the low Mach number setting, the structure factor
for density is
\begin{equation}
S_{\rho}=\av{\left(\widehat{\d{\rho}}\right)\left(\widehat{\d{\rho}}\right)^{\star}}=\rho^{4}\sum_{i,j=1}^{N}\frac{S_{w}^{\left(i,j\right)}}{\bar{\rho}_{i}\bar{\rho}_{j}}.\label{eq:S_rho}
\end{equation}

\section{\label{sec:NumericalAlgorithm}Numerical Algorithm}

In this section we give some details about our numerical algorithms,
and then present some validation studies that verify the deterministic
and stochastic order of accuracy of our schemes. In particular, we
confirm that we can accurately model equilibrium and non-equilibrium
concentration fluctuations in multispecies ternary mixtures. In Section
\ref{sec:Instabilities} we use the algorithms described here to model
the development of instabilities during diffusive mixing in ternary
mixtures.

\subsection{Low Mach Integrator}

The numerical algorithms we use to solve the multispecies low Mach
number equations (\ref{eq:momentum_eq},\ref{eq:div_v_constraint},\ref{eq:species_eq})
are closely based on the binary mixture algorithms described in detail
in Ref. \cite{LowMachImplicit}. In particular, the spatial discretization
of the quasi-incompressible flow EOS constraint (\ref{eq:div_v_constraint})
and the velocity equation (\ref{eq:momentum_eq}), as well as the
temporal integration algorithms, are identical to the binary case
\cite{LowMachImplicit}. Some of the key features of the algorithms
developed in Refs. \cite{LowMachExplicit,LowMachImplicit} are: 
\begin{enumerate}
\item We employ a uniform staggered-grid finite-volume (flux-based) spatial
discretization because of the ease of enforcing the constraint on
the velocity divergence (note that our compressible algorithm \cite{MultispeciesCompressible}
uses a collocated grid) and incorporating thermal fluctuations \cite{LLNS_Staggered}. 
\item Our spatial discretization strictly preserves mass and momentum conservation,
as well as the equation of state (EOS) constraint \cite{LowMachExplicit}
(but see Section \ref{sub:DriftCorrection}).
\item By using the high-resolution Bell-Dawson-Shubin (BDS) scheme \cite{BDS}
for mass advection we can robustly handle the case of no mass diffusion
(no dissipation in (\ref{eq:species_eq})). 
\item Our temporal discretization uses a predictor-corrector integrator
that treats all terms except momentum diffusion (viscosity) explicitly.
We have developed two different temporal integrators, one for the
\emph{inertial} momentum equation (\ref{eq:momentum_eq}), and one
for the viscous-dominated or \emph{overdamped} limit \cite{MultiscaleIntegrators}
in which the velocity equation becomes the steady Stokes equation
\cite{LowMachImplicit}.
\item We treat viscosity implicitly without splitting the pressure update,
relying on a recently-developed variable-coefficient multigrid-preconditioned
Stokes solver \cite{StokesKrylov}. This makes our algorithms efficient
and accurate over a broad range of Reynolds number, including the
zero Reynolds number limit, even in the presence of nontrivial boundary
conditions.
\end{enumerate}
The key difference between binary mixtures \cite{LowMachExplicit,LowMachImplicit}
and multispecies mixtures is the handling of the density equation
\eqref{eq:rho_eq} and the computation of the diffusive and stochastic
mass fluxes. In the binary case, the conserved variables we use are
$\rho$ and $\rho_{1}$, with the corresponding primitive variables
being $\rho$ and the mass fraction $c\equiv w_{1}=\rho_{1}/\rho$.
In the multispecies case, our conserved variables are the partial
densities $\rho_{k}$; the total density $\rho=\sum_{i=1}^{N}\rho_{i}$
is computed from those as needed. The corresponding primitive variables
are $\rho$ and $w_{k}$. In the binary case we expressed all of the
diffusive fluxes in terms of gradients of mass fractions, but in the
multispecies case we rely on the more traditional formulation in terms
of gradients of number (mole) fractions, and we also include $x_{k}$
as primitive variables. Further details on the computation of the
multispecies diffusive and stochastic mass fluxes are given in Section
\ref{sub:DiffusionNumerics}.

\subsubsection{\label{sub:DriftCorrection}EOS drift}

Our low Mach number algorithms are specifically designed to ensure
that the evolution remains on the EOS constraint, i.e., that the partial
densities or equivalently the density and the composition in each
grid cell strictly satisfy (\ref{eq:EOS}) \cite{LowMachExplicit}.
Nevertheless, due to roundoff error and finite solver tolerance in
the fluid Stokes solver, a slow drift off the EOS constraint occurs
over multiple time steps. To correct this, we occasionally need to
project the state (partial densities) back onto the constraint \cite{LowMachExplicit}.
A similar projection onto the EOS is required in the BDS advection
scheme for average states extrapolated to the faces of the grid \cite{LowMachImplicit}.

For binary mixtures, we used a simple $L_{2}$ projection onto the
EOS. For mixtures of many species, some of the species may be trace
species or not present at all, and in this case it seems more appropriate
to use a mass-fraction-weighted $L_{2}$ projection step. Given a
state $\left(\rho,\V w\right)$ that does not necessarily obey (\ref{eq:EOS}),
the weighted $L_{2}$ projection consists of correcting $\rho_{k}$
as follows,
\[
\rho_{k}\leftarrow\rho_{k}-\D{\rho}_{k},
\]
where the correction is
\[
\D{\rho}_{k}=\frac{w_{k}}{\bar{\rho}_{k}}\left(\sum_{i=1}^{N}\frac{w_{i}}{\bar{\rho}_{i}^{2}}\right)^{-1}\left(\sum_{j=1}^{N}\frac{\rho_{j}}{\bar{\rho}_{j}}-1\right),
\]
which vanishes for species not present ($w_{k}=0$). When performing
a global projection onto the EOS one should additionally re-distribute
the total change in the mass of species $k$ over all of the grid
cells to ensure that the projection step does not change the total
mass of any species \cite{LowMachExplicit}.

\subsection{\label{sub:DiffusionNumerics}Diffusive and stochastic mass fluxes}

The computation of the diffusive deterministic and stochastic mass
fluxes for binary mixtures is described in detail in Ref. \cite{LowMachExplicit}.
We follow a similar but slightly different procedure for multispecies
mixtures, primarily guided by the desire to make the algorithm efficient
for mixtures of many species.

\subsubsection{Mixture Model}

The user input to our fluid dynamics code, i.e., the \emph{mixture
model}, is a specification of the required thermodynamic (e.g., non-ideality
factors) and transport properties (e.g., shear viscosity) of the mixture
as a function of state. The state of the mixture is described by the
variables $\left(\V w,P,T\right)$, or, equivalently, $\left(\V x,P,T\right)$,
where we recall that in our low Mach number model the pressure and
temperature are specified and not modeled explicitly, and the density
is not an independent variable since it is determined from the EOS
constraint \eqref{eq:EOS}. Therefore, the mixture model in our low
Mach code consists of specifying the thermodynamic and transport properties
as a function of the composition $\V w$.

In the multispecies case, the mixture model requires specifying binary
Maxwell-Stefan diffusion coefficients for each pair of species, i.e.,
the lower triangle of the matrix $\M D$. Additional input is the
vector of thermodiffusion coefficients $\V D^{(T)}$ (recall that
only $N-1$ of these are independent since an arbitrary constant can
be added to this vector), and the Hessian of the excess free energy
per particle $\M H$. MS diffusion coefficients can be interpolated
as a function of composition using Vignes- or Darken-type formulas
\cite{Diffusion_InfiniteDilution,Darken_MS_diffusion,MS_diffusion_NMR},
based on data obtained experimentally \cite{DCMIX2} or from molecular
dynamics simulations \cite{Ternary_MD,Ternary_MD_correction}. The
thermodynamics can be parametrized using Wilson, NTLR, or UNIQUAC
models, and Hessian matrices $\M H$ can be computed from the formulas
presented in Appendix D of the book by Krishna and Taylor \cite{MulticomponentBook_KT},
based on experimental or molecular dynamics data \cite{ThermodynamicsFluctuations_MD}.
We are not aware of any models for parameterizing the thermodiffusion
coefficients as a function of composition in liquid mixtures. We note,
however, that despite the availability of various mixture models,
experimental efforts to obtain the parameters required in these models
and compare various models are very recent. We are not aware of any
mixture of more than two species for which there is reliable and reproducible
data for the mass and thermal diffusion and thermodynamic coefficients
even in the vicinity of a reference state, yet alone over a broad
range of compositions.

From the mixture model input, i.e., $\eta$, $\M D$, $\V D^{(T)}$
and $\M H$ , we compute the following quantities. First, we obtain
the matrix $\M{\Lambda}$ using \eqref{eq:Lambda_ij}, and then from
$\M{\Lambda}$ we compute the diffusion matrix $\M{\chi}$ using \eqref{eq:chi_iterative},
as discussed in more detail in Appendix \ref{sec:DiffusionMatrix}.
We also compute the matrix of thermodynamic factors $\M{\Gamma}$
using \eqref{eq:Gamma_F}, as well as the vector of thermal diffusion
ratios $\V{\zeta}$ using \eqref{eq:zeta_def}. These computations
provide all of the matrices and vectors required to compute the non-advective
mass fluxes in \eqref{eq:species_eq}. We remind the reader that the
species volume fractions $\V{\phi}$ are easily computable for our
model of a mixture of incompressible components, $\varphi_{k}=\rho_{k}\theta_{k}=\rho_{k}/\bar{\rho}_{k}$.

\subsubsection{Spatial Discretization}

The basic spatial discretization of the fluid equations and mass advection
is unchanged from our previous work on binary mixtures \cite{LowMachExplicit,LowMachImplicit}
and we do not discuss it further here. Here we explain how we handle
the diffusive and stochastic mass fluxes in the multispecies setting.
The deterministic and stochastic mass fluxes are computed on the faces
of the grid, and the divergence of the flux is computing using a conservative
difference, in two dimensions
\begin{equation}
(\grad\cdot\V F)_{i,j}=\Delta x^{-1}\left[\V F_{i+\myhalf,j}^{(x)}-\V F_{i-\myhalf,j}^{(x)}\right]+\Delta y^{-1}\left[\V F_{i,j+\myhalf}^{(y)}-\V F_{i,j-\myhalf}^{(y)}\right],\label{eq:discrete_diffusion}
\end{equation}
where $\V F=\overline{\V F}+\widetilde{\V F}$.

Our spatial discretization of the deterministic diffusive fluxes \eqref{eq:diff_fluxes}
closely mimics the one described in Section IV.A of \cite{LowMachExplicit},
and is based on centered differences and centered averaging. In order
to avoid division by zero in the absence of certain species in some
parts of the domain, in each cell (i.e., for each cell center) we
modify the densities $\rho_{k}\leftarrow\max\left(\epsilon,\,\rho_{k}\right)$
to be no-smaller than a small constant $\epsilon$ on the order of
the roundoff tolerance; this modification is only done for the purpose
of the diffusive flux calculation. In each cell, we compute the primitive
variables $\rho$, $\V w$ and $\V x$ and then use the user-provided
mixture model to compute $\M{\Gamma}$ (if the mixture is non-ideal),
$\M{\chi}$, $\V{\zeta}$ and $\V{\phi}$. Next, in each cell $(i,j)$,
we compute the matrices $\rho\M W\M{\chi}$ and $\M{\Gamma}$ and
the vectors $\V{\zeta}/T$, and $\left(\V{\phi}-\V w\right)/\left(nk_{B}T\right)$.
Then, we average (interpolate) these matrices and vectors to the faces
of the grid using arithmetic averaging, for example,
\[
\left(\rho\M W\M{\chi}\right)_{i+\myhalf,j}=\frac{\left(\rho\M W\M{\chi}\right)_{i,j}+\left(\rho\M W\M{\chi}\right)_{i+1,j}}{2},
\]
and compute gradients of composition, pressure, and temperature using
centered differences, for example,
\[
\left(\grad\V x\right)_{i+\myhalf,j}^{(x)}=\frac{\V x_{i+1,j}-\V x_{i,j}}{\Delta x}.
\]
Note that the key property $\sum_{i=0}^{N}\grad x_{i}=\V 0$ is preserved,
for example, 
\[
\V 1^{T}\left(\grad\V x\right)_{i+\myhalf,j}^{(x)}=\frac{\V 1^{T}\V x_{i+1,j}-\V 1^{T}\V x_{i,j}}{\Delta x}=0.
\]
Finally, we compute the deterministic fluxes using \eqref{eq:diff_fluxes}
by a matrix-vector product, for example, 
\[
\left(\rho\M W\M{\chi}\M{\Gamma}\grad\V x\right)_{i+\myhalf,j}^{(x)}=\left(\rho\M W\M{\chi}\right)_{i+\myhalf,j}\M{\Gamma}_{i+\myhalf,j}\left(\grad\V x\right)_{i+\myhalf,j}^{(x)}.
\]

Note that the important properties of \eqref{eq:diff_fluxes} discussed
in Section \ref{sub:Fickian} are maintained by this discretization.
To ensure mass conservation, it is crucial that the mass fluxes for
different species add up to zero, for example, it must be that $\V 1^{T}\V F_{i+\myhalf,j}^{(x)}=0$
on every $x$ face of the grid. In the continuum formulation this
is true because $\V 1^{T}\M W\M{\chi}=\V w^{T}\M{\chi}=0$; it is
not hard to show that the arithmetic averaging procedure used above
preserves this property,
\[
\V 1^{T}\left(\rho\M W\M{\chi}\right)_{i+\myhalf,j}^{(x)}=\frac{1}{2}\left(\V 1^{T}\left(\rho\M W\M{\chi}\right)_{i,j}+\V 1^{T}\left(\rho\M W\M{\chi}\right)_{i+1,j}\right)=\frac{1}{2}\left(\rho_{i,j}\M w_{i,j}^{T}\M{\chi}_{i,j}+\rho_{i+1,j}\M w_{i+1,j}^{T}\M{\chi}_{i+1,j}\right)=0,
\]
since $\M{\chi}\V w=0$ in each cell. Similarly, the continuum properties
$\V 1^{T}\V{\zeta}=0$, $\V 1^{T}\M{\Gamma}\grad\V x=0$ and $\V 1^{T}\left(\V{\phi}-\V w\right)/\left(nk_{B}T\right)=0$
are preserved discretely due to their linearity and the linearity
of the averaging process. This shows the importance of the linearity
of the interpolation from the cell centers to the cell faces. 

Upon spatial discretization, the stochastic fluxes acquire a prefactor
of $\D V^{-\myhalf}$ due to the delta function correlation of white-noise,
where $\D V$ is the volume of a grid cell \cite{DFDB}. This converts
the spatio-temporal white-noise process $\V{\mathcal{Z}}\left(\V r,t\right)$
into a collection of independent temporal white-noise processes $\V{\mathcal{Y}}\left(t\right)$,
one process for each face of the grid, for example,
\[
\left(\widetilde{\V F}\right)_{i+\myhalf,j}^{(x)}=\sqrt{\frac{2k_{B}}{\D V}}\,\left(\M L_{\frac{1}{2}}\right)_{i+\myhalf,j}^{(x)}\V{\mathcal{Y}}_{i+\myhalf,j}^{(x)}.
\]
In our code, we compute the Onsager matrix $\M L$ in every cell and
then compute $\M L_{\frac{1}{2}}$ by Cholesky factorization; an equally
good alternative is to compute $\M{\chi}_{\frac{1}{2}}$ by Cholesky
factorization %
\footnote{Note that it is straightfoward to modify the standard Cholesky factorization
algorithm to work for semi-definite matrices by simply avoiding division
by zero pivot entries; the factorization process remains numerically
stable and works even when some of the species vanish.%
}. We then use arithmetic averaging to compute face-centered Cholesky
factors $\M L_{\frac{1}{2}}$. An alternative procedure, which is
likely better at maintaining discrete fluctuation-dissipation balance
\cite{LLNS_S_k} but is the number of dimensions times more expensive,
is to average $\M L$ to the faces, and then perform a Cholesky factorization
on each face of the grid. Note, however, that achieving strict discrete
fluctuation-dissipation balance requires expressing the fluxes in
terms of the discrete gradients of the chemical potential, which is
rather inconvenient and not numerically well-behaved. In this work
we chose to work with gradients of number fractions and thus only
achieve discrete fluctuation-dissipation balance approximately.

\subsection{Numerical Tests}

In the deterministic setting, we have confirmed the second-order accuracy
of our numerical method by repeating the lid-driven cavity test used
in our previous work on binary mixtures \cite{LowMachImplicit}. The
essential difference is that the bubble being advected through a pure
liquid of a first species in the lid-driven cavity is now composed
of a mixture of two other species, making this a ternary mixture test.
Our numerical results show little to no difference between the ternary
and binary mixture cases, and show second-order pointwise deterministic
convergence for our low Mach number scheme.

In this section we focus on tests in the context of fluctuating hydrodynamics,
in particular, we examine the matrix of dynamic structure factors
$\M S_{w}\left(\V k,t\right)$ defined in (\ref{eq:S_k_w}), for a
ternary mixture. We use the computer algebra system Maple to evaluate
\eqref{eq:S_w} and \eqref{eq:S_k_w_sol} and obtain explicit formulas
to which we compare our numerical results below. We also examine $S_{\rho}$,
since, according to (\ref{eq:S_rho}), by examining the fluctuations
in density we are examining the correlations among all pairs of species.

\subsubsection{Equilibrium Fluctuations}

One of the key quantities used to characterize the intensity of \emph{equilibrium}
thermal fluctuations is the static structure factor or static spectrum
of the fluctuations. We perform these tests in the steady Stokes regime
since the velocity fluctuations decouple from density fluctuations
at equilibrium; the only purpose of the fluid solver at uniform equilibrium
is to ensure that the density remains consistent with the composition.

In this equilibrium test we use a ternary mixture with Stefan-Maxwell
diffusion matrix and non-ideality matrix 
\[
\M D=\begin{pmatrix}0 & 0.5 & 1.0\\
0.5 & 0 & 1.5\\
1.0 & 1.5 & 0
\end{pmatrix},\quad\mbox{and}\quad\M H=\begin{pmatrix}4.0 & 1.5 & 2.5\\
1.5 & 3.0 & 0.5\\
2.5 & 0.5 & 2.0
\end{pmatrix},
\]
in some arbitrary units in which $k_{B}=1$. The molecular masses
for the ternary mixture are $m_{1}=1.0,\, m_{2}=2.0,\, m_{3}=3.0$,
and the pure component densities are $\bar{\rho}_{1}=2.0,\,\bar{\rho}_{2}=3.0,\,\bar{\rho}_{3}=3.857$.
The system is a two-dimensional periodic system at equilibrium with
equilibrium densities $\rho_{1}=0.6,\,\rho_{2}=1.05,\,\rho_{3}=1.35$.
At these conditions the equilibrium density variance is $\D V\,\av{\left(\d{\rho}\right)^{2}}=S_{\rho}=0.3,$
where $\D V$ is the volume of a grid cell. We employ a square grid
of $32\times32$ cells with grid spacing $\D x=\D y=1$ for these
investigations, with the thickness in the third direction set to give
a large $\D V=10^{6}$ and thus ensure consistency with linearized
fluctuating hydrodynamics. A total of $2\times10^{4}$ time steps
are skipped in the beginning to allow equilibration of the system,
and statistics are then collected for an additional $10^{6}$ steps.

\begin{figure}
\begin{centering}
\includegraphics[width=0.49\textwidth]{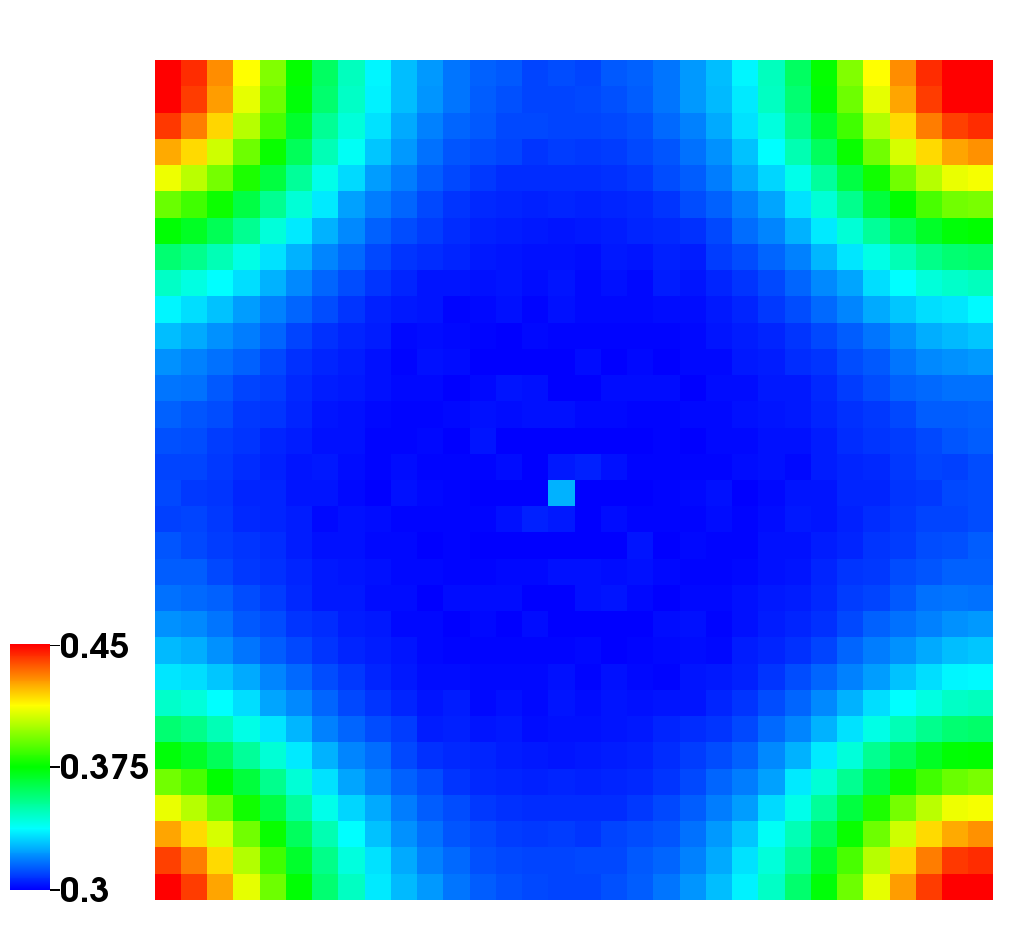}\includegraphics[width=0.49\textwidth]{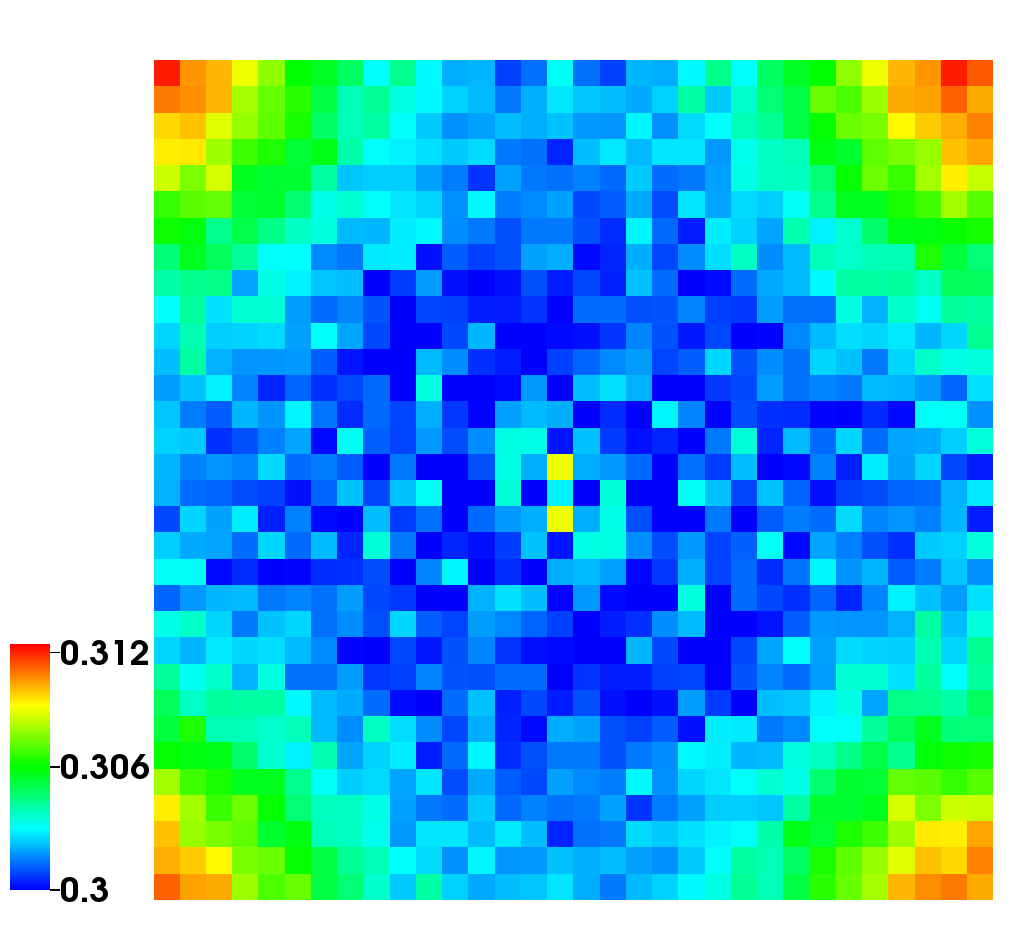}
\par\end{centering}

\caption{\label{fig:S_k}Equilibrium static structure factors $S_{\rho}\left(k_{x},k_{y}\right)$
as a function of wavevector (zero being at the center of the figures)
for a ternary mixture. The correct result, which is recovered in the
limit $\D t\rightarrow0$, is $S_{\rho}=0.3$ independent of wavenumber.
(Left) Clear artifacts are seen at grid scales for $\D t=0.1$, which
is $85\%$ of the stability limit. (Right) The artifacts decrease
by a factor of 8 as the time step is reduced in half. Note that the
statistical errors are now nearly comparable to the numerical error.}
\end{figure}

At equilibrium, the static structure factors are independent of the
wavenumber due to the local nature of the correlations, $S_{w}^{\left(i,j\right)}\left(\V k\right)=S_{\text{eq},\, w}^{\left(i,j\right)}=\mbox{const.}$
Since we include mass diffusion using an explicit temporal integrator,
for finite time step sizes $\D t$ we expect to see some deviation
from a flat spectrum at the largest wavenumbers (i.e., for $k\sim\D x^{-1}$)
\cite{LLNS_S_k,DFDB}. In Fig. \ref{fig:S_k} we show the spectrum
of density fluctuations at equilibrium for two different time step
sizes, a large time step size $\D t=0.1$ (left panel), and a smaller
time step size $\D t=0.05$ (right panel). Since the largest eigenvalue
of the diffusion matrix is around $\chi\approx2$, the largest stable
time step size is $\D t_{\max}\approx0.12$. As seen in the figure,
for $\D t=0.1$, which is close to the stability limit, we see a significant
enlargement of the fluctuations at the corners of the Fourier grid;
when we reduce the time step by a factor of 2 we reduce the error
by a factor of around 8, consistent with the fact that the explicit
midpoint method used in our overdamped algorithm \cite{LowMachImplicit}
is third-order accurate for static covariances \cite{DFDB}. Therefore,
in the limit of sufficiently small time steps we will recover the
correct flat spectrum, demonstrating that our equations and our numerical
scheme obey a fluctuation-dissipation principle.

\begin{figure}
\begin{centering}
\includegraphics[width=0.49\textwidth]{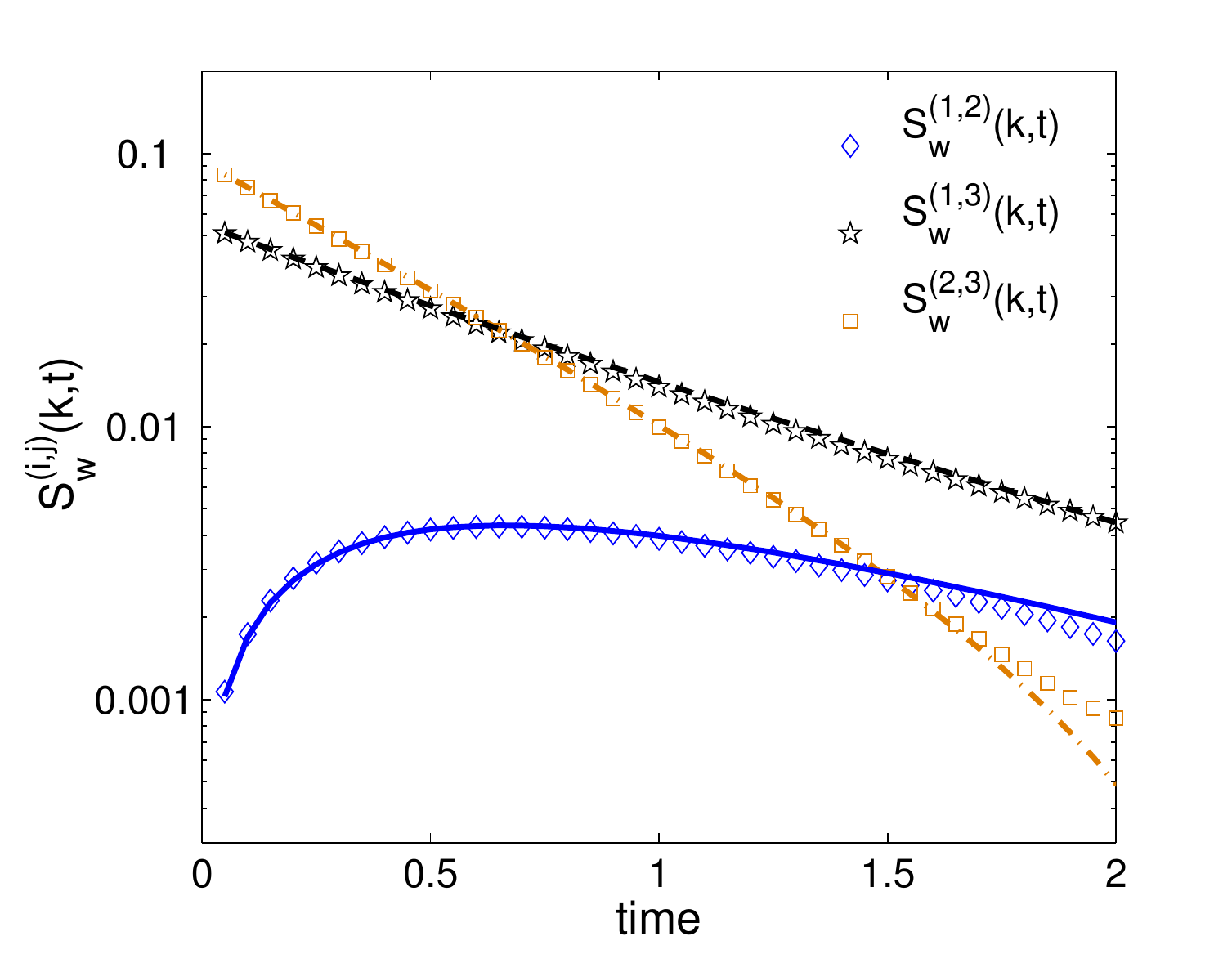}\includegraphics[width=0.49\textwidth]{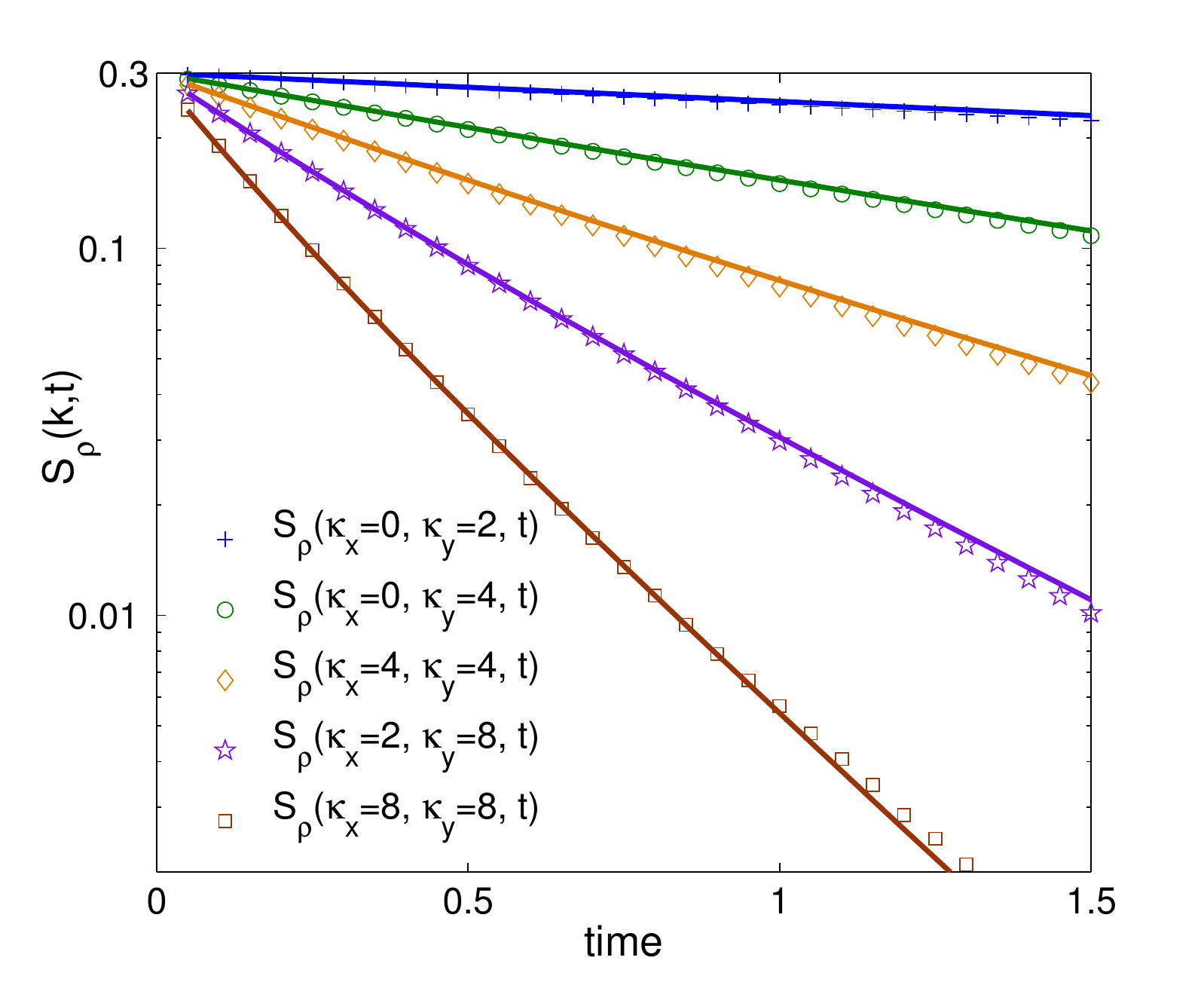}
\par\end{centering}

\caption{\label{fig:S_k_t}Equilibrium dynamic structure factors $\M S_{w}\left(\V k,t\right)$
for a ternary mixture, as a function of time for wavenumber $\V k=\left(\kappa_{x},\kappa_{y}\right)\cdot2\pi/L$.
Numerical results are shown with symbols and theoretical predictions
are shown with solid lines of the same color as the corresponding
symbols. (Left) $S_{w}^{(i,j)}\left(\V k,t\right)$ for $\kappa_{x}=\kappa_{y}=4$
for $i\neq j$. (Right) $S_{\rho}\left(\V k,t\right)$ for several
wavenumbers. Note that at large times statistical noise begins to
dominate the signal. }
\end{figure}

In the left panel of Fig. \ref{fig:S_k_t} we show numerical results
for the dynamic structure factors $S_{w}^{(i,j)}\left(\V k,t\right)$
for several $i\neq j$ for $\V k=\left(4,4\right)\cdot2\pi/L$, where
$L=32$ is the length of the square domain. Note that the factors
for $i=j$ are not statistically independent due to the constraint
that mass fractions sum to unity, and are thus not shown. In the right
panel of Fig. \ref{fig:S_k_t} we show numerical results for $S_{\rho}\left(\V k,t\right)$,
given by \eqref{eq:S_rho}, for several different wavenumbers $\V k=\left(\kappa_{x},\kappa_{y}\right)\cdot2\pi/L$.
We compare the numerical results to the theoretical prediction \eqref{eq:S_k_w_sol},
which is a sum of two exponentially-decaying functions. Excellent
agreement is seen between simulation and theory, demonstrating that
our numerical method correctly reproduces both the statics and dynamics
of the compositional fluctuations.

\subsubsection{Non-Equilibrium Fluctuations}

Fluctuations in systems out of equilibrium are known to be long-range
correlated and significantly enhanced compared to equilibrium. In
particular, in the presence of an imposed (macroscopic) concentration
gradient, concentration fluctuations exhibit a characteristic power-law
static structure factor $\sim k^{-4}$ \cite{FluctHydroNonEq_Book}.
In Section IV.C in Ref. \cite{MultispeciesCompressible}, we studied
the long-ranged (giant) concentration fluctuations in a ternary mixture
in the presence of a gradient imposed via the boundary conditions,
and confirmed that our multispecies compressible algorithm correctly
reproduced theoretical predictions; here we repeat this test but for
a mixture of three incompressible liquids.

In order to simplify the theoretical calculations, see Appendix B
in Ref. \cite{MultispeciesCompressible}, we take the first two of
the three species to be dynamically identical (indistinguishable),
and take the molecular masses to be equal, $m_{1}=m_{2}=m_{3}=1.0$
(this makes mass and mole fractions identical). The Stefan-Maxwell
diffusion matrix is taken to be 
\[
\M D=\begin{pmatrix}0 & 2.0 & 1.0\\
2.0 & 0 & 1.0\\
1.0 & 1.0 & 0
\end{pmatrix}
\]
and the mixture is assumed to be ideal, $\M H=0$, and isothermal,
$\grad T=0$. In order to focus our attention on the nonequilibrium
fluctuations we set the stochastic mass flux to zero, $\widetilde{\V F}=0$;
this ensures that all concentration fluctuations come from the coupling
to the velocity fluctuations via the gradient and eliminate the statistical
errors coming from a finite background spectrum. The pure component
densities are $\bar{\rho}_{1}=\bar{\rho}_{2}=\bar{\rho}_{3}=1.0$,
giving an incompressible fluid, $\grad\cdot\V v=0$, consistent with
the theoretical calculations. A weak concentration gradient is imposed
by enforcing Dirichlet (reservoir \cite{LowMachExplicit}) boundary
conditions for the mass fractions at the top and bottom boundaries,
$\V w\left(y=0,\, t\right)=\left(0.2493,0.245,0.5057\right)$ and
$\V w\left(y=L,\, t\right)=\left(0.250729,0.255,0.494271\right)$.
These values are chosen so that the deterministic diffusive flux of
the first species vanishes at $y=L/2$, $\bar{F}_{1}\left(y=L/2,t\right)=0$,
leading to a diffusion barrier for the first species, as in Ref. \cite{MultispeciesCompressible}.

The computational grid has $128\times64$ grid cells with grid spacing
$\D x=\D y=1$, with the thickness in the third direction set to give
$\D V=10^{6}$, and time step $\D t=0.1$. In order to study the spectrum
of the giant concentration fluctuations, we compute the Fourier spectrum
of the mass fractions averaged along the direction of the gradient;
this corresponds to $k_{y}=0$ and thus $k=k_{x}$. A total of $1.8\times10^{5}$
time steps are preformed at the beginning of the simulation to allow
the system to equilibrate. Statistics are then collected for $5\times10^{5}$
steps.

In this nonequilibrium example the coupling to the velocity equation
is crucial and is the cause of the giant fluctuations. The normal
component of the velocity at the two physical boundaries follows from
the EOS and the diffusive fluxes through the boundary, see Eq. (15)
in Ref. \cite{LowMachExplicit}. For the tangential ($x$) component
of velocity we use either no-slip (zero velocity) or free-slip (zero
shear stress) boundary conditions. In the limit of infinite Schmidt
number, $\nu=\eta/\rho\gg\chi$, where $\chi$ is a typical mass diffusion
coefficient, the overdamped equations apply and $\nu\, S_{w}^{(i,j)}\left(\V k\right)$
approaches a limit independent of the actual value of the Schmidt
number. For finite Schmidt numbers, however, the actual value of the
Schmidt number affects the spectrum, see Appendix B in Ref. \cite{MultispeciesCompressible}
for the explicit formulas.

\begin{figure}
\begin{centering}
\includegraphics[width=0.49\textwidth]{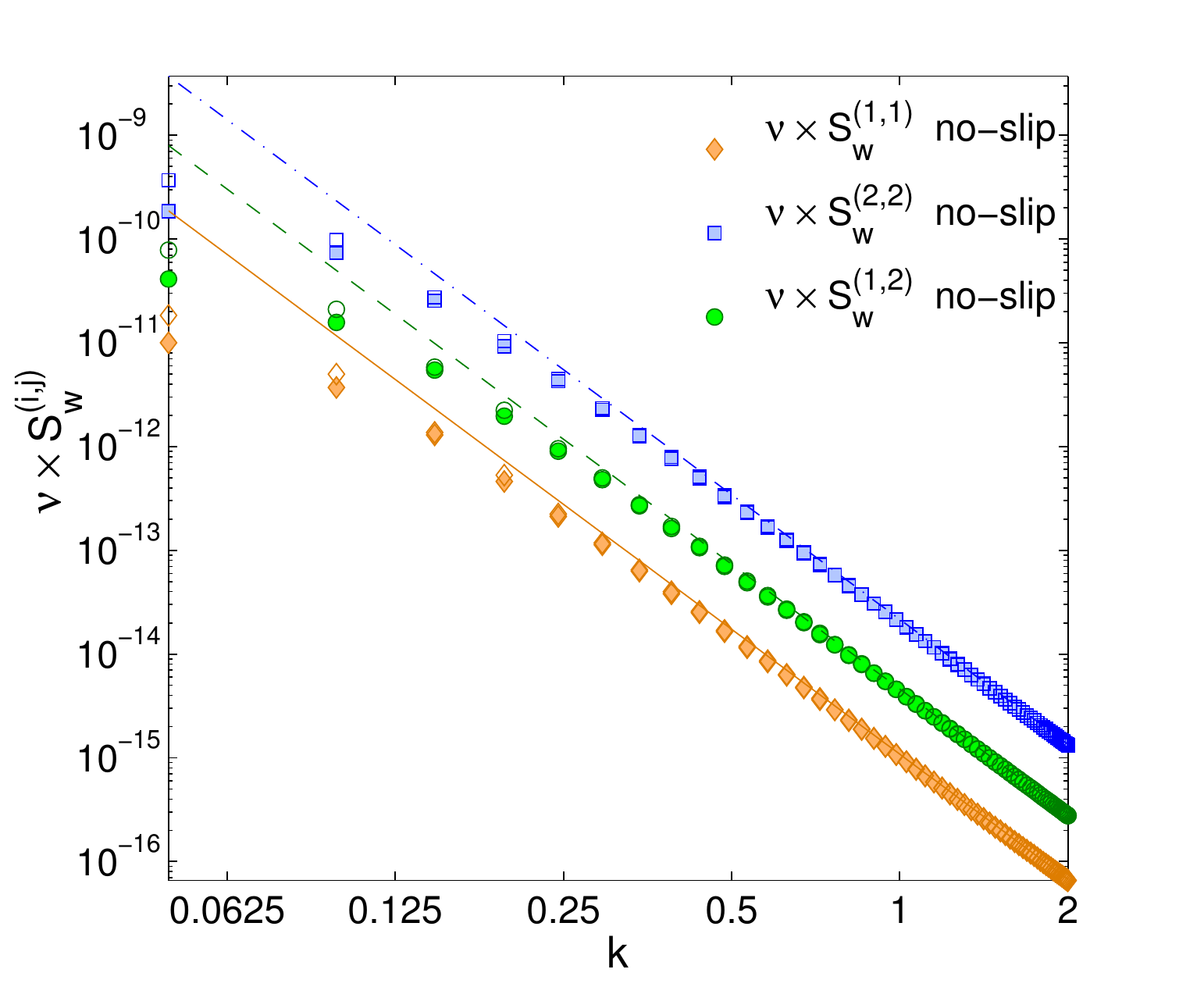}\includegraphics[width=0.49\textwidth]{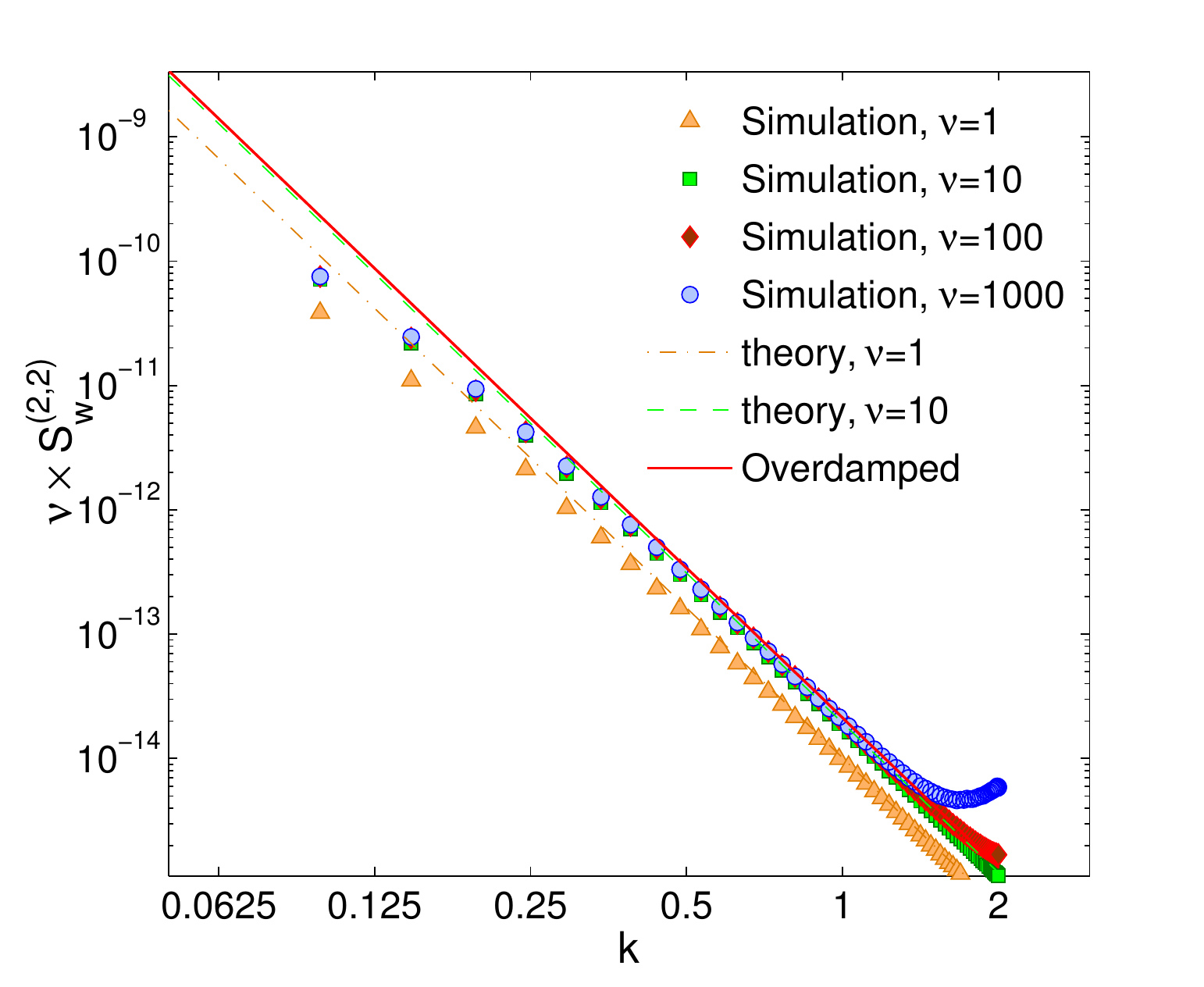}
\par\end{centering}

\caption{\label{fig:S_k_giant}Static structure factors of the mass fractions
averaged along the direction of the gradient, exhibiting giant $\sim k^{-4}$
fluctuations. (Left) Cross-correlations for the overdamped equations.
Filled symbols are for no-slip boundary conditions for velocity, while
empty ones are for free-slip boundaries. Lines of the same color show
the theoretical prediction in a ``bulk'' system (no boundaries).
(Right) Spectrum of fluctuations of $w_{2}$ for several different
values of the viscosity for the inertial equations. The overdamped
theory is shown for comparison. For $\nu=1$ (Schmidt number $\sim1$)
a difference between the inertial and overdamped results is seen and
reproduced well by the numerical scheme. For $\nu\gtrsim10$ there
is very little difference between overdamped and inertial, and the
two algorithms produce similar results. For very large viscosities,
however, one should use the overdamped integrator, as evidenced by
the notable departure from the theory at large wavenumbers for $\nu=1000$.}
\end{figure}

For the overdamped integrator, the actual value of the viscosity does
not matter beyond simply rescaling the amplitude of the fluctuations,
since the velocity equation is a time-independent (steady) Stokes
equation in the viscous-dominated limit. In the left panel of Fig.
\ref{fig:S_k_giant} we show numerical results for $\nu\, S_{w}^{(i,j)}\left(k\right)$
for $i\neq j$, obtained using our \emph{overdamped} algorithm \cite{LowMachImplicit}.
Excellent agreement with the theoretical prediction in Appendix B
in Ref. \cite{MultispeciesCompressible} is seen for wavenumbers larger
than $L^{-1}$; for small wavenumbers the confinement suppresses the
giant fluctuations in a manner that depends on the specific boundary
conditions imposed \cite{GiantFluctFiniteEffects}. In the right panel
of Fig. \ref{fig:S_k_giant} we show $\nu\, S_{w}^{(2,2)}\left(k\right)$
for several values of the kinematic viscosity, as obtained using our
\emph{inertial} algorithm \cite{LowMachImplicit}. The implicit-midpoint
(Crank-Nicolson) scheme used to treat viscosity in the inertial algorithm
is unconditionally stable and allows an arbitrary time step size to
be used. It is, however, well-known that this kind of scheme can produce
unphysical results for very large viscous Courant numbers due to fact
it is not $L$-stable (see discussion in Appendix B in Ref. \cite{DFDB}).
This is seen in the results in the right panel of Fig. \ref{fig:S_k_giant}
for the largest viscosity $\nu=1000$ (corresponding to viscous Courant
number $\nu\D t/\D x^{2}=100$) at the larger wavenumbers. It is actually
quite remarkable that we can use the inertial integrator with rather
large time step sizes and get very good results over most of the wavenumbers
of interest; this is a property that stems from a specific fluctuation-dissipation
balance in the implicit midpoint scheme \cite{DFDB}. These results
demonstrate that both our overdamped and inertial methods are able
to reproduce the correct spectrum of the nonequilibrium concentration
fluctuations.

\subsubsection{\label{sub:ThermoBaro}Thermodiffusion and Barodiffusion}

In the giant fluctuation example shown in Fig. \ref{fig:S_k_giant},
the system was kept out of equilibrium by imposed concentrations on
the boundaries, which is difficult to realize in experiments. Instead,
experiments that measure giant fluctuations in liquid mixtures typically
rely on the Soret effect to induce a concentration gradient via an
imposed temperature gradient \cite{SoretDiffusion_Croccolo}. A concentration
gradient can also be induced via barodiffusion in the presence of
large gravitational accelerations, as used in ultracentrifuges for
the purposes of separation of macromolecules and isotopes \cite{MulticomponentBook_KT}.
Barodiffusion and thermodiffusion enter in the density equations (\ref{eq:species_eq})
in the same manner, however, the key difference is that barodiffusion
requires gravity which also enters via the buoyancy term in the velocity
equation (\ref{eq:momentum_eq},\ref{eq:div_v_constraint}). Furthermore,
the steady state gradient induced by barodiffusion is determined by
equilibrium thermodynamics only and does not involve any kinetic transport
coefficients. 

It is well known that there is no nonequilibrium enhancement of the
fluctuations at steady state for a system in a gravitational field
\cite{GiantFluct_Barodiffusion} in the absence of external forcing
(see Eq. (28) in \cite{GiantFluctuations_Theory}) %
\footnote{Note, however, that the dynamics of the fluctuations is affected by
gravity and by barodiffusion \cite{GiantFluctuations_Theory}%
}. This is because the system is still in thermodynamic equilibrium,
despite the presence of spatial nonuniformity (sedimentation). In
particular, without doing any calculations we know that the equilibrium
distribution of the fluctuations is the Gibbs-Boltzmann distribution,
with a local free-energy functional that now includes a gravitational
energy contribution. In this section we demonstrate that our low Mach
number approach captures this important distinction between (ordinary)
equilibrium fluctuations in the presence of barodiffusion, and (giant)
nonequilibrium fluctuations in the presence of thermodiffusion. 

We consider a solution of potasium salt and sucrose in water (see
Section \ref{sec:Instabilities} for more details) in an ultracentrifuge.
The physical parameters of this ternary mixture are given in Section
\ref{sub:PhysicalParameters}, and a brief theoretical analysis is
given in Appendix \ref{sec:Sedimentation}. We perform two dimensional
simulations of a system of physical dimensions $0.8\times0.8\times0.1$cm
divided into $64\times64\times1$ finite-volume cells. Periodic boundary
conditions are used in the $x$ direction and impermeable no-slip
boundaries are used in the $y$ direction. The average mass fractions
over the domain are set to $\V w_{\text{av}}=\left(0.0492,\,0.0229,\,0.9279\right).$
A total of $0.5\cdot10^{6}$ time steps are performed at the beginning
of the simulation to allow the system to equilibrate before statistics
are collected for $10^{6}$ steps.

In order to induce a strong sedimentation in this mixture we need
to increase the ratio $m_{2}g/\left(k_{B}T\right)$ by six orders
of magnitude relative to its reference value on Earth (see (\ref{eq:sediment_dilute})).
In actual experiments this would be accomplished by increasing the
effective gravity (i.e., centrifugal acceleration) in an ultracentrifuge;
however, increasing gravity by such a large factor makes the system
of equations (\ref{eq:momentum_eq},\ref{eq:div_v_constraint},\ref{eq:species_eq})
numerically too stiff for our semi-implicit temporal integrator. This
is because buoyancy changes the time scale for relaxation of large-scale
(small wavenumber) concentration fluctuations from the usual slow
diffusive relaxation to a very fast non-diffusive relaxation \cite{GiantFluctuations_Cannell}.
Therefore, instead of increasing $g$ we artificially decrease $k_{B}$
by six orders of magnitude, and apply Earth gravity $g=-981$ along
the negative $y$ direction. With these parameters our inertial temporal
integrator is stable with time step size up to about $\D t=0.5$s;
the results reported below are for $\D t=0.25$s.

\begin{figure}
\begin{centering}
\includegraphics[width=0.49\textwidth]{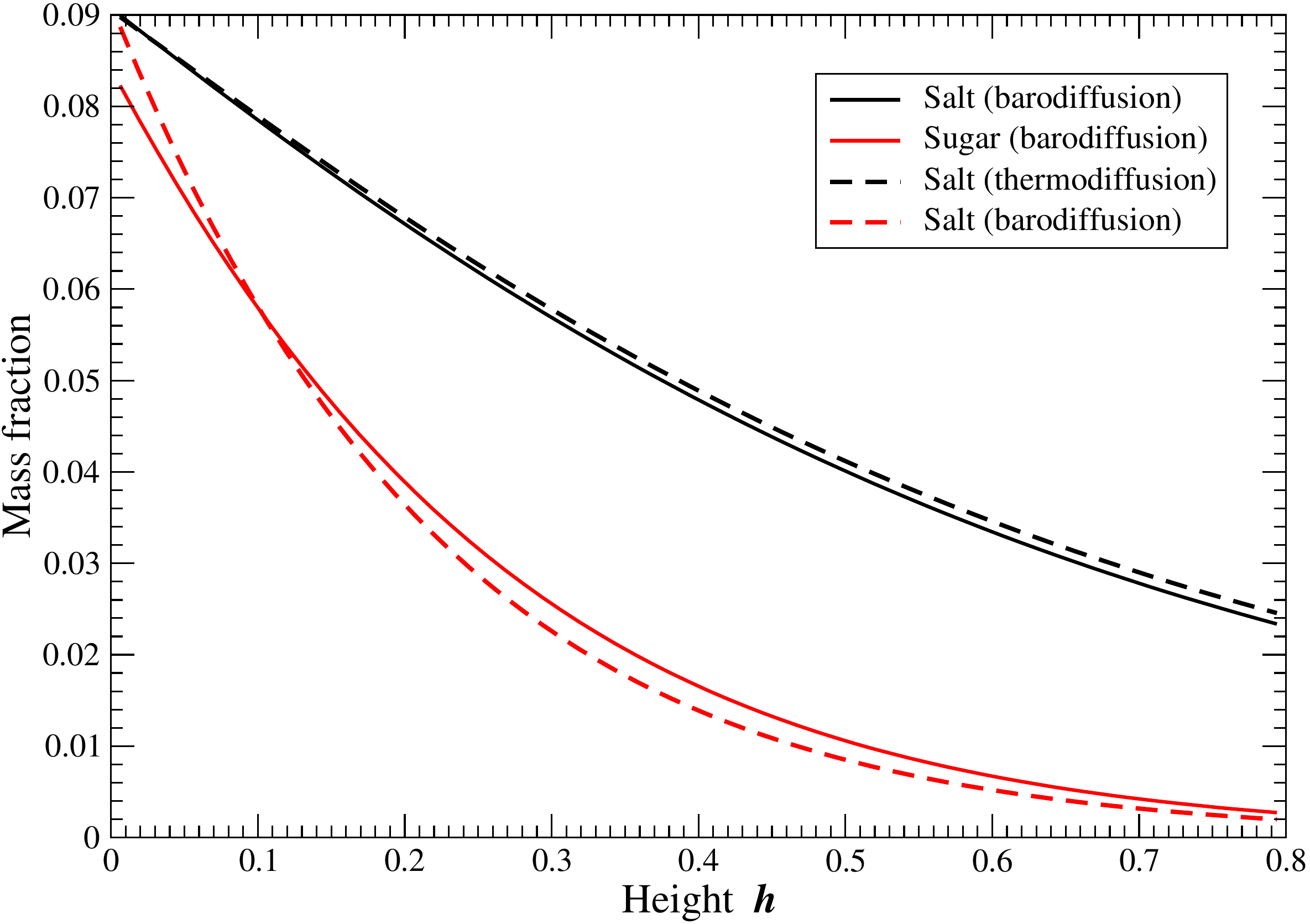}\includegraphics[width=0.49\textwidth]{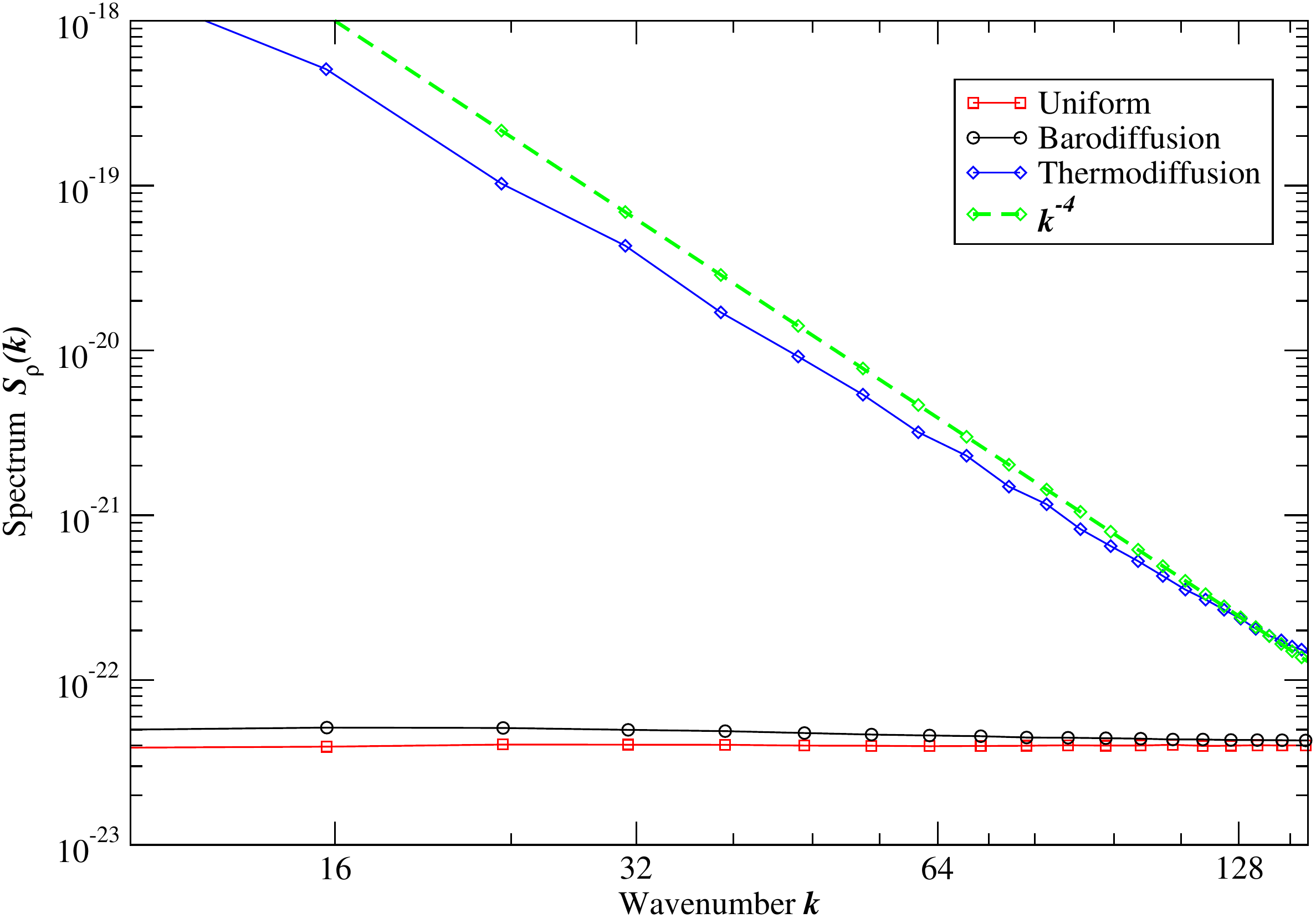}
\par\end{centering}

\caption{\label{fig:S_k_baro}Comparison of fluctuations in the presence of
a strong concentration gradient induced by baro and thermo diffusion.
(Left) The steady-state vertical concentration profile $w_{1}(h)$
and $w_{2}(h)$ as a function of height. (Right) Static structure
factors of density averaged along the direction of the gradient, exhibiting
giant $\sim k^{-4}$ fluctuations for gradients induced by thermodiffusion.
By contrast, fluctuations in the system with strong sedimentation
induced by barodiffusion exhibit a similar spectrum as those in a
homogeneous bulk equilibrium system in the absence of gravity.}
\end{figure}

For comparison, we use the same parameters but turn gravity off and
induce a concentration gradient via thermodiffusion. Specifically,
we set the temperature at the bottom wall to $293$K and $300$K at
the top wall, and set the thermodiffusion constants to the artifical
values $\V D^{(T)}=\left(-5\cdot10^{-4},-2\cdot10^{-4},7\cdot10^{-4}\right)$.
These values ensure that the steady state vertical profiles of the
mass fraction of salt and sugar are very similar between the barodiffusion
and thermodiffusion simulations (see (\ref{eq:thermodiff_dilute})),
as shown in the left panel of Fig. \ref{fig:S_k_baro}. In the right
panel of the figure we show the spectrum $S_{\rho}\left(k\right)$
of the vertically-averaged density. For comparison, in addition to
the cases of gradients induced by baro and thermo diffusion, we also
show the spectrum for a spatially-uniform system at thermodynamic
equilibrum, in the absence of gravity and at a constant temperature
of 293K. We see that the spectrum for barodiffusion is very similar
to that for uniform thermodynamic equilibrium, while the spectrum
for thermodiffusion shows the $k^{-4}$ power-law behavior as in Fig.
\ref{fig:S_k_giant}. This demonstrates that our numerical code correctly
reproduces equilibrium fluctuations even in the presence of strong
sedimentation.

\section{\label{sec:Instabilities}Diffusion-Driven Gravitational Instabilities}

In this section we use our numerical methods to study the development
of diffusion-driven gravitational instabilities in ternary mixtures.
In Section IV.D in Ref. \cite{MultispeciesCompressible} we studied
the development of a diffusion-driven Rayleigh-Taylor (RT) instability
in a ternary gas mixture. Here we simulate similar instabilities in
a ternary mixture of incompressible liquids, using realistic parameters
corresponding to recent experimental measurements, and study the effect
of (nonequilibrium) thermal fluctuations on the development of the
instabilities. Our investigations are inspired by the body of work
by Anne De Wit and collaborators on diffusion- and buyoancy-driven
instabilities \cite{DiffusiveInstability_Chemistry_PRL,DiffusiveInstability_Porous,MixedDiffusiveInstability}.
In particular, in Ref. \cite{MixedDiffusiveInstability} a classification
of these instabilities in a ternary mixture are proposed, and several
of the instabilities are investigated experimentally. In the first
part of this section we perform simulations of the experimental measurements
of a mixed-mode instability (MMI). In the second part we investigate
diffusive layer convection (DLC) (just as we did in Section IV.D in
Ref. \cite{MultispeciesCompressible} for gases), in a hypothetical
shadowgraphy or light scattering experiment that could, in principle,
be performed in the laboratory.

We begin with a brief summary of the experimental setup of Carballido-Landeira
\emph{et al.} \cite{MixedDiffusiveInstability} for getting a diffusion-driven
gravitational instability in a simple ternary mixture: a solution
of salt in water on top of a solution of sugar in water. The concentrations
of salt and sugar are small so that even though this is a ternary
mixture in this very dilute limit one can think of salt and sugar
diffusing in water without significant interaction. The key here is
the difference in the diffusion coefficient between sugar (a larger
organic molecule diffusing slower) and salt (a smaller ion diffusing
faster) in water. Both sugar and salt solutions have a density that
grows with the concentration of the solute.

In the experiments, one starts with an almost (to within experimental
controls) flat and almost sharp interface between the two solutions.
Even if one starts in a stable configuration, with the denser solution
on the bottom, the differential diffusion effects can create a local
minimum in density below the contact line and a local maximum above
the contact line. This leads to an unstable configuration and the
development of DLC at symmetric distances above and below the contact
line. If one starts with an unstable configuration of the denser solution
on top, before the RT instability has time to develop and perturb
the interface, differential diffusion effects can lead to the development
of local extrema in the density above and below the contact line that
are outside the range of the initial densities. The dynamics is then
a combination of RT and DLC giving rise to a mixed mode instability
(MMI). The DLC leads to characteristic \textquotedblleft{}Y shaped\textquotedblright{}
convective structures developing around the interface at the locations
of the local adverse density gradients, which evolve around an interface
that is slowly perturbed by the RT growth to a finite amplitude modulation.
See Section III in Ref. \cite{MixedDiffusiveInstability} for more
details, and the bottom row of panels in Fig. 1 in Ref. \cite{MixedDiffusiveInstability},
as well as our numerical results in Fig. \ref{fig:MMI_development},
for an illustration of the development of the instability.

\subsection{\label{sub:PhysicalParameters}Physical Parameters}

We use CGS units in what follows (centimeters for length, seconds
for time, grams for mass). Following the experiments of Carballido-Landeira
\emph{et al.}, we consider a ternary mixture of potassium salt (KCl,
species 1, molar mass $M_{1}=74.55$, denoted by A in \cite{MixedDiffusiveInstability}),
sugar (sucrose, species 2, molar mass $M_{2}=342.3$, denoted by B
in \cite{MixedDiffusiveInstability}) and water (species 3, molar
mass $M_{3}=18.02$), giving molecular masses $\V m=(1.238\cdot10^{-22},5.684\cdot10^{-22},2.99\cdot10^{-23})$.
The initial configuration is salt solution on top of sugar solution.
In Ref. \cite{MixedDiffusiveInstability}, it is assumed that the
density dependence on the concentration can be captured by (this is
a good approximation for very dilute solutions)
\begin{equation}
\rho=\rho_{0}\left(1+\alpha_{1}Z_{1}+\alpha_{2}Z_{2}\right)=\rho_{0}\left(1+\frac{\alpha_{1}}{M_{1}}\rho_{1}+\frac{\alpha_{2}}{M_{2}}\rho_{2}\right),\label{eq:eos_2}
\end{equation}
where $\rho_{0}=\bar{\rho}_{3}=1.0$ is the density of water, $\alpha_{1}=48$
for KCl and $\alpha_{2}=122$ for sucrose, and $Z_{k}$ is the molar
density of each component, related to the partial density via $\rho_{k}=Z_{k}M_{k},$
where $M_{k}$ is the molar mass. Noting that we can write our EOS
(\ref{eq:EOS}) in the form
\begin{equation}
\frac{\rho_{1}}{\bar{\rho}_{1}}+\frac{\rho_{2}}{\bar{\rho}_{2}}+\frac{\rho-\rho_{1}-\rho_{2}}{\bar{\rho}_{3}}=1,\label{eq:eos_1}
\end{equation}
and comparing (\ref{eq:eos_1}) and (\ref{eq:eos_2}), we get
\[
1-\frac{\rho_{0}}{\bar{\rho}_{k}}=\rho_{0}\frac{\alpha_{k}}{M_{k}},
\]
which gives us the EOS parameters 
\[
\bar{\rho}_{1}=2.81\text{ and }\bar{\rho}_{2}=1.55.
\]
Note that here these should not be thought of as pure component densities
since the solubility of the solvents in water is finite, rather, they
are simply parameters that enable us to match our model EOS \eqref{eq:EOS}
to the empirical density dependence in the dilute regime.

For the dilute solutions we consider here it is sufficient to assume
that $\M D$ is constant for the range of compositions of interest,
and the mixtures are essentially ideal, $\M H=\V 0$. We also assume
isothermal conditions with the ambient temperature to $T=293$K, and
assume constant viscosity $\eta=0.01$. Since the ternary mixture
under consideration is what can be considered ``infinite dilution'',
we rely on the approximation proposed in Ref. \cite{Diffusion_InfiniteDilution},
\[
D_{13}=D_{1},\, D_{23}=D_{2},\, D_{12}=\frac{D_{1}D_{2}}{D_{3}},
\]
where from table I in Ref. \cite{MixedDiffusiveInstability} we read
the diffusion coefficient of low-dilution KCl in water as $D_{1}=1.91\cdot10^{-5}$,
and the diffusion coefficient of dilute sucrose in water as $D_{2}=0.52\cdot10^{-5}$.
Here $D_{3}$ is the self-diffusion coefficient of pure water, $D_{3}=2.3\cdot10^{-5}$,
giving $D_{12}\approx4.32\cdot10^{-6}$.

In the simulations reported below, we have neglected barodiffusion
and assumed that the pressure is equal to the atmospheric pressure
throughout the system. If barodiffusion were included we would obtain
partial sedimentation in the final state of the mixing experiment
(where the system has completely mixed), as discussed in more detail
in Section \ref{sub:ThermoBaro} and Appendix \ref{sec:Sedimentation}.
In the final steady state, (\ref{eq:sediment_dilute}) suggests that
the ratio of the mass fraction of sugar between the top and bottom
of a cell of height $H=1$ will be 
\[
\frac{w_{2}\left(y_{\max}\right)}{w_{2}\left(y_{\min}\right)}=\exp\left(\left(1-\frac{\bar{\rho}_{3}}{\bar{\rho}_{2}}\right)\frac{m_{2}gH}{k_{B}T}\right)\approx\exp\left(-0.35\cdot\frac{5.7\cdot10^{-22}\cdot980}{4.2\cdot10^{-14}}\right)\approx1-4.7\cdot10^{-6},
\]
which is indeed negligible and likely not experimentally measurable.

\subsection{Mixed-Mode Instability in a Hele-Shaw Cell}

\begin{figure}[hb]
\centering{}\includegraphics[width=2in]{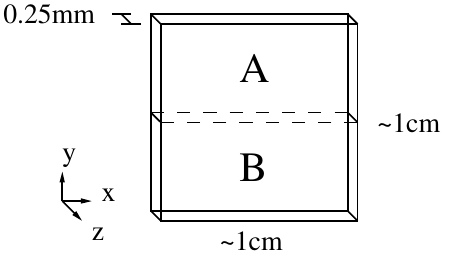}\hspace{0.5cm}\includegraphics[width=2in]{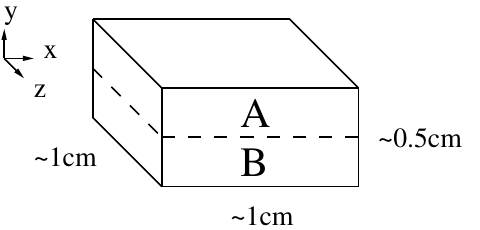}
\caption{Initial configuration for the development of diffusion-driven gravitational
instabilities in a ternary mixture consisting of a solution of salt
(A) initially on top of a solution of sugar (B). (Left) \label{fig:mmi_setup}
MMI setup leading to a quasi-two-dimensional instability illustrated
in Fig. \ref{fig:MMI_development}. (Right) \label{fig:dlc_setup}DLC
setup leading to a three-dimensional instability illustrated in Fig.
\ref{fig:DLC}.}
\end{figure}

In this section we examine the mixed-mode instability illustrated
in the bottom row of panels in Fig. 1 in Ref. \cite{MixedDiffusiveInstability}.
The geometry of the system is a Hele-Shaw cell, i.e., two parallel
glass plates separated by a narrow gap, as illustrated in the left
panel of Fig. \ref{fig:mmi_setup}. In our simulation we model a domain
of length $0.8\times0.8\times0.025$, with gravity $g=-981$ along
the negative $y$ direction. It is well-known that the flow averaged
along the $z$ axes in a Hele-Shaw setup can be \emph{approximated
}with a two-dimensional Darcy law; this is used in the simulations
reported in Ref. \cite{MixedDiffusiveInstability}. Here we do not
rely on any approximations but rather model the actual three-dimensional
structure of the flow in all directions. We divide our domain into
$256\times256\times8$ cells, and impose periodic boundary conditions
in the $x$ direction and no-slip walls in the $z$ direction. In
the y direction, we impose a reservoir (Dirichlet) boundary conditions
for concentration to match the initial concentrations in the top and
bottom half of the domain, and impose a free-slip boundary condition
on velocity to model an open reservoir of salt solution on the top
and sugar solution at the bottom of the domain.

Note that the momentum diffusion time across a domain of length $L\sim1$
is $\tau_{L}\sim L^{2}/(2\nu)\approx50$. The experimental snapshots
in Fig. 1 of Ref. \cite{MixedDiffusiveInstability} indicate that
this is comparable to the time it takes for the instability to fully
develop. It is therefore not safe to rely on the overdamped approximation,
so we use the inertial integrator described in Ref. \cite{LowMachImplicit}
in our simulations. Nevertheless the overdamped limit is a relatively
good approximation in practice since the characteristic length scale
of the $Y$-shaped DLC fingers observed in the experiments is $L\sim0.013$,
corresponding to viscous diffusion time $\tau_{L}\sim8.5\cdot10^{-3}$.
We have compared numerical simulations using the overdamped and inertial
integrators and found little difference. We simulate the development
of the instability to $t\approx63.9$ with a fixed time step size
of $\Delta t=6.39\cdot10^{-2}$, which corresponds to $75$\% of the
stability limit dictated by our explicit treatment of mass diffusion.
We use a centered discretization of advection; in these well-resolved
simulations little difference is observed between centered advection
and the more sophisticated BDS advection scheme \cite{BDS} summarized
in Ref. \cite{LowMachImplicit}.

The initial concentrations, denoted with superscript zero in what
follows, on the top and bottom are determined from the dimensionless
ratio reported in \cite{MixedDiffusiveInstability}, 
\begin{equation}
R=\frac{\alpha_{2}Z_{2}^{0}}{\alpha_{1}Z_{1}^{0}}=\frac{\alpha_{2}w_{2}/M_{2}}{\alpha_{1}w_{1}/M_{1}}=0.89.\label{eq:R_def}
\end{equation}
Specifically, we set the initial mass fractions of salt and sugar
to $w_{1}^{0}=0.0864$ and $w_{2}^{0}=0.1368$ respectively, more
precisely,
\begin{eqnarray*}
\V w_{\text{top}}^{0} & = & (0.0864,\,0,\,0.9136)\\
\V w_{\text{bottom}}^{0} & = & (0,\,0.1368,\,0.8632)
\end{eqnarray*}
which, from the EOS \eqref{eq:EOS} gives a density difference of
about 0.8\%. Similar results are observed for other values of the
concentrations if the dimensionless ratio $R$ in \eqref{eq:R_def}
is kept fixed, with the main difference being that lower concentrations
lead to a slower development and growth of the instability.

\begin{figure*}
\begin{centering}
\includegraphics[width=0.49\textwidth]{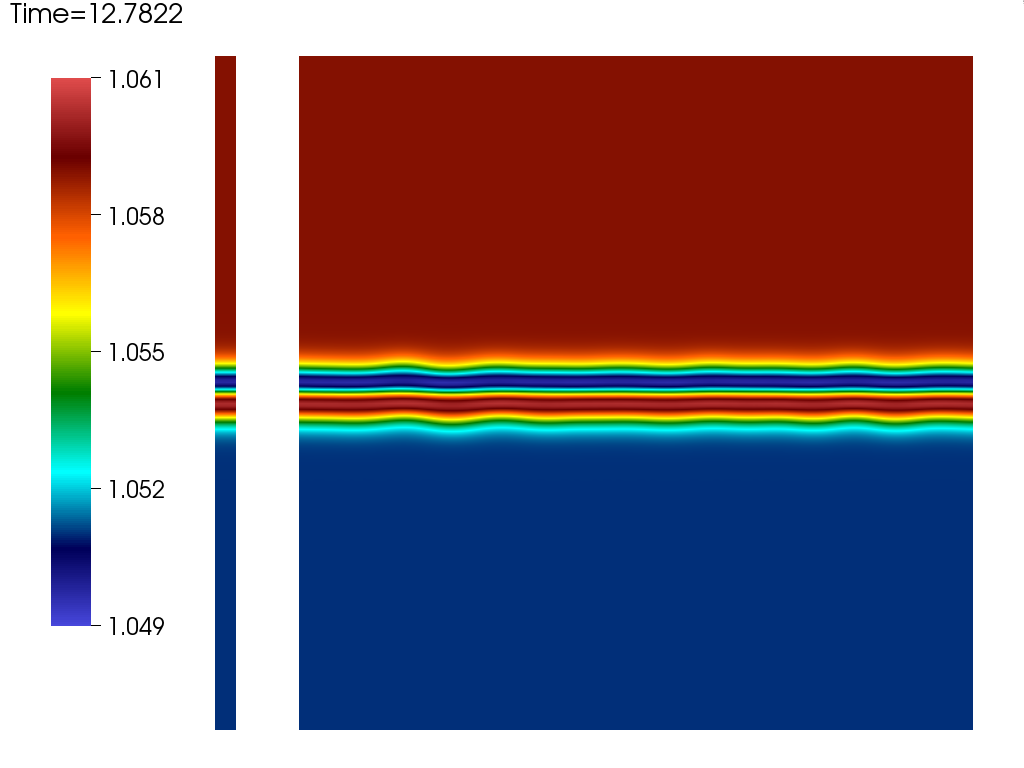}\includegraphics[width=0.49\textwidth]{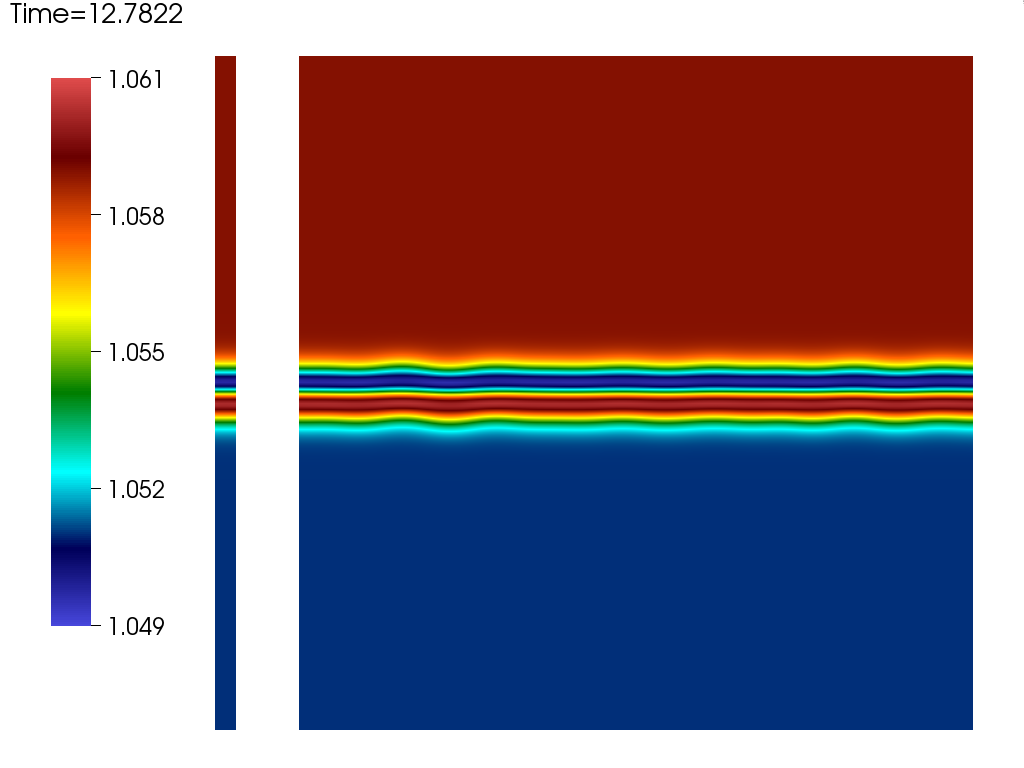}
\par\end{centering}

\begin{centering}
\includegraphics[width=0.49\textwidth]{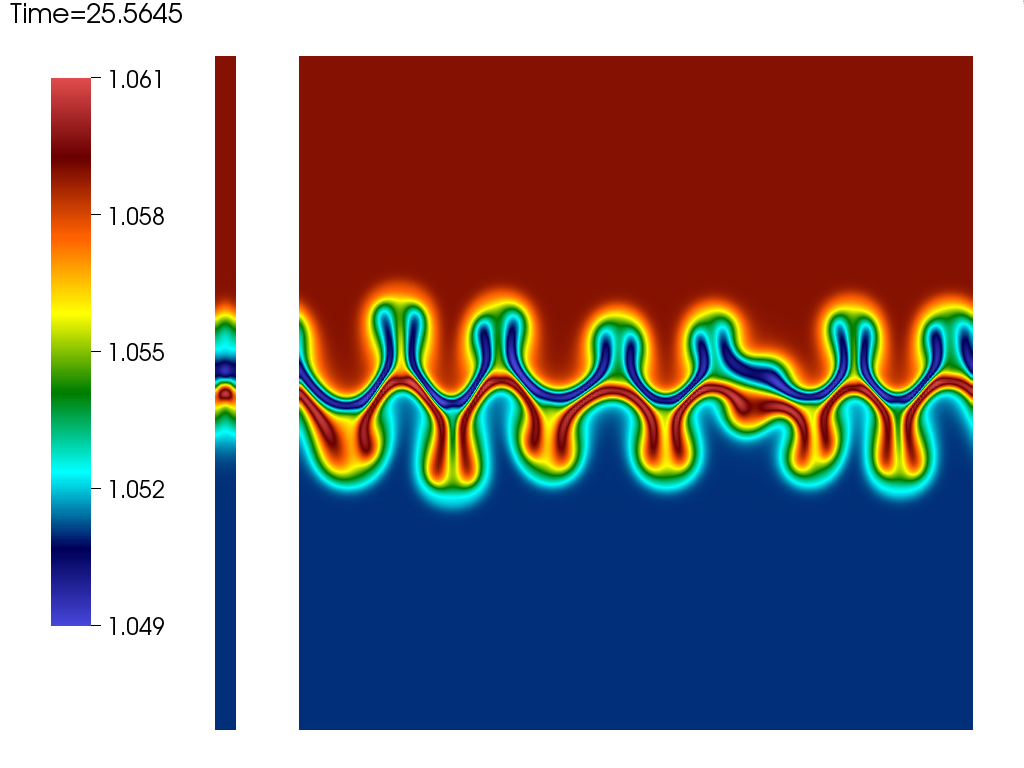}\includegraphics[width=0.49\textwidth]{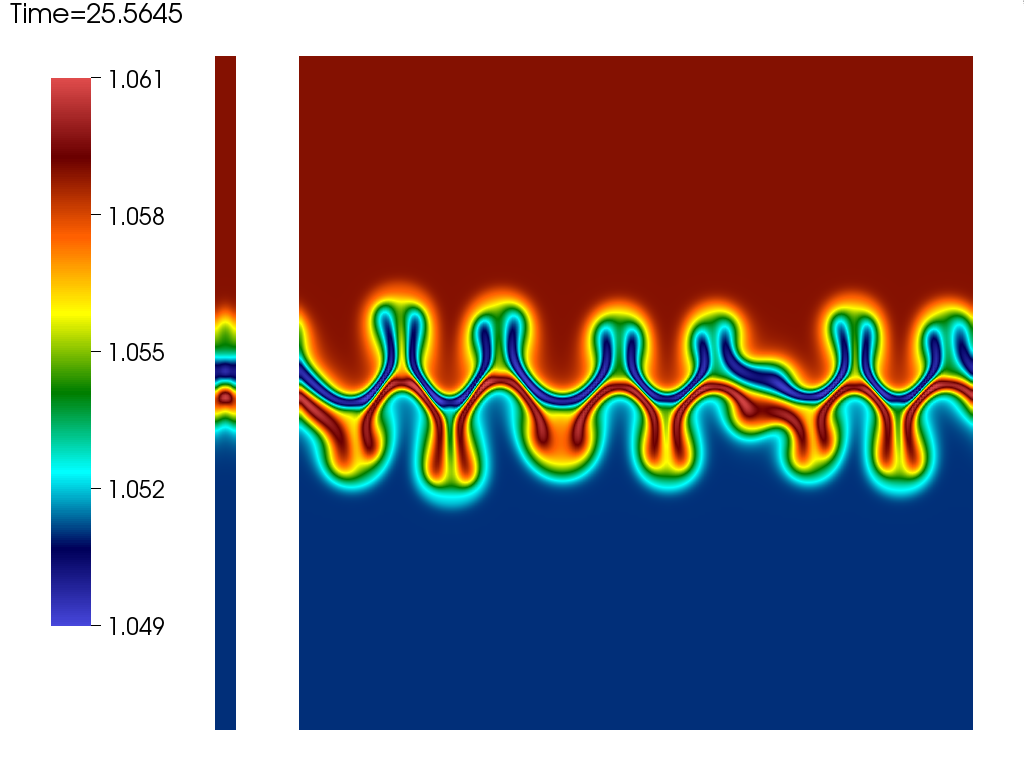}
\par\end{centering}

\begin{centering}
\includegraphics[width=0.49\textwidth]{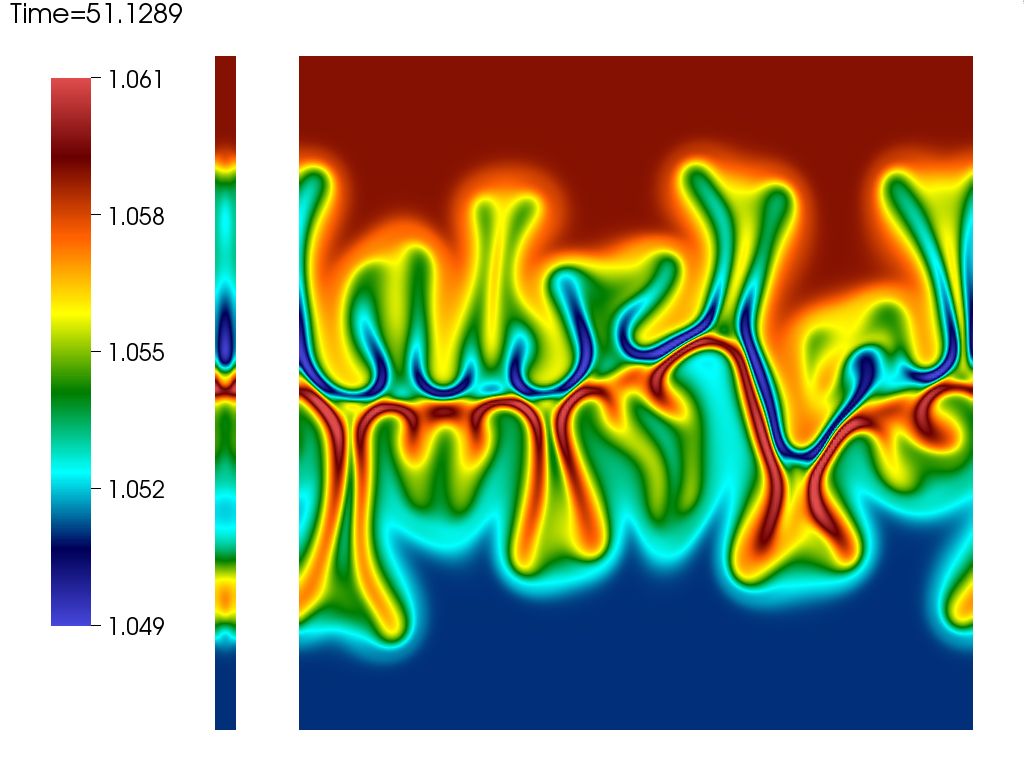}\includegraphics[width=0.49\textwidth]{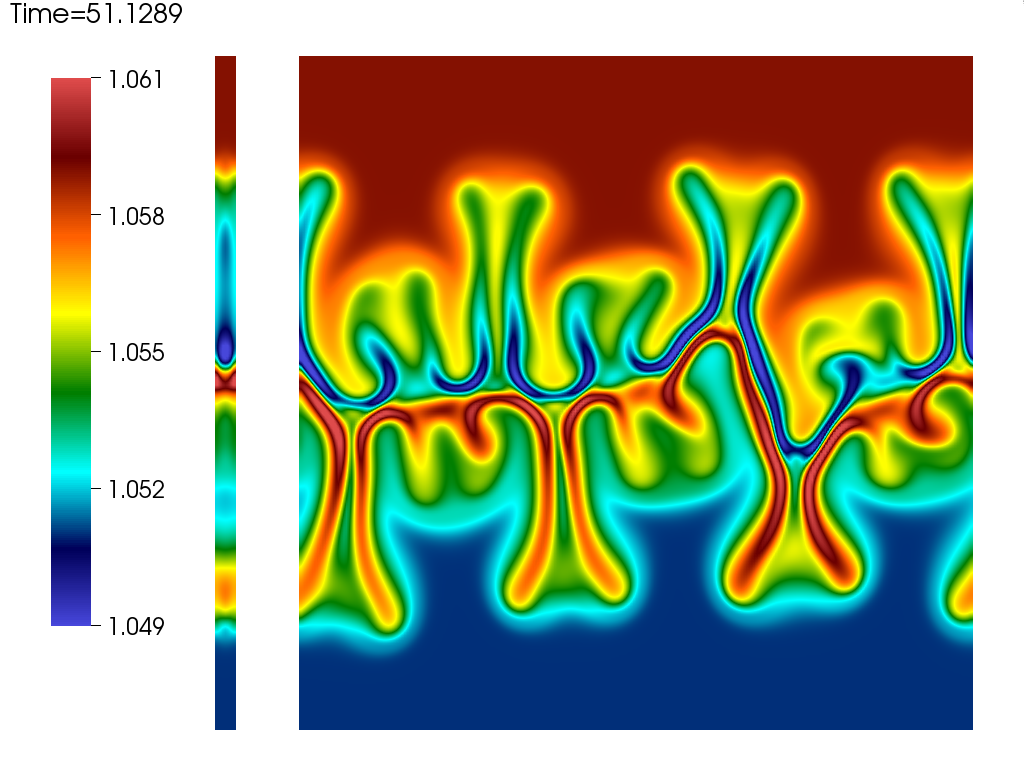}
\par\end{centering}

\caption{\label{fig:MMI_development}Development and growth of a mixed-mode
instability in a Hele-Shaw setup. Two-dimensional slices of the three-dimensional
density field are shown as color plots at times $t\approx13$ (top
row), $t\approx26$ (middle row) and $t\approx51$ (bottom row). The
square images show $\rho\left(x,y,z=0.0125\right)$ (halfway between
the glass plates) and the thin vertical images show the corresponding
slice $\rho\left(x=0,y,z\right)$ (corresponding to the left edge
of the square images). Compare these to the experimental results shown
in the bottom row of panels in Fig. 1 in Ref. \cite{MixedDiffusiveInstability}.
(Left) With random initial perturbation and thermal fluctuations.
(Right) Deterministic simulation starting with the same random initial
perturbation as for the left panels, but \emph{no} thermal fluctuations.}
\end{figure*}

In the experiments, the initial interface between the two solutions
is not, of course, perfectly flat. To model this effect, it is common
in the literature to add a small random perturbation to the initial
conditions. Assuming that the growth of the unstable modes is exponential
in time, the time it will take for the instability to reach a certain
point in its development (e.g., to first split the Y-shaped fingers)
will depend on the amplitude of the initial perturbation. Since the
initial condition is not known to us and is impossible to measure
experimentally to high accuracy, it is not possible to directly compare
snapshots of the instability at ``the same point in time'' between
simulations and experiments, or between simulations that use different
initial perturbations. Instead, we compare a simulation in which thermal
fluctuations are accounted for with one in which thermal fluctuations
are not accounted for, both starting from the \emph{same} randomly
perturbed initial interface.

Specifically, we randomly perturb the concentrations in the layer
of cells just above the interface, setting the mass fractions to $\V w=r\V w_{\text{bottom}}^{0}+(1-r)\V w_{\text{top}}^{0}$
next to the interface, with $r$ being a uniformly distributed random
number between $0$ and $0.1$. As we explained above, a direct comparison
between simulations and experiments is not possible. Nevertheless,
our numerical results shown in Fig. \ref{fig:MMI_development} are
very similar, in both time development and visual appearance, to the
experimental images shown in the bottom row of panels in Fig. 1 in
Ref. \cite{MixedDiffusiveInstability}. Here we use density $\rho$
as an indicator of the instability even though in actual light scattering
or shadowgraph experiments it is the index of refraction rather than
density that is observed; both density and index of refraction are
(approximately) linear combinations of the concentrations and should
behave similarly. More detailed comparisons between simulations and
experiments require a careful coordination of the two and will not
be attempted here. Instead, we focus our attention on examining the
role (if any) played by the thermal fluctuations in the triggering
and evolution of the instability.

Comparing the left and right column of panels in Fig. \ref{fig:MMI_development}
shows only minor differences between the deterministic and the fluctuating
hydrodynamics calculation. This indicates that the dynamics is dominated
by the unstable growth of the initial perturbation, with thermal fluctuations
adding a weak perturbation, which, though weak, does to some extent
affect the details of the patterns formed at late times but not their
generic features. By tuning the strength of the initial perturbation
one can also tune the impact of the fluctuations on the dynamics;
after all, if one starts with a perfectly flat interface the instability
will be triggered by thermal fluctuations only. Nevertheless, we can
conclude that once the instability develops sufficiently (e.g., the
Y-shaped fingers form), the dynamics becomes dominated by the deterministic
growth of the unstable modes. This can be seen from the fact that
stochastic simulations (not shown) starting from a perfectly flat
interface develop the same features at long times as deterministic
simulations with a random initial perturbation.

\begin{figure}
\begin{centering}
\includegraphics[width=0.75\textwidth]{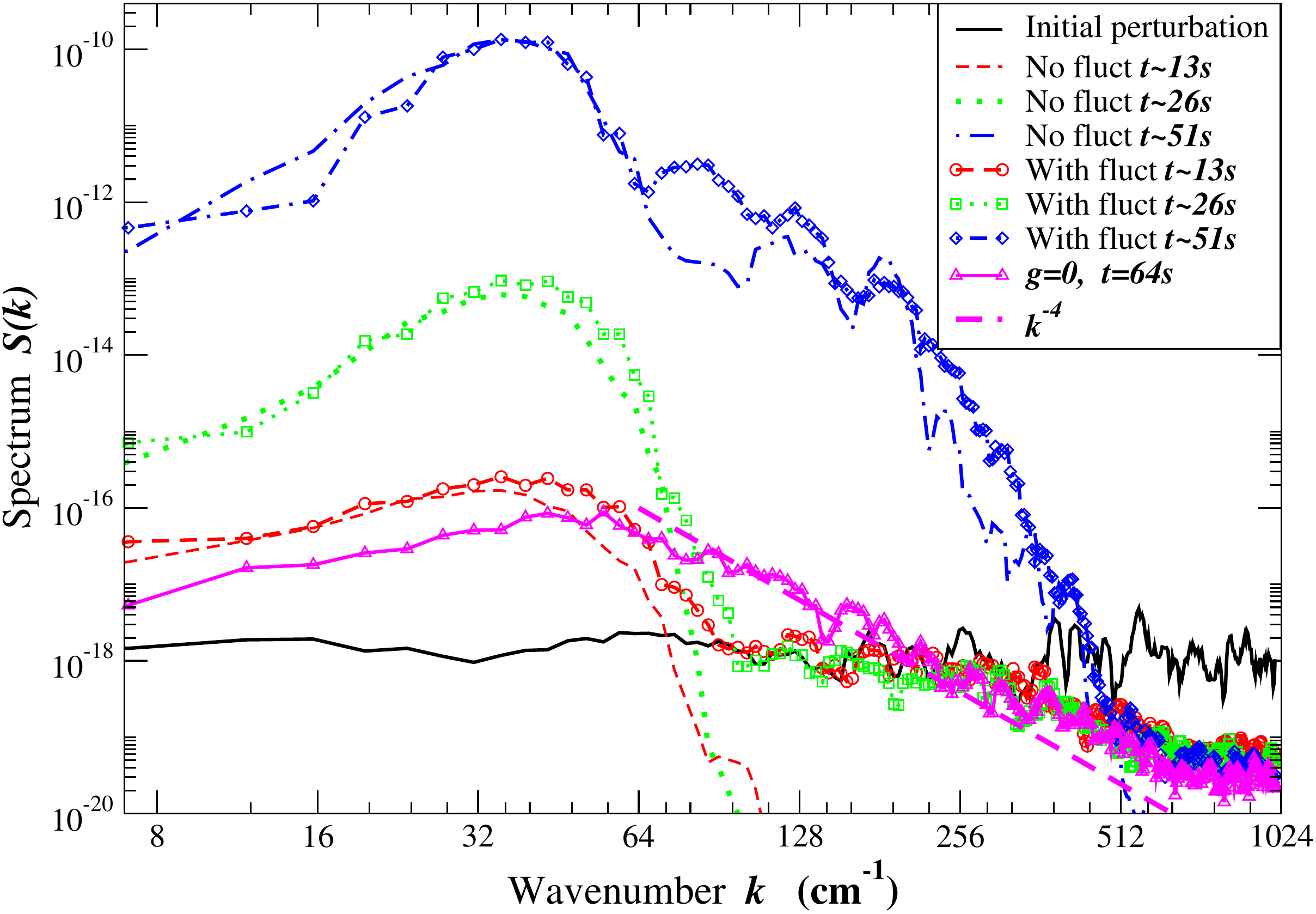}
\par\end{centering}

\caption{\label{fig:MMI-early}Early time development of a mixed-mode instability
instability in a Hele-Shaw cell. The spectrum $S(k)$ of the density
averaged along the vertical direction and the thickness of the sample
is shown as a function of wavenumber, at several points in time. The
same random initial perturbation is added to the initial condition
(black solid line) in all simulations. A first simulation is deterministic,
and clearly shows an unstable (exponential) growth for modes with
$k\lesssim64$, while larger wavenumbers decay in time. A second simulation
accounts for nonequilibrium thermal fluctuations, and displays giant
fluctuations in addition to the unstable growth. At late times nonlinearities
trigger all modes to show a nontrivial spectrum, not affected significantly
by the thermal fluctuations. Note that the most unstable (growing)
mode has wavelength of about $1.4$mm which compares very favorably
to the experimentally-reported 1.3mm \cite{MixedDiffusiveInstability}.
A third simulation is in microgravity ($g=0$) and shows a slow development
of a stable $k^{-4}$ giant fluctuation spectrum.}
\end{figure}

To get a more quantitative understanding of the development of the
instability, in Fig. \ref{fig:MMI-early} we show the Fourier spectrum
of the density $\bar{\rho}\left(x\right)$ averaged along the $y$
and $z$ directions, which is a measure of the fluctuations of the
diffusive ``interface.'' In these investigations we used a simulation
box of size $1.6\times0.2\times0.05$ (grid size $512\times64\times16$
cells), and set the initial mass fractions to be four times smaller
(note that this does not change the dimensionless number $R$ in \eqref{eq:R_def}),
$w_{1}^{0}=0.0216$ and $w_{2}^{0}=0.0342$, in order to slow down
the development of the instability and allow (giant) nonequilibrium
thermal fluctuations to develop. The remaining parameters were identical
to those reported above.

For a gravitationally-stable configuration, the spectrum of $\bar{\rho}\left(x\right)$,
called the static structure factor $S(k)$, exhibits a characteristic
giant fluctuation $k^{-4}$ power-law decay at large wavenumbers $k$
at long times \cite{FluctHydroNonEq_Book}, as illustrated in Fig.
\ref{fig:S_k_giant} by showing $S(k)$ for a simulation in which
gravity is switched off. At small wavenumbers, gravity and confinement
damp the amplitude and affect the dynamics of the giant fluctuations,
as is well-understood for stable steady states \cite{FluctHydroNonEq_Book,GiantFluctuations_Microgravity,GiantFluctuations_Cannell,GiantFluctuations_Theory,GiantFluctFiniteEffects}.
We emphasize that the thermal fluctuations in these investigations
are completely dominated by the nonequilibrium (giant) fluctuations
induced by the presence of large concentration gradients at the interface;
the equilibrium concentration fluctuations are many orders of magnitude
weaker in comparison. This is seen by the fact that we observe identical
results if we completely turn off the stochastic mass fluxes, and
keep \emph{only} the stochastic momentum flux (velocity fluctuations).
Similarly, turning off the stochastic momentum flux but keeping the
stochastic mass flux, in this particular setup, leads to very small
fluctuations that are below the tolerance of our linear solvers.

To our knowledge, nonequilibrium fluctuations have not been examined
in an unstable situation like the one we study here; some studies
have been carried out near the onset of the Rayleigh-Bernard thermal
convection instability \cite{RayleighBernard_Fluctuations,RayleighBernard_LLNS}.
It is important to emphasize that the very validity of fluctuating
hydrodynamics has \emph{not} been established for unstable flows;
one can justify linearized fluctuating hydrodynamics \cite{FluctHydroNonEq_Book,LebowitzHydroReview,MicroToSPDE_Review},
but in a linearization the fluctuations do \emph{not} affect the deterministic
flow and thus cannot trigger the instability. The results in Fig.
\ref{fig:MMI-early} show that in the absence of thermal fluctuations,
there is a clear band of wavenumbers $k\lesssim64$ whose amplitude
grows in time due to the instability, while the remaining modes decay
to zero rapidly. It is important to note that the most unstable wavelength
of $\lambda\approx2\pi/45\approx0.14$cm is very close to the experimentally
reported value of $0.13$cm \cite{MixedDiffusiveInstability}. When
thermal fluctuations are present, the same band of wavenumbers grows
in a similar manner, but at the same time, larger wavenumbers show
a nontrivial power-law spectrum. Eventually, however, the unstable
modes completely dominate the dynamics and there is essentially no
difference between the simulations with and without thermal fluctuations;
to make a more precise quantitative statement multiple Monte Carlo
simulations are required to perform ensemble averages and reduce the
statistical error, as well as to eliminate the effects of the boundary
conditions along the horizontal and vertical directions. Note that
at later times the fluid flow becomes chaotic and further information
may be gained by also examining the spectrum of the fluid velocity
and not just the vertical projection of the density; we leave such
detailed investigations for future studies.

\subsection{Diffusive-Layer Convection (DLC) instability}

In this section we use our algorithm to model a hypothetical experiment
in which the gravity points in the direction perpendicular to the
fluid-fluid interface and do the confining glass slips, as illustrated
in the right panel of Fig. \ref{fig:dlc_setup}. This kind of geometry
has already been used to experimentally investigate isothermal free
diffusive mixing in a \emph{binary} liquid mixture \cite{GiantFluctuations_Cannell}
in a \emph{stable} configuration %
\footnote{Note that for a binary mixture the only alternative to a stable configuration
is to have an RT-unstable configuration.%
}. To our knowledge, similar experiments have not been performed for
ternary mixtures, although it is feasible they could be if the initial
configuration can be prepared with sufficient control. 

To be specific, we select the initial concentrations of the sugar
and salt solutions to get a diffusive layer convection (DLC) instability;
all other parameters are the same as for the MMI setup summarized
above. The dimensionless parameter used for Fig. 1(c) in Ref. \cite{MixedDiffusiveInstability}
is 
\[
R=\frac{\alpha_{2}Z_{2}^{0}}{\alpha_{1}Z_{1}^{0}}=1.25,
\]
giving the required ratio of mass fractions,
\[
\frac{w_{1}^{0}}{w_{2}^{0}}=\frac{\alpha_{1}Z_{1}^{0}}{\alpha_{2}Z_{2}^{0}}\cdot\frac{\alpha_{2}M_{1}}{\alpha_{1}M_{2}}=\frac{1}{R}\cdot\frac{\alpha_{2}M_{1}}{\alpha_{1}M_{2}}=0.44.
\]
We set $w_{1}^{0}=0.022$ and $w_{2}^{0}=0.05$, which gives an initial
density of 1.014375 on the top and 1.018062 on the bottom, for a density
difference of about -0.4\% (recall that the sign of the difference
is opposite for DLC and MMI configurations).

The dimensions of the domain in our simulations are $1\times0.5\times1$,
with periodic boundary conditions in $x$ and $z$, and reservoir
(Dirichlet) boundary condition for concentration together with free-slip
boundary condition on velocity in the $y$ direction. The grid is
$256\times128\times256$ grid cells, and the time step size is fixed
at $\Delta t=0.025$, which corresponds to a maximum advective Courant
number of $\sim0.5$. We treat advection using the unlimited bilinear
BDS high-resolution scheme \cite{SemiLagrangianAdvection_3D} summarized
in Ref. \cite{LowMachImplicit}. The initial interface is now perfectly
flat, so that the instability is triggered by the nonequilibrium (giant)
thermal fluctuations. We focus our investigation here on the differences
in the spectrum of the thermal fluctuations with (unstable) and without
(marginally stable) gravity. Note that experimental measurements of
giant concentration fluctuations in microgravity are feasible and
have been performed for binary liquid mixtures \cite{FractalDiffusion_Microgravity};
microgravity measurements on ternary mixtures are ongoing or in the
planning stage \cite{DCMIX2}.

\begin{figure}[hb]
\includegraphics[width=3in]{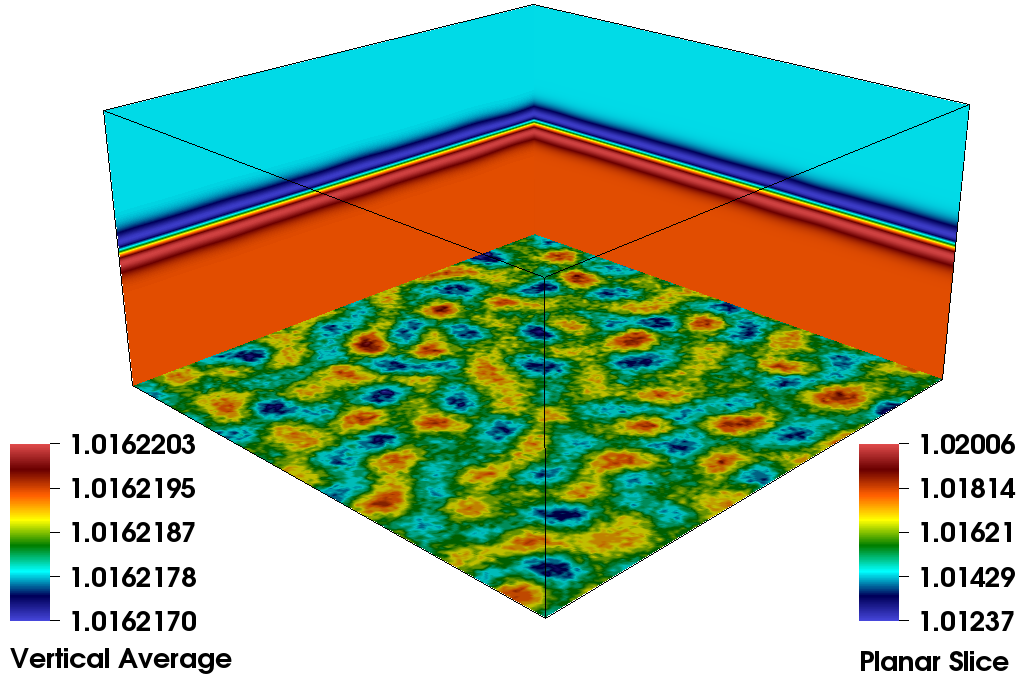}
\includegraphics[width=3in]{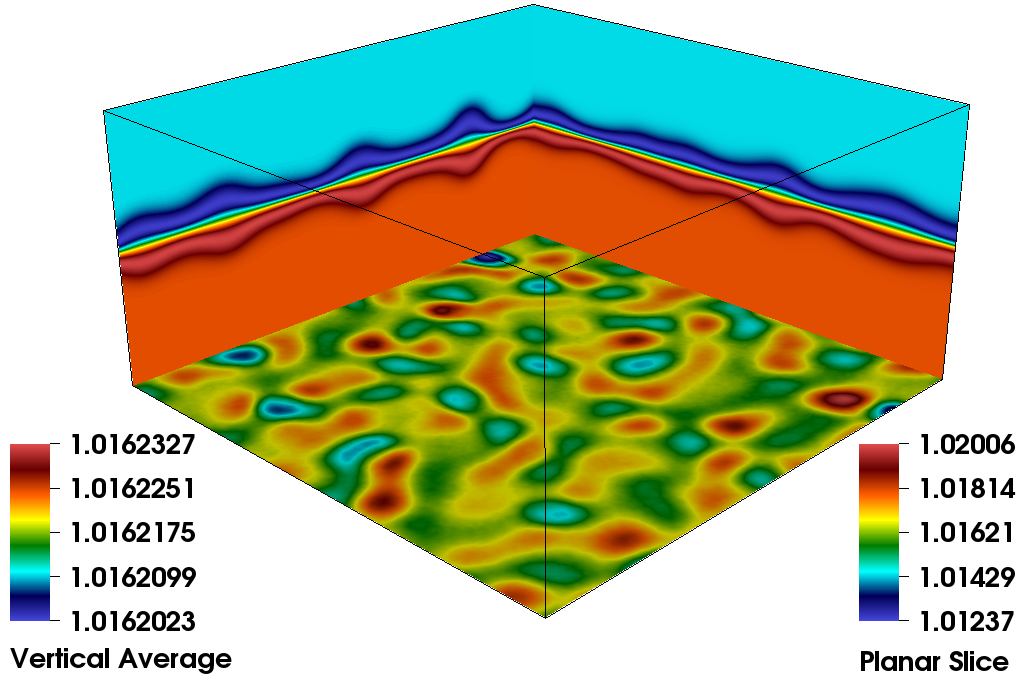}

\includegraphics[width=3in]{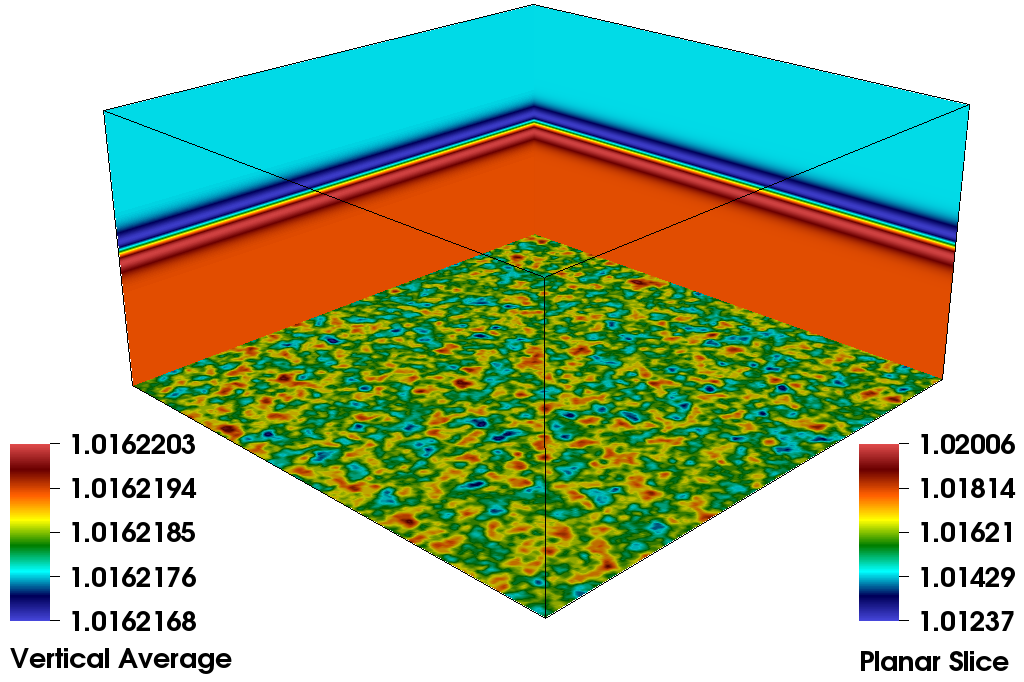}
\includegraphics[width=3in]{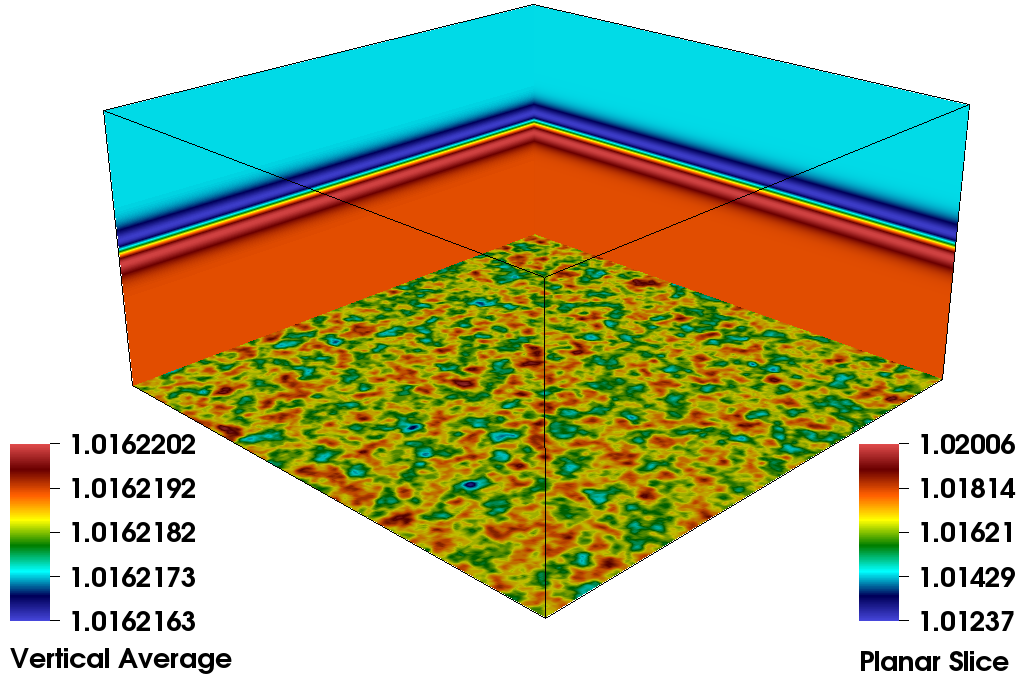}

\caption{\label{fig:DLC}Development of a fully three-dimensional diffusive
layer convection (DLC) instability as a layer of less-dense salty
water is placed on top of a horizontal layer of denser sweet water.
The color plots show the vertically averaged density $\rho$ (horizontal
planes) and planar slices of $\rho$ (vertical planes). The left panels
correspond to time $t=15$ since the initial contact between the miscible
liquids, and the right panels correspond to $t=19$. The top row of
panels is for a flat initial interface driven unstable by thermal
fluctuations in the presence of gravity, and the bottom row is in
microgravity ($g=0$).}
\end{figure}

In this geometry, the interface between the two fluids cannot be visualized,
rather, one sees the average index of refraction (a linear combination
of the average concentrations) along the thickness of the sample (direction
of the gradient and of gravity). In Fig. \ref{fig:DLC} we show color
plots for the vertically-averaged density at two points in time, one
as the instability is just beginning to dominate the dynamics, and
the other as the instability has fully developed. In the presence
of gravity, at early times there are pronounced giant thermal fluctuations
in addition to the fluctuations coming from the growth of the unstable
modes, which dominate the dynamics at late times similarly to the
MMI case in Fig. \ref{fig:MMI_development}. The giant nonequilibrium
fluctuations are more clearly visualized by considering stable free
diffusive mixing in microgravity ($g=0,$ bottom panels). 

\begin{figure}
\begin{centering}
\includegraphics[width=0.5\textwidth]{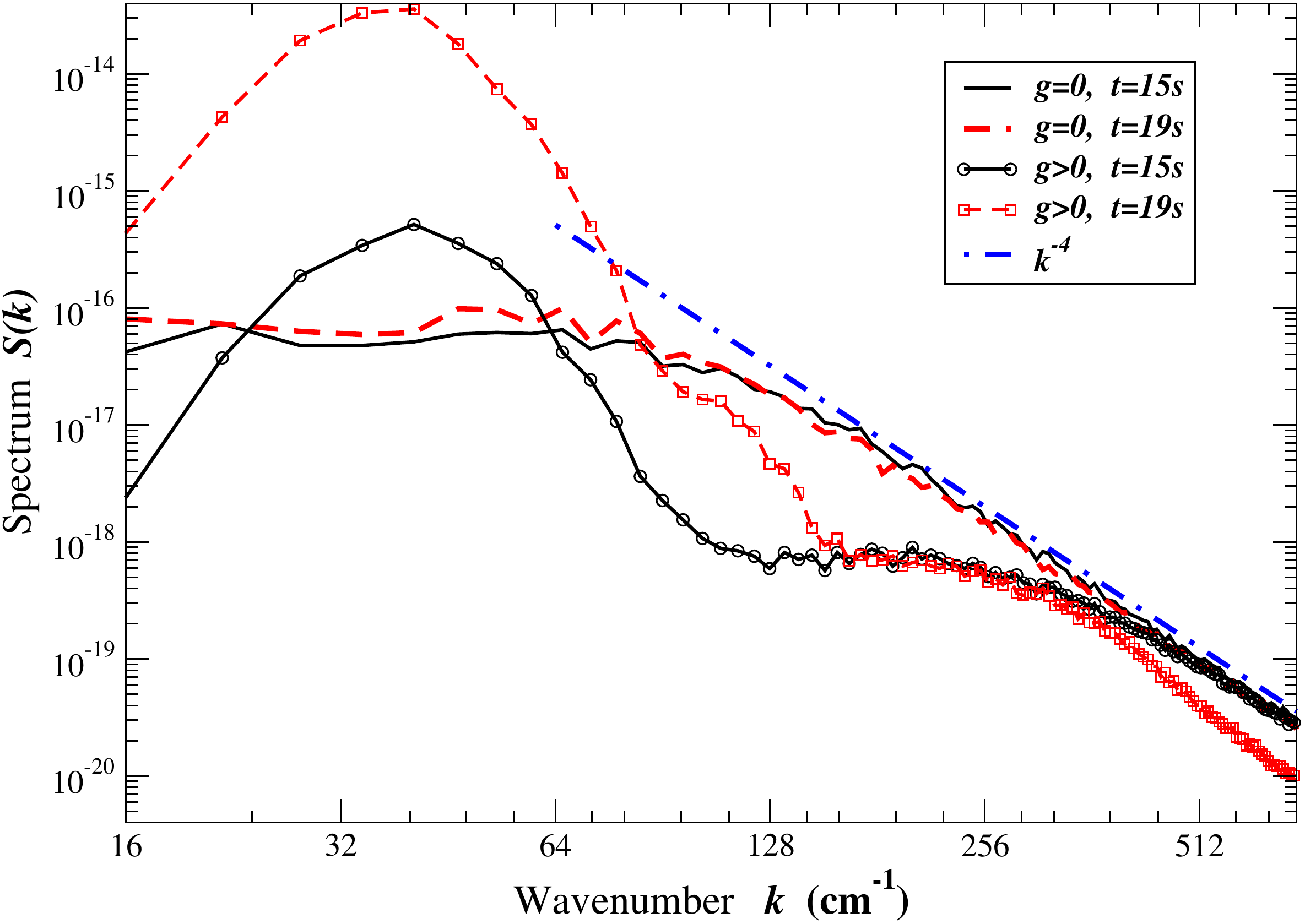}
\par\end{centering}

\caption{\label{fig:DLC_early}Radially averaged spectrum $S(k)$ of the vertically-averaged
density shown in the four panels in Fig. \ref{fig:DLC}. In Earth
gravity ($g>0$), the most unstable (growing) mode is seen to have
a wavelength of about $2\pi/32\approx0.2$cm, but there is in fact
a rather broad range of wavenumbers that are unstable. In microgravity
($g=0$) the deterministic state is stable but the fluctuations grow
to be ``giant'' with a characteristic $k^{-4}$ power spectrum.}
\end{figure}

In Fig. \ref{fig:DLC_early}, we show the radially-averaged power
spectrum $S(k=\norm{\V k})$ corresponding to the vertically-averaged
density $\bar{\rho}\left(x,z\right)$. Behavior similar to the MMI
instability shown in Fig. \ref{fig:MMI_development} is observed.
The band of wavenumbers between $16\lesssim k\lesssim128$ is seen
to grow in time, and the characteristic $k^{-4}$ giant fluctuation
spectrum is seen at large wavenumbers, especially clear in the results
for microgravity. Deterministic simulations (not shown) starting from
a randomly-perturbed interface show similar unstable growth and are
eventually indistinguishable from simulations in which the instability
is triggered and enhanced by thermal fluctuations.

We defer detailed studies of these phenomena for future work, in the
hope that our work will stimulate careful experimental investigations
that can directly be compared to our computer simulations. By adjusting
the initial concentrations and the choice of the two solutes one can
change the range of unstable wavenumbers and the time scale for the
development of the instability; this is best done by performing a
linear stability analysis \cite{DiffusiveInstability_Porous}. It
is feasible that for some choice of parameters a more substantial
interaction between the spectrum at large wavenumbers, dominated by
giant fluctuations, and the spectrum at small wavenumbers, dominated
by the instability, will occur. If this is the case one may be able
to observe a measurable influence of nonequilibrium concentration
fluctuations on the development and grown of the instability, allowing
us, for the first time, to experimentally access the validity of nonequilibrium
\emph{nonlinear} fluctuating hydrodynamics.

\section{Conclusions}

We have developed a low Mach number fluctuating hydrodynamics formulation
of momentum and mass transport in non-ideal mixtures of incompressible
liquids with given pure-component densities. In the present low Mach
number model energy transport is not modeled explicitly and the temperature
is assumed constant in time. Momentum transport is accounted for in
a quasi-incompressible framework in which pressure is in mechanical
(hydrostatic) equilibrium and fast sound waves are eliminated from
the system of equations because density instantaneously follows the
local composition. The thermodynamics of the mixture is described
by the Hessian of the normalized excess Gibbs energy per particle.
The transport properties are given by the shear viscosity, the Maxwell-Stefan
diffusion coefficients, and the thermal diffusion coefficients or
ratios, as a function of the composition of the mixture. 

We used our low Mach number algorithm to study the development of
gravitational instabilities during diffusive mixing in ternary mixtures.
These mixed mode and diffusive convection instabilities are specific
to mixtures of more than two species and occur because of an interaction
between the familiar buoyancy-driven Rayleigh-Taylor instability with
differential diffusion effects. Our simulations of the mixed-mode
instability closely mimic recent experiments \cite{MixedDiffusiveInstability}
performed in a Hele-Shaw setup, and we found good qualitative agreement
between experimental and computational results. A more quantitative
comparison is not possible at this stage and should be the subject
of future work. A particularly challenging aspect is to experimentally
control or measure the initial conditions with sufficient accuracy
to be able to directly compare simulations to experiments.

Because of the presence of sharp gradients at the interface between
the two diffusively mixing solutions, giant concentration fluctuations
develop in the form of power-law tails in the spectrum of the concentration
fluctuations. These nonequilibrium fluctuations are much larger than
equilibrium ones and have the potential to trigger and feed the instability
and affect the growth of the unstable structures. It is important
to observe that such a coupling of the fluctuations back to the macroscopic
dynamics requires nonlinearity, and is not possible in linearized
fluctuating hydrodynamics. While some nonlinear effects in fluctuating
hydrodynamics have been verified to occur in real liquids, it is not
clear whether our nonlinear fluctuating hydrodynamic simulations can
account for the effect of the fluctuations on the evolution of the
mean flow. In particular, the nonlinear fluctuating hydrodynamic equations
are ill-posed and some regularization, implicit in our finite-volume
discretization, is required to even give meaning to the equations
\cite{DiffusionJSTAT}. The conclusions of the simulations performed
here is that fluctuations, despite being ``giant'', are quickly
overwhelmed by the deterministic growth of the unstable modes. This
suggests that in actual experiments the development of the instability
is primarily triggered by the imperfections in the initial condition
or external fluctuations (e.g., vibrations).

In the Hele-Shaw geometry studied experimentally in \cite{MixedDiffusiveInstability},
the no-slip boundaries strongly damp the velocity fluctuations and
reduce the giant fluctuations. Furthermore, the quasi-two dimensional
geometry limits the possibilities for interactions between the fluctuations
and the instability. For this reason, we also reported simulations
of diffusive layer convection for a three-dimensional geometry, as
used in existing experiments in binary mixtures. Because giant fluctuations
develop on a slow diffusive time scale for large and intermediate
wavenumbers, it takes some time after the initial contact between
the two fluids for the power-law spectrum to develop. At the same
time, the unstable smaller wavenumbers have an exponentially-growing
amplitude, so that the instability can develop much faster than the
giant fluctuations. In our simulations we reduced the concentrations
of solutes to make the density difference very small and thus slow
down the unstable growth; it remains to be seen what is possible and
can be observed experimentally.

An important challenge for experimental physicists is to devise ways
to measure the thermophysical properties in multispecies mixtures
\cite{DCMIX2}. While our formulation can handle mixtures of arbitrary
numbers of species with essentially complete physical fidelity, it
is not possible to use our codes for realistic non-dilute mixtures
because many of the parameters, notably the thermodynamic factors
and the Maxwell-Stefan diffusion coefficients, are missing. We believe
that it will be necessary to use computer simulations using the types
of methods described here, together with Monte Carlo sampling of parameters,
and potentially also molecular dynamics calculations \cite{Ternary_MD,ThermodynamicEquations_MD},
\emph{in addition} to experimental measurement of observation functions,
in order to obtain robust estimates of thermophysical parameters with
a quantified uncertainty. The traditional approach of tabulating values
with error bars, that has worked for binary mixtures, fails for multispecies
mixtures due to proliferation of a larger number of parameters that
are not all independent but are constrained by thermodynamic symmetries
and stability conditions.

In future work, we will consider the inclusion of chemical reactions
in our low Mach number models. It is also possible to include thermal
effects in our formulation by accounting for the effects of thermal
expansion in the quasi-incompressibility constraint. Two key difficulties
are constructing a spatial discretization that ensures preservation
of an appropriately generalized EOS, as well as developing suitable
temporal integrators that can handle the multitude of time scales
that appear due the presence of slow mass, fast momentum, and intermediate
energy diffusion, and, potentially, ultrafast chemical reactions.
Another challenge for future work on low Mach number fluctuating hydrodynamics
is to account for the effects of surface tension in mixtures of immiscible
or partially miscible liquids. While some thermodynamically-consistent
constructions of diffuse-interface models exist for ternary mixtures
\cite{CHNS_Ternary}, we believe that much more work is required to
formulate a complete multispecies system of equations in the presence
of square-gradient terms in the free energy functional, even in the
isothermal setting.
\begin{acknowledgments}
We would like to thank Anne De Wit and Jorge Carballido Landeira for
numerous discussions regarding their experiments on gravitational
instabilities in ternary mixtures. This material is based upon work
supported by the U.S. Department of Energy Office of Science, Office
of Advanced Scientific Computing Research, Applied Mathematics program
under Award Number DE-SC0008271 and under contract No. DE-AC02-05CH11231.
\end{acknowledgments}
\appendix

\section*{Appendix}

\section{\label{sec:DiffusionMatrix}Computing the Diffusion Matrix}

An essential complication with treating all species equally, is that
the $N$ mass or mole fractions are not all independent but must sum
to unity. For example, the equations (\ref{eq:d_eq}) are not independent,
and one must supplement them with the condition that the total mass
flux is zero. Similarly, the rows and columns of $\M{\Lambda}$ sum
to zero, $\M{\Lambda}\V 1=\V 0$ where $\V 1$ denotes a vector of
ones, and therefore $\M{\Lambda}$ is not invertible; the range of
$\M{\Lambda}$ are vectors that sum to zero. To deal with these complications,
let us now introduce the projection matrix \cite{MulticomponentBook_Giovangigli}
\[
\M Q=\M I-\V 1\V w^{T},
\]
which satisfies $\M Q^{T}\V w=\M Q\V 1=\V 0$; therefore, pre-multiplying
a matrix by $\M Q^{T}$ ensures that the matrix has a range consisting
of vectors that sum to zero, and post-multiplying by $\M Q^{T}$ ensures
that the matrix has $\M w$ in its null-space. Now, let us define
a diffusion matrix as
\begin{equation}
\M{\chi}=\M Q\left[\M{\Lambda}+\alpha\V w\V w^{T}\right]^{-1}\M Q^{T},\label{eq:chi_def}
\end{equation}
where $\alpha\neq0$ is an arbitrary constant, for example, $\alpha=\text{Trace}\left(\M{\Lambda}\right)$;
an alternative equivalent formula is (\ref{eq:Lambda_to_chi}).

Here is a quick summary of some useful relations derived in \cite{IterativeChi_GG}:
\begin{enumerate}
\item $\M{\Lambda}\V 1=\V 0$ and $\M{\Lambda}$ is symmetric positive (semi)definite
(SPD) on $\V w^{\perp}$.
\item $\M{\chi}\V w=\V 0$ and $\M{\chi}$ is SPD on $\V 1^{\perp}$.
\item $\M{\chi}$ and $\M{\Lambda}$ are generalized inverses of each other,
$\M{\chi}\M{\Lambda}\M{\chi}=\M{\chi}$ and $\M{\Lambda}\M{\chi}\M{\Lambda}=\M{\Lambda}$,
more specifically, $\M{\Lambda}\M{\chi}=\M Q^{T}$ and $\M{\chi}\M{\Lambda}=\M Q$.
\item The SPD matrices
\[
\tilde{\M{\Lambda}}=\M{\Lambda}+\alpha\V w\V w^{T}=\tilde{\M{\chi}}^{-1}=\left(\M{\chi}+\alpha^{-1}\V 1\V 1^{T}\right)^{-1}
\]
are inverses of each other, and $\tilde{\M{\Lambda}}\equiv\M{\Lambda}$
on the subspace $\V 1^{\perp}$ and $\tilde{\M{\chi}}=\M{\chi}$ on
the subspace $\V w^{\perp}$.
\end{enumerate}
In order to numerically compute $\M{\chi}$ from $\M{\Lambda}$, one
could use a matrix inverse, as in (\ref{eq:Lambda_to_chi}). This
fails, however, when some of the species vanish since the matrix $\M{\Lambda}+\alpha\V w\V w^{T}$
is not invertible; this can be fixed by using a pseudoinverse, computed
via the singular value decomposition (SVD) of $\M{\Lambda}$. However,
both matrix inversion and SVD involve $O(N^{3})$ operations and can
be expensive for large numbers of species. An alternative is to use
an iterative numerical procedure \cite{IterativeChi_GG}
\begin{equation}
\M{\chi}=\lim_{M\rightarrow\infty}\M{\chi}_{M}=\lim_{M\rightarrow\infty}\left[\sum_{j=0}^{M}\left(\M Q\M N\right)^{j}\right]\M Q\M M^{-1}\M Q^{T},\label{eq:chi_iterative}
\end{equation}
where $\M N=\M I-\M M^{-1}\M{\Lambda}$, and $\M M$ is a diagonal
matrix with entries
\[
M_{ii}=\frac{x_{i}}{1-w_{i}}\sum_{j\neq i=1}^{N}\frac{x_{j}}{D_{ij}}.
\]
The sum converges rapidly so only a few terms in the sum are needed
to compute $\M{\chi}_{M}$ without having to do matrix inverses, but
the speed of convergence is hard to access \emph{a priori} and in
practice we set $M$ to a fixed integer such as $M=5$ or $M=10$.
It is important to note that, as proven in \cite{MulticomponentBook_Giovangigli},
the truncated sum to $M$ terms gives an approximation $\M{\chi}_{M}\approx\M{\chi}$
that is symmetric positive semi-definite, and satisfies the properties
$\M{\chi}_{M}\V w=\V 0$ as it must. It is therefore perfectly consistent
with thermodynamics to just approximate $\M{\chi}$ with $\M{\chi}_{M}$.
After all, since the Maxwell-Stefan diffusion coefficients are only
known to at most two decimal places in practice, it makes little sense
to invert $\M{\Lambda}$ exactly or compute its (expensive) SVD. Instead,
$\M{\chi}_{M}$ for small $M$ is in practice an equally good approximation
to the true diffusion matrix.

\section{\label{sec:Sedimentation}Static Equilibrium of a Dilute Solution
in Gravity}

Barodiffusion is often neglected for liquid mixtures since its contribution
to the diffusive fluxes is typically negligible compared to those
due to compositional or temperature gradients, unless there are very
large pressure gradients due to large accelerations, as in ultracentrifuges.
Nevertheless, including barodiffusion is crucial in order to obtain
the correct equilibrium steady state of a mixture in the presence
of gravity. This is a direct consequence of the fact that barodiffusion
has thermodynamic origin. If barodiffusion is neglected, at thermodynamic
equilibrium the mixture would mix uniformly even in the presence of
gravity, reaching the state of maximal entropy. However, physically,
we know that heavier species will migrate toward the bottom, minimizing
the free energy by balancing the gain in potential energy with the
loss of entropy. Here we show that once barodiffusion is accounted
for the hydrodynamic equations correctly reproduce the statistical
mechanics of systems in a gravitational field.

As a simple case in which we know the ``correct'' answer, let us
focus on the case when the mixture is a dilute suspension of a number
of solutes such as colloids or a macromolecule (e.g., solution of
sugar in water). In the dilute limit, the different solutes do not
affect each other, and we can, in fact, focus on one of the solutes
only and consider a binary solution. Let us take the first species
to be the solute and the second species to be the solvent. For a dilute
solution, $w_{1}\ll1$, $\M{\Gamma}\approx\M I$, $\bar{m}\approx m_{2}$,
$\rho\approx\rho_{2}\approx\bar{\rho}_{2}$, giving $x_{1}\approx\left(m_{2}/m_{1}\right)w_{1}$
and $\phi_{1}=w\rho/\bar{\rho}_{1}\approx w\bar{\rho}_{2}/\bar{\rho}_{1}$.
Note that hydrostatic equilibrium is built into the low Mach number
equations through the reference pressure, which satisfies $dP/dh=\rho g$,
where $h$ is the height. Thermodynamic equilibrium, i.e., a vanishing
of the chemical potential gradients, corresponds to a vanishing of
the diffusion driving force for the solute,
\begin{equation}
d_{1}=\frac{dx_{1}}{dh}+\frac{\left(\phi_{1}-w_{1}\right)}{nk_{B}T}\frac{dP}{dh}=0\approx\frac{m_{2}}{m_{1}}\frac{dw_{1}}{dh}+\frac{m_{2}g}{k_{B}T}\left(\frac{\bar{\rho}_{2}}{\bar{\rho}_{1}}-1\right)w_{1},\label{eq:thermo_eq_1}
\end{equation}
which directly gives the gravitational sedimentation profile of an
ideal gas with molecular mass $m_{e}$,
\begin{equation}
w_{1}=w_{1}^{0}\exp\left(\frac{m_{e}gh}{k_{B}T}\right),\label{eq:sediment_dilute}
\end{equation}
where $m_{e}$ is the excess mass of the colloids over that of the
fluid,
\[
m_{e}=\left(\bar{\rho}_{1}-\bar{\rho}_{2}\right)\frac{m_{1}}{\bar{\rho}_{1}}=\left(1-\frac{\bar{\rho}_{2}}{\bar{\rho}_{1}}\right)m_{1}.
\]
The result \eqref{eq:sediment_dilute} is in agreement with the statistical-mechanical
notion that the solute subsystem can be thought of as an ideal gas
of particles subject to the Archimedean gravity force $m_{e}g$.

A similar calculation for the case of thermodiffusion gives
\begin{equation}
w_{1}=w_{1}^{0}\exp\left(-\frac{D_{1}^{(T)}-D_{2}^{(T)}}{D_{12}}\,\frac{\grad T}{T}\, h\right),\label{eq:thermodiff_dilute}
\end{equation}
which is similar in form but does not have an origin in equilibrium
statistical mechanics.

\section{\label{sec:MS_Fluct}Fluctuating Maxwell-Stefan Description}

In this Appendix, we show that thermal fluctuations can be directly
added to the Maxwell-Stefan formulation in a manner that is intuitive
and consistent with the one we presented here, derived from the principles
of nonequilibrium thermodynamics.

In the Maxwell-Stefan description of diffusion, one considers a frictional
force between species $i$ and $j$ proportional to the difference
in the particular velocities of the two species, with friction coefficient
$\gamma_{ij}=x_{i}x_{j}/D_{ij}$. It is natural to include thermal
fluctuations in this description by including a fluctuating component
to the frictional force with covariance proportional to $\gamma_{ij}$,
following the traditional Langevin approach. This leads us to the
MS equation with fluctuations,
\begin{equation}
\V d_{i}=\sum_{j\neq i=1}^{N}\left[\gamma_{ij}\left(\V v_{i}-\V v_{j}\right)+\left(\frac{2}{n}\gamma_{ij}\right)^{\frac{1}{2}}\widetilde{\mathcal{Z}}_{ij}\right],\label{eq:d_SM-K-fluct}
\end{equation}
where $\widetilde{\mathcal{Z}}_{ij}\left(\V r,t\right)=-\widetilde{\mathcal{Z}}_{ji}\left(\V r,t\right)$
are space-time white-noise processes, one process per pair of species,
for a total of $N\left(N-1\right)/2$ random forces. From the above
we may augment \eqref{eq:d_def_MS} to account for fluctuations as
\[
\V d=-\rho^{-1}\M{\Lambda}\M W^{-1}\overline{\V F}-\frac{\grad T}{T}\V{\zeta}+\sqrt{\frac{2}{n}}\,\M K\,\widetilde{\V{\mathcal{Z}}},
\]
where $\V{\mathcal{Z}}$ is a vector of $N\left(N-1\right)/2$ independent
white noise processes and $\M K$ is a matrix that can directly be
read from \eqref{eq:d_SM-K-fluct}. Following the same linear algebra
steps as before, we can solve for the fluxes to obtain the additional
fluctuating contribution
\[
\widetilde{\V F}=-\sqrt{\frac{2}{n}}\,\rho\M W\M{\chi}\M K\,\widetilde{\V{\mathcal{Z}}}.
\]
The covariance of this stochastic flux is
\[
\av{\widetilde{\V F}\left(\V r,t\right)\widetilde{\V F}\left(\V r^{\prime},t^{\prime}\right)}=2\bar{m}\rho\left(\M W\M{\chi}\left(\M K\M K^{\star}\right)\M{\chi}\M W\right)\,\delta\left(\V r-\V r^{\prime}\right)\delta\left(t-t^{\prime}\right).
\]
It can be shown that $\M{\chi}\left(\M K\M K^{\star}\right)\M{\chi}=\M{\chi}$,
which shows that the covariance
\[
\av{\widetilde{\V F}\left(\V r,t\right)\widetilde{\V F}\left(\V r^{\prime},t^{\prime}\right)}=2\bar{m}\rho\left(\M W\M{\chi}\M W\right)\,\delta\left(\V r-\V r^{\prime}\right)\delta\left(t-t^{\prime}\right)=2k_{B}\M L\,\delta\left(\V r-\V r^{\prime}\right)\delta\left(t-t^{\prime}\right)
\]
is proportional to the Onsager matrix \eqref{eq:Onsager_L}, and therefore
the stochastic mass fluxes obtained from the fluctuating MS description
are statistically identical to \eqref{eq:stoch_flux}. This justifies
the prefactor $\sqrt{2/n}$ in the noise in \eqref{eq:d_SM-K-fluct};
one can also justify this factor based on kinetic theory considerations.

In this fluctuating MS description the stochastic mass flux is constructed
explicitly without using Cholesky factorization of the Onsager matrix.
That is, this construction gives a ``square root'' of the Onsager
matrix as an explicit construct,
\[
\M L_{\frac{1}{2}}=\left(\frac{\bar{m}\rho}{k_{B}}\right)^{\frac{1}{2}}\,\M W\M{\chi}\M K,
\]
which is different from \eqref{eq:sqrt_L}. Note, however, that this
formulation requires using $N\left(N-1\right)/2$ random processes
instead of only $N-1$ random processes required when \eqref{eq:sqrt_L}
is used; this increase in the number of random numbers is typically
not justified by the savings of a Cholesky factorization.

\section{\label{sub:S_k}Static and Dynamic Structure Factors}

In this secton we compute the equilibrium static \eqref{eq:S_k} and
dynamic \eqref{eq:S_k_w} structure factors for a mixture at equilibrium.
In linearized fluctuating hydrodynamics the coupling between the mass
(concentration) equations and the velocity equation disappears at
equilibrium, and we can focus our attention on the mass equations
\eqref{eq:species_eq}. When \eqref{eq:species_eq} is linearized
around a uniform state at thermodynamic equilibrium, we can omit the
advective term $\grad\cdot\left(\rho\V w\V v\right)$, and we can
treat density as constant.

Since we want to compute fluctuations in the mass fractions, we first
convert \eqref{eq:species_eq} to use gradients of mass fractions
instead of gradients of mole fractions, by using the chain rule
\[
\M{\Gamma}\grad\V x=\M{\Gamma}\left(\frac{\partial\V x}{\partial\V w}\right)\grad\V w=\frac{\bar{m}}{k_{B}T}\M W\left(\frac{\partial\V{\mu}}{\partial\V x}\right)\left(\frac{\partial\V x}{\partial\V w}\right)\grad\V w=\frac{\bar{m}}{k_{B}T}\M W\M{\Upsilon}\grad\V w.
\]
The symmetric Hessian matrix
\[
\M{\Upsilon}=\left(\frac{\partial\V{\mu}}{\partial\V x}\right)\left(\frac{\partial\V x}{\partial\V w}\right)=\left(\frac{\partial\V{\mu}}{\partial\V w}\right)=\frac{\partial^{2}g}{\partial\V w^{2}}
\]
can be related to $\V H$ using (\ref{eq:Gamma},\ref{eq:Gamma_F},\ref{eq:dx_dw}),
\[
\M{\Upsilon}=\frac{k_{B}T}{\bar{m}}\M W^{-1}\left[\left(\M X-\V x\V x^{T}\right)+\left(\M X-\V x\V x^{T}\right)\M H\left(\M X-\V x\V x^{T}\right)\right]\M W^{-1}.
\]
We can therefore write the mass flux due to composition gradients
in terms of mass fraction gradients as
\begin{equation}
\overline{\V F}=-\frac{\bar{m}\rho}{k_{B}T}\M W\M{\chi}\M W\M{\Upsilon}\grad\V w=-\rho\M M\grad\V w,\label{eq:F_comp_w}
\end{equation}
where
\[
\M M=\M W\M{\chi}\left[\left(\M X-\V x\V x^{T}\right)+\left(\M X-\V x\V x^{T}\right)\M H\left(\M X-\V x\V x^{T}\right)\right]\M W^{-1}.
\]

At thermodynamic equilibrium, the fluctuating diffusion equation (\ref{eq:species_eq})
can be expressed entirely in terms of the fluctuations of mass fractions
by using (\ref{eq:F_comp_w}),
\begin{equation}
\partial_{t}\left(\d{\V w}\right)=\M M\grad^{2}\left(\d{\V w}\right)+\sqrt{\frac{2}{n}}\,\M W\M{\chi}^{\frac{1}{2}}\grad\cdot\V{\mathcal{Z}}=\M M\grad^{2}\left(\d{\V w}\right)+\M N\grad\cdot\V{\mathcal{Z}},\label{eq:OU_w}
\end{equation}
where $\M N=\sqrt{2/n}\,\M W\M{\chi}^{\frac{1}{2}}$. After taking
a Fourier transform (\ref{eq:OU_w}) becomes a multi-dimensional Orstein-Uhlenbeck
(OU) equation, one system of $N$ equations for each wavenumber. The
dynamic structure factor is
\begin{equation}
\M S_{w}\left(\V k,t\right)=\exp\left(-\M Mk^{2}t\right)\,\M S_{w}\left(\V k\right).\label{eq:S_k_w_sol}
\end{equation}

A standard result for OU processes \cite{GardinerBook,LLNS_S_k} states
that at steady state the covariances satisfy the linear system of
equations
\[
\M M\M S_{w}+\M S_{w}\M M^{\star}=\M N\M N^{\star}.
\]
This equation needs to be supplemented by the conditions that the
static structure factor $\M S_{w}\left(\V k\right)$ is symmetric,
and that the row and column sums of $\M S_{w}$ are zero because the
mass fractions sum to one,
\[
\M S_{w}\V 1=\V 0.
\]
This system of two equations for $\M S_{w}$ has a unique solution;
after some algebraic manipulations the solution can be written in
the evidently symmetric form \eqref{eq:S_w}. For ideal mixtures $\M H=0$
and \eqref{eq:S_w} can be simplified to
\begin{equation}
\M S_{w}^{\text{id}}=\rho^{-1}\left(\M I-\V w\V 1^{T}\right)\M W\M M\left(\M I-\V 1\V w^{T}\right),\label{eq:S_w_id}
\end{equation}
which matches the corresponding expression for a mixture of ideal
gases \cite{Bell:09}.

In actual experiments, what can be measured through dynamic light
scattering or shadowgraphy is the spectrum of refractive index $n_{r}\left(\V w;P_{0},T_{0}\right)$.
A related quantity is the density structure factor $S_{\rho}=\av{\left(\widehat{\d{\rho}}\right)\left(\widehat{\d{\rho}}\right)^{\star}}$,
which can easily be obtained from $\M S_{w}$ in the low Mach setting
as follows. First, observe that in the low Mach number limit, density
fluctuations do not include the contribution from pressure fluctuations
(sound peaks in the static structure factor), rather, they only include
a contribution due to fluctuations in mass fractions (central peak)
\cite{LowMachExplicit}. Physically this corresponds to observing
density fluctuations at a longer time scale, i.e., averaging over
the fast pressure fluctuations on the sonic time scale. If we expand
the EOS constraint \eqref{eq:EOS} to account for small thermal fluctuations,
$\rho\leftarrow\rho+\d{\rho}$ and $w_{i}\leftarrow w_{i}+\d w_{i}$,
we get 
\[
\left(\rho+\d{\rho}\right)^{-1}-\rho^{-1}=\sum_{i}\frac{\left(w_{i}+\d w_{i}\right)}{\bar{\rho}_{i}}-\rho^{-1}\approx-\rho^{-2}\d{\rho}\approx\sum_{i}\frac{\d w_{i}}{\bar{\rho}_{i}},
\]
giving the density fluctuations
\begin{equation}
\d{\rho}\approx\rho^{2}\sum_{i}\frac{\d w_{i}}{\bar{\rho}_{i}}.\label{eq:drho_LM}
\end{equation}
From this relation we can derive all properties of the density fluctuations,
such as static and dynamic structure factors, from the corresponding
values for the mass fractions. For example, the dynamic or static
structure factor of density can be obtained from the corresponding
result for the mass fractions via \eqref{eq:S_rho}.


\begin{thebibliography}{10}

\bibitem{FluctHydroNonEq_Book}
J.~M.~O.~De Zarate and J.~V. Sengers.
\newblock {\em {Hydrodynamic fluctuations in fluids and fluid mixtures}}.
\newblock Elsevier Science Ltd, 2006.

\bibitem{Bell:09}
J.B. Bell, A.~Garcia, and S.~Williams.
\newblock Computational fluctuating fluid dynamics.
\newblock {\em ESAIM: M2AN}, 44(5):1085--1105, 2010.

\bibitem{DCMIX2}
Valentina Shevtsova, Cecilia Santos, Vitaliy Sechenyh, Jean~Claude Legros, and
  Aliaksandr Mialdun.
\newblock Diffusion and soret in ternary mixtures. preparation of the dcmix2
  experiment on the iss.
\newblock {\em Microgravity Science and Technology}, 25(5):275--283, 2014.

\bibitem{DiffusiveInstability_Porous}
Philip~MJ Trevelyan, Christophe Almarcha, and Anne De~Wit.
\newblock Buoyancy-driven instabilities of miscible two-layer stratifications
  in porous media and hele-shaw cells.
\newblock {\em Journal of fluid mechanics}, 670:38--65, 2011.

\bibitem{MixedDiffusiveInstability}
Jorge Carballido-Landeira, Philip~MJ Trevelyan, Christophe Almarcha, and Anne
  De~Wit.
\newblock Mixed-mode instability of a miscible interface due to coupling
  between rayleigh-taylor and double-diffusive convective modes.
\newblock {\em Physics of Fluids}, 25(2):024107, 2013.

\bibitem{DiffusiveInstability_Chemistry_PRL}
Christophe Almarcha, Philip~MJ Trevelyan, Patrick Grosfils, and Anne De~Wit.
\newblock Chemically driven hydrodynamic instabilities.
\newblock {\em Phys. Rev. Lett.}, 104(4):044501, 2010.

\bibitem{GiantFluctuations_Nature}
A.~Vailati and M.~Giglio.
\newblock {Giant fluctuations in a free diffusion process}.
\newblock {\em Nature}, 390(6657):262--265, 1997.

\bibitem{FractalDiffusion_Microgravity}
A.~Vailati, R.~Cerbino, S.~Mazzoni, C.~J. Takacs, D.~S. Cannell, and M.~Giglio.
\newblock {Fractal fronts of diffusion in microgravity}.
\newblock {\em Nature Communications}, 2:290, 2011.

\bibitem{MultispeciesCompressible}
K.~{Balakrishnan}, A.~L. {Garcia}, A.~{Donev}, and J.~B. {Bell}.
\newblock Fluctuating hydrodynamics of multispecies nonreactive mixtures.
\newblock {\em Phys. Rev. E}, 89:013017, 2014.

\bibitem{GiantFluctFiniteEffects}
Jose~Maria Ortiz~de Zarate, Jose~Antonio Fornes, and Jan~V. Sengers.
\newblock {Long-wavelength nonequilibrium concentration fluctuations induced by
  the Soret effect}.
\newblock {\em Phys. Rev. E}, 74:046305, Oct 2006.

\bibitem{LowMachExplicit}
A.~Donev, A.~J. Nonaka, Y.~Sun, T.~G. Fai, A.~L. Garcia, and J.~B. Bell.
\newblock {Low Mach Number Fluctuating Hydrodynamics of Diffusively Mixing
  Fluids}.
\newblock {\em Communications in Applied Mathematics and Computational
  Science}, 9(1):47--105, 2014.

\bibitem{LowMachImplicit}
A.~J. Nonaka, Y.~Sun, J.~B. Bell, and A.~Donev.
\newblock {Low Mach Number Fluctuating Hydrodynamics of Binary Liquid
  Mixtures}.
\newblock Submitted to CAMCOS, ArXiv preprint 1410.2300, 2015.

\bibitem{IrrevThermoBook_Mazur}
S.~R. DeGroot and P.~Mazur.
\newblock {\em Non-Equilibrium Thermodynamics}.
\newblock North-Holland Publishing Company, Amsterdam, 1963.

\bibitem{MulticomponentBook_KT}
Ross Taylor and R~Krishna.
\newblock {\em Multicomponent mass transfer}.
\newblock Wiley, New York, 1993.

\bibitem{OttingerBook}
H.~C. {\"O}ttinger.
\newblock {\em Beyond equilibrium thermodynamics}.
\newblock Wiley Online Library, 2005.

\bibitem{IrrevThermoBook_Kuiken}
Gerard D.~C. Kuiken.
\newblock {\em Thermodynamics of Irreversible Processes: Applications to
  Diffusion and Rheology}.
\newblock Wiley, 1994.

\bibitem{NonEqThermo_Bedeaux}
S.~Kjelstrup, D.~Bedeaux, E.~Johannessen, and J.~Gross.
\newblock {\em Non-Equilibrium Thermodynamics for Engineers}.
\newblock World Scientific Pub Co Inc, 2010.

\bibitem{MulticomponentBook_Giovangigli}
V.~Giovangigli.
\newblock {\em Multicomponent Flow Modeling}.
\newblock Birkhauser Boston, 1999.

\bibitem{MulticomponentNonideal_GG}
Vincent Giovangigli and Lionel Matuszewski.
\newblock Mathematical modeling of supercritical multicomponent reactive
  fluids.
\newblock {\em Mathematical Models and Methods in Applied Sciences},
  23(12):2193--2251, 2013.

\bibitem{GENERIC_Mixtures}
H.C. {\"O}ttinger.
\newblock {Constraints in nonequilibrium thermodynamics: General framework and
  application to multicomponent diffusion}.
\newblock {\em J. Chem. Phys.}, 130:114904, 2009.

\bibitem{Cahn-Hilliard_QuasiIncomp}
J.~Lowengrub and L.~Truskinovsky.
\newblock {Quasi-incompressible Cahn-Hilliard fluids and topological
  transitions}.
\newblock {\em Proceedings of the Royal Society of London A}, 454(1978):2617,
  1998.

\bibitem{AnelasticApproximation}
D.R. Durran.
\newblock Improving the anelastic approximation.
\newblock {\em J. Atmos. Sci}, 46(11):1453--1461, 1989.

\bibitem{LowMachAdaptive}
R.B. Pember, L.H. Howell, J.B. Bell, P.~Colella, W.Y. Crutchfield, W.A.
  Fiveland, and J.P. Jessee.
\newblock {An adaptive projection method for unsteady, low-Mach number
  combustion}.
\newblock {\em Combustion Science and Technology}, 140(1-6):123--168, 1998.

\bibitem{CHNS_Ternary}
Junseok Kim, Kyungkeun Kang, John Lowengrub, et~al.
\newblock Conservative multigrid methods for ternary cahn-hilliard systems.
\newblock {\em Communications in Mathematical Sciences}, 2(1):53--77, 2004.

\bibitem{Landau:Fluid}
L.D. Landau and E.M. Lifshitz.
\newblock {\em Fluid Mechanics}, volume~6 of {\em Course of Theoretical
  Physics}.
\newblock Pergamon Press, Oxford, England, 1959.

\bibitem{MS_diffusion_ionic}
Xin Liu, Thijs~JH Vlugt, and Andr{\'e} Bardow.
\newblock Maxwell--stefan diffusivities in binary mixtures of ionic liquids
  with dimethyl sulfoxide (dmso) and h2o.
\newblock {\em The Journal of Physical Chemistry B}, 115(26):8506--8517, 2011.

\bibitem{Diffusion_InfiniteDilution}
Xin Liu, Andre{\'e} Bardow, and Thijs~JH Vlugt.
\newblock Multicomponent maxwell- stefan diffusivities at infinite dilution.
\newblock {\em Industrial \& Engineering Chemistry Research}, 50(8):4776--4782,
  2011.

\bibitem{Darken_MS_diffusion}
Xin Liu, Thijs~JH Vlugt, and Andre{\'e} Bardow.
\newblock {Predictive Darken equation for Maxwell-Stefan diffusivities in
  multicomponent mixtures}.
\newblock {\em Industrial \& Engineering Chemistry Research},
  50(17):10350--10358, 2011.

\bibitem{MS_diffusion_NMR}
Andr{\'e} Bardow, Ernesto Kriesten, Mihai~Adrian Voda, Federico Casanova,
  Bernhard Bl{\"u}mich, and Wolfgang Marquardt.
\newblock Prediction of multicomponent mutual diffusion in liquids: Model
  discrimination using nmr data.
\newblock {\em Fluid Phase Equilibria}, 278(1):27--35, 2009.

\bibitem{DiffusionMatrix_Gases}
VI~Kurochkin, SF~Makarenko, and GA~Tirskii.
\newblock Transport coefficients and the onsager relations in the kinetic
  theory of dense gas mixtures.
\newblock {\em Journal of Applied Mechanics and Technical Physics},
  25(2):218--225, 1984.

\bibitem{Flames_Giovangigli}
Vincent Giovangigli, Lionel Matuszewski, and Francis Dupoirieux.
\newblock {Detailed modeling of planar transcritical $H_2-O_2-N_2$ flames}.
\newblock {\em Combustion Theory and Modelling}, 15(2):141--182, 2011.

\bibitem{TernaryEquilibriumFluct}
Jos{\'e}~M Ortiz~de Z{\'a}rate, Jorge~Luis Hita, and Jan~V Sengers.
\newblock Fluctuating hydrodynamics and concentration fluctuations in ternary
  mixtures.
\newblock {\em Comptes Rendus M{\'e}canique}, 2013.

\bibitem{LLNS_Staggered}
F.~Balboa Usabiaga, J.~B. Bell, R.~Delgado-Buscalioni, A.~Donev, T.~G. Fai,
  B.~E. Griffith, and C.~S. Peskin.
\newblock {Staggered Schemes for Fluctuating Hydrodynamics}.
\newblock {\em SIAM J. Multiscale Modeling and Simulation}, 10(4):1369--1408,
  2012.

\bibitem{BDS}
J.~B. Bell, C.~N. Dawson, and G.~R. Shubin.
\newblock An unsplit, higher order {G}odunov method for scalar conservation
  laws in multiple dimensions.
\newblock {\em J. Comput. Phys.}, 74:1--24, 1988.

\bibitem{MultiscaleIntegrators}
S.~Delong, Y.~Sun, , B.~E. Griffith, E.~Vanden-Eijnden, and A.~Donev.
\newblock {Multiscale temporal integrators for fluctuating hydrodynamics}.
\newblock {\em Phys. Rev. E}, 90:063312, 2014.

\bibitem{StokesKrylov}
M.~Cai, A.~J. Nonaka, J.~B. Bell, B.~E. Griffith, and A.~Donev.
\newblock {Efficient Variable-Coefficient Finite-Volume Stokes Solvers}.
\newblock {\em Comm. in Comp. Phys. (CiCP)}, 16(5):1263--1297, 2014.

\bibitem{Ternary_MD}
Xin Liu, Ana Mart{\'\i}n-Calvo, Erin McGarrity, Sondre~K Schnell, Sof{\'\i}a
  Calero, Jean-Marc Simon, Dick Bedeaux, Signe Kjelstrup, Andre{\'e} Bardow,
  and Thijs~JH Vlugt.
\newblock Fick diffusion coefficients in ternary liquid systems from
  equilibrium molecular dynamics simulations.
\newblock {\em Industrial \& Engineering Chemistry Research},
  51(30):10247--10258, 2012.

\bibitem{Ternary_MD_correction}
Xin Liu, Sondre~K Schnell, Jean-Marc Simon, Dick Bedeaux, Signe Kjelstrup,
  Andre{\'e} Bardow, and Thijs~JH Vlugt.
\newblock Correction to fick diffusion coefficients of liquid mixtures directly
  obtained from equilibrium molecular dynamics.
\newblock {\em The Journal of Physical Chemistry B}, 116(20):6070--6070, 2012.

\bibitem{ThermodynamicsFluctuations_MD}
Sondre~K Schnell, Xin Liu, Jean-Marc Simon, Andre{\'e} Bardow, Dick Bedeaux,
  Thijs~JH Vlugt, and Signe Kjelstrup.
\newblock Calculating thermodynamic properties from fluctuations at small
  scales.
\newblock {\em The Journal of Physical Chemistry B}, 115(37):10911--10918,
  2011.

\bibitem{DFDB}
S.~Delong, B.~E. Griffith, E.~Vanden-Eijnden, and A.~Donev.
\newblock {Temporal Integrators for Fluctuating Hydrodynamics}.
\newblock {\em Phys. Rev. E}, 87(3):033302, 2013.

\bibitem{LLNS_S_k}
A.~Donev, E.~Vanden-Eijnden, A.~L. Garcia, and J.~B. Bell.
\newblock {On the Accuracy of Explicit Finite-Volume Schemes for Fluctuating
  Hydrodynamics}.
\newblock {\em Communications in Applied Mathematics and Computational
  Science}, 5(2):149--197, 2010.

\bibitem{SoretDiffusion_Croccolo}
F.~Croccolo, H.~Bataller, and F.~Scheffold.
\newblock {A light scattering study of non equilibrium fluctuations in liquid
  mixtures to measure the Soret and mass diffusion coefficient}.
\newblock {\em J. Chem. Phys.}, 137:234202, 2012.

\bibitem{GiantFluct_Barodiffusion}
PN~Segr{\`e} and JV~Sengers.
\newblock Nonequilibrium fluctuations in liquid mixtures under the influence of
  gravity.
\newblock {\em Physica A: Statistical Mechanics and its Applications},
  198(1):46--77, 1993.

\bibitem{GiantFluctuations_Theory}
A.~Vailati and M.~Giglio.
\newblock {Nonequilibrium fluctuations in time-dependent diffusion processes}.
\newblock {\em Phys. Rev. E}, 58(4):4361--4371, 1998.

\bibitem{GiantFluctuations_Cannell}
F.~Croccolo, D.~Brogioli, A.~Vailati, M.~Giglio, and D.~S. Cannell.
\newblock Nondiffusive decay of gradient-driven fluctuations in a
  free-diffusion process.
\newblock {\em Phys. Rev. E}, 76(4):041112, 2007.

\bibitem{GiantFluctuations_Microgravity}
A.~Vailati, R.~Cerbino, S.~Mazzoni, M.~Giglio, G.~Nikolaenko, C.J. Takacs, D.S.
  Cannell, W.V. Meyer, and A.E. Smart.
\newblock {Gradient-driven fluctuations experiment: fluid fluctuations in
  microgravity}.
\newblock {\em Applied Optics}, 45(10):2155--2165, 2006.

\bibitem{RayleighBernard_Fluctuations}
M.~Wu, G.~Ahlers, and D.S. Cannell.
\newblock Thermally induced fluctuations below the onset of
  {Rayleigh}-{B\'enard} convection.
\newblock {\em Phys. Rev. Lett.}, 75(9):1743--1746, 1995.

\bibitem{RayleighBernard_LLNS}
J.~M. Ortiz~de Z{\'a}rate, F.~Peluso, and J.~V. Sengers.
\newblock {Nonequilibrium fluctuations in the Rayleigh-B{\'e}nard problem for
  binary fluid mixtures}.
\newblock {\em Euro. Phys. J. E}, 15(3):319--333, 2004.

\bibitem{LebowitzHydroReview}
J.~L. Lebowitz, E.~Presutti, and H.~Spohn.
\newblock {Microscopic models of hydrodynamic behavior}.
\newblock {\em J. Stat. Phys.}, 51(5):841--862, 1988.

\bibitem{MicroToSPDE_Review}
Giambattista Giacomin, Joel~L Lebowitz, and Errico Presutti.
\newblock Deterministic and stochastic hydrodynamic equations arising from
  simple microscopic model systems.
\newblock In {\em Stochastic Partial Differential Equations: Six Perspectives},
  number~64 in Mathematical Surveys and Monographs, page 107. American
  Mathematical Soc., 1999.

\bibitem{SemiLagrangianAdvection_3D}
Andy Nonaka, S~May, Ann~S Almgren, and John~B Bell.
\newblock A three-dimensional, unsplit godunov method for scalar conservation
  laws.
\newblock {\em SIAM Journal on Scientific Computing}, 33(4):2039--2062, 2011.

\bibitem{DiffusionJSTAT}
A.~Donev, T.~G. Fai, and E.~Vanden-Eijnden.
\newblock {A reversible mesoscopic model of diffusion in liquids: from giant
  fluctuations to Fick's law}.
\newblock {\em Journal of Statistical Mechanics: Theory and Experiment},
  2014(4):P04004, 2014.

\bibitem{ThermodynamicEquations_MD}
S.~Kjelstrup, D.~Bedeaux, I.~Inzoli, and J.M. Simon.
\newblock {Criteria for validity of thermodynamic equations from
  non-equilibrium molecular dynamics simulations}.
\newblock {\em Energy}, 33(8):1185--1196, 2008.

\bibitem{IterativeChi_GG}
Vincent Giovangigli.
\newblock Convergent iterative methods for multicomponent diffusion.
\newblock {\em IMPACT of Computing in Science and Engineering}, 3(3):244--276,
  1991.

\bibitem{GardinerBook}
C.~W. Gardiner.
\newblock {\em {Handbook of stochastic methods: for physics, chemistry \& the
  natural sciences}}, volume Vol. 13 of {\em Series in synergetics}.
\newblock Springer, third edition, 2003.

\end{thebibliography}

\end{document}